\documentclass[fleqn,usenatbib]{mnras}

\usepackage{newtxtext,newtxmath}
\usepackage[T1]{fontenc}

\DeclareRobustCommand{\VAN}[3]{#2}
\let\VANthebibliography\thebibliography
\def\thebibliography{\DeclareRobustCommand{\VAN}[3]{##3}\VANthebibliography}

\usepackage{graphicx}	% Including figure files
\usepackage{amsmath}	% Advanced maths commands
\usepackage{multirow}   % multirow tables
\usepackage{soul}
\RequirePackage{color}\definecolor{GREEN}{rgb}{0,0.45,0} %DIF PREAMBLE

\title[IGRINS Doppler imaging of WISE 1049AB]{Global weather map reveals persistent top-of-atmosphere features on the nearest brown dwarfs}

\author[X. Chen et al.]{
Xueqing Chen,$^{1,2}$\thanks{E-mail: xueqing.chen@ed.ac.uk}
Beth A. Biller,$^{1,2}$
Johanna M. Vos,$^{3}$
Ian J. M. Crossfield,$^{4}$
Gregory N. Mace,$^{5}$
\newauthor
Callie E. Hood,$^{6}$
Xianyu Tan,$^{7}$
Katelyn N. Allers$^{8}$,
Emily C. Martin$^{6}$,
Emma Bubb,$^{1,2}$
Jonathan J. Fortney,$^{6}$
\newauthor
Caroline V. Morley,$^{5}$
Mark Hammond$^{9}$
\\
$^{1}$Institute for Astronomy, University of Edinburgh, Blackford Hill, Edinburgh EH9 3HJ, UK\\
$^{2}$Centre for Exoplanet Science, University of Edinburgh, Edinburgh EH8 9YL, UK\\
$^{3}$School of Physics, Trinity College Dublin, The University of Dublin, Dublin 2, Ireland\\
$^{4}$Department of Physics and Astronomy, University of Kansas, 1082 Malott, 1251 Wescoe Hall Dr., Lawrence, KS 66045, USA\\
$^{5}$Department of Astronomy, University of Texas at Austin, Austin, TX 78712, USA\\
$^{6}$Department of Astronomy and Astrophysics, University of California, Santa Cruz, CA 95064, USA\\
$^{7}$Tsung-Dao Lee Institute \& School of Physics and Astronomy, Shanghai Jiao Tong University, Shanghai 201210, China\\
$^{8}$Department of Physics and Astronomy, Bucknell University, Lewisburg, PA 17837, USA\\
$^{9}$Atmospheric, Oceanic and Planetary Physics, Department of Physics, University of Oxford, Oxford, UK\\
}

\date{Accepted XXX. Received YYY; in original form ZZZ}

\pubyear{2024}

\begin{document}
\label{firstpage}
\pagerange{\pageref{firstpage}--\pageref{lastpage}}
\maketitle

\begin{abstract}
Brown dwarfs and planetary-mass companions display rotationally modulated photometric variability, especially those near the L/T transition. This variability is commonly attributed to top-of-atmosphere (TOA) inhomogeneities, with proposed models including patchy thick and thin clouds, planetary-scale jets, or chemical disequilibrium.
Surface mapping techniques are powerful tools to probe their atmospheric structures and distinguish between models. 
One of the most successful methods for stellar surface mapping is Doppler imaging, where the existence of TOA inhomogeneities can be inferred from their varying Doppler shifts across the face of a rotating star.
We applied Doppler imaging to the nearest brown dwarf binary WISE 1049AB (aka Luhman 16AB) using time-resolved, high-resolution spectroscopic observations from Gemini IGRINS, and obtained for the first time H and K band simultaneous global weather map for brown dwarfs. 
Compared to the only previous Doppler map for a brown dwarf in 2014 featuring a predominant mid-latitude cold spot on WISE 1049B and no feature on WISE 1049A, our observations detected persistent spot-like structures on WISE 1049B in the equatorial to mid-latitude regions on two nights, and revealed new polar spots on WISE 1049A. 
Our results suggest stability of atmospheric features over timescale of days and possible long-term stable or recurring structures.
H and K band maps displayed similar structures in and out of CO bands, indicating the cold spots not solely due to chemical hotspots but must involve clouds.
Upcoming 30-m extremely large telescopes (ELTs) will enable more sensitive Doppler imaging of dozens of brown dwarfs and even a small number of directly-imaged exoplanets.
\end{abstract}

\begin{keywords}
planets and satellites: atmospheres --
techniques: spectroscopic --
stars: variables --
brown dwarfs
\end{keywords}

%%%%%%%%%%%%%%%%%%%%%%%%%%%%%%%%%%%%%%%%%%%%%%%%%%

%%%%%%%%%%%%%%%%% BODY OF PAPER %%%%%%%%%%%%%%%%%%

\section{Introduction}

%L/T transition brown dwarfs

As our understanding of exoplanets expands, a key focus has been to investigate the structures and dynamics of their atmospheres. With current and forthcoming telescopes such as the James Webb Space Telescope (JWST) and 30-m class telescopes, direct imaging spectroscopy has the potential to directly measure exoplanetary atmospheres. Currently, brown dwarfs serve as analogs to directly-imaged exoplanets due to their comparable observed properties such as mass, gravity, temperature, and near-IR color. Since there is no need to overcome light from the bright host star, isolated brown dwarfs have been observed in much greater quality and quantity than high-contrast exoplanet companions. These substellar objects lack the mass to sustain nuclear fusion, so they cool along spectral types M, L, T, and Y as they age, resulting in the formation of condensate clouds and complex chemistry in their atmospheres \citep{Kirkpatrick2005}. 

Photometric and spectroscopic observations have revealed that brown dwarfs and directly-imaged planetary-mass objects undergo an abrupt color change when they transition from the redder L type to the bluer T type within a narrow range of effective temperatures at the L/T transition (e.g. \citealt{Dupuy2012}). 
Additionally, rotationally modulated variability in photometric light curves is commonly observed in time-resolved observations of brown dwarfs and directly-imaged planetary-mass objects (e.g. \citealt{Artigau2009, Radigan2012, Heinze2013, Buenzli2014, Buenzli2015, Lew2016, Apai2017, Biller2018, Vos2019}), with an enhanced level of variability observed specifically during the L/T transition \citep{Radigan2014, Liu2023}. This suggests the presence of heterogeneous top-of-atmosphere (TOA) structures, particularly at the L/T transition.
The most widely accepted explanation is the presence of thick and thin patchy silicate clouds \citep{Ackerman2001, Marley2010, Apai2013}, leading to variations in the TOA brightness as the object rotates. Temperature variations driven by convective perturbations \citep{zhang2014,showman2019,tan2022,hammond2023} and large-scale waves/vortices driven by cloud radiative feedback \citep{Tan2021b,Tan2021a} have also been proposed as dynamical mechanisms responsible for the surface inhomogeneities and their time evolution. 
However, alternative cloudless models have also been proposed which consider non-equilibrium chemistry and temperature variations (e.g. \citealt{Tremblin2016, Tremblin2020}) as potential drivers of the observed variability.

% spectral mapping
Surface mapping techniques offer a new tool for constraining the top-of-atmosphere (TOA) structures from the rotationally modulated variability.
\cite{Apai2013} developed the 
\textsc{Stratos} mapping routine that models the 2D TOA structure of brown dwarfs by fitting elliptical spots to rotationally modulated light curves. They found that at least three spots covering 20\%-30\% of the surface are required to explain the HST light curves of the L/T transition dwarfs 2MASS 2139 and SIMP 0136. 
\cite{Karalidi2015} developed \textsc{Aeolus}, a Markov-chain Monte Carlo code that maps the surface of a brown dwarf with parametrized elliptical spots. The code was validated by successfully reproducing the main spot features on Jupiter with HST light curves, and was also applied on 2M2139 and SIMP0136, achieving similar results as \cite{Apai2013}. \cite{Karalidi2016} further used \textsc{Aeolus} to produce a map of the benchmark brown dwarf binary WISE 1049AB with HST light curves and retrieved 3-4 spots in the TOA of both A and B component with $\sim$200 K temperature difference from the background. 
Apart from maps with only spot-like features, maps including both spots and banded structures such as planetary-scale waves have also been explored in \cite{Apai2017, Apai2021}.

Doppler imaging is one of the most powerful techniques to infer surface maps of rotating objects. Methods based on just photometric light curves lose all latitudinal information, hence often suffer from high degeneracy. Doppler imaging, on the other hand, utilizes individual line shapes in a time series 
of high-resolution spectra, which encodes more information about the TOA brightness distribution. 
This technique takes advantage of the changes in absorption line shapes due to varying Doppler shifts across the face of a rotating object.
When inhomogeneous features (e.g. a dark spot) on the TOA rotate in and out of view, the corresponding line from that patch will first be blue-shifted and then red-shifted. This causes the disk-integrated, Doppler-broadened spectral lines to change shape over time, which can be captured with a time series of high-resolution spectra (e.g. R $\gtrsim$ 50,000). This technique has long been used to map stellar spots (e.g. \citealt{Vogt1987, CollierCameron1995, Hatzes1998, Strassmeier2009, Roettenbacher2017}), but to date, it has only been successfully applied to one brown dwarf, WISE 1049B \citep{Crossfield2014}. The only Doppler imaging attempt in the literature for another object, 2MASSW J0746425+200032AB, resulted in a non-detection of coherent TOA structure \citep{Wang2017}. In the upcoming era of 30-meter telescopes, where detection sensitivity will be high enough to directly map more brown dwarfs and even a few giant exoplanets, Doppler imaging is expected to become one of the most powerful techniques to characterize their atmospheres and constrain our general circulation models of these objects \citep{Crossfield2014a, Snellen2014, Plummer2023}. 

% target description
WISE J104915.57-531906.1AB (also known as Luhman 16AB, \citealt{Luhman_2013}; hereafter WISE 1049AB) is a benchmark binary brown dwarf system for Doppler imaging studies. With a distance of 1.998±0.0004 pc \citep{Sahlmann2015}, they are the closest and brightest brown dwarfs to Earth. The A and B components span the L/T transition, with spectral types L7.5±1 and T0.5±1 respectively \citep{Burgasser2013}. Since both A and B appear to be typical L/T transition objects given their spectra, they offer a unique case for studying heterogeneous atmospheric structures and comparing different phases in the transition between spectral types.

Due to its proximity and brightness, WISE 1049AB has been the subject of many in-depth studies that allowed a reasonably good constraint on the mass, rotation period, and inclination of both components (e.g. \citealt{Gillon2013, Biller2013, Bedin2017, Garcia2017}). \cite{Lazorenko2018} measured their dynamical mass as 33.5±0.3 M$_\mathrm{Jup}$ for A and 28.6±0.3 M$_\mathrm{Jup}$ for B. Periodogram analysis of long-term variability by \cite{Apai2021} showed that WISE 1049B's period peaks around 5.2h, while the period for A is around 7h. Based on previously measured projected rotational velocity, the inclination of WISE 1049B is within a few degrees from edge-on, and A is inclined more than 62$^{\circ}$ \citep{Apai2021} (with edge-on defined as 90$^{\circ}$). The rotational speed and geometric configuration are both favorable for Doppler imaging observations since a relatively high $v\sin i$ allows sufficient broadening of spectral lines for their shape changes to be observed, and a nearly equator-on geometry allows the full variability amplitude to be measured without geometric dilution \citep{Vos2017}. The target properties are summarized in Table \ref{tab:target}.

\begin{table}
    \centering
    \caption{Selected literature properties for WISE 1049AB.}
    \label{tab:target}
    \renewcommand{\arraystretch}{1.2}
    \begin{tabular}{c|c|c|c}
    \hline
    Property                & WISE 1049A            & WISE 1049B           & Reference   \\
    \hline
    SpecType                & L7.5                  & T0.5                 & (1)   \\
    Mass (M$_\mathrm{Jup}$) & 34.2 $^{+1.3}_{-1.1}$ & 27.9$^{+1.1}_{-1.0}$ & (2)   \\
                         & or 33.5 ± 0.3            & 28.6 ± 0.3           & (3)   \\
    Period (hr)             & 6.94                  & 5.28                 & (4)   \\
                            & or 4.5-5.5            & or 4.87 ± 0.01  & (5), (6)   \\
                            & or 8                  & or 5.1 ± 0.1         & (7)   \\
    Inclination             & >62$^{\circ}$         & >$\sim$80$^{\circ}$  & (4)   \\
            & or 34$^{\circ}$-72$^{\circ}$ & or 64$^{\circ}$ ± 8$^{\circ}$ & (8)   \\
    NIR Var Amp             & $\sim$4\%             & 7\% - 11\%      & (5), (9)   \\
    \hline
    \end{tabular}
    \\
    References. (1) \cite{Burgasser2013}, (2) \cite{Garcia2017},  (3) \cite{Lazorenko2018} (4) \cite{Apai2021}, (5) \cite{Buenzli2015a}, (6) \cite{Gillon2013}, (7) \cite{Mancini2015}, (8) \cite{Karalidi2016}, (9) \cite{Biller2013}
\end{table}

\begin{table*}
\centering
\caption{Table of observation parameters for WISE 1049AB. The SNRs reported are values per spectral pixel averaged over all spectral pixels, spectral orders, and time steps.}
\label{tab:obs}
\begin{tabular}{cccccccc}
\hline
\multicolumn{1}{c}{Date} & \multicolumn{1}{c}{Target} & \multicolumn{1}{c}{Exposure time} & \multicolumn{1}{c}{Telluric A0V star} & \multicolumn{1}{c}{\begin{tabular}[c]{@{}c@{}}H band SNR\\ (per exposure)\end{tabular}} & \multicolumn{1}{c}{\begin{tabular}[c]{@{}c@{}}K band SNR\\ (per exposure)\end{tabular}} & \multicolumn{1}{c}{\begin{tabular}[c]{@{}c@{}}H band SNR \\ (binned)\end{tabular}} & \multicolumn{1}{c}{\begin{tabular}[c]{@{}c@{}}K band SNR\\ (binned)\end{tabular}} \\ \hline
\multirow{2}{*}{2020-02-09} & WISE 1049B & \multirow{2}{*}{287 s $\times$ 56} & \multirow{2}{*}{HIP 45977} & 44 & 55 & 89 & 109 \\
 & WISE 1049A &  &  & 37 & 61 & 74 & 122 \\
\multirow{2}{*}{2020-02-11} & WISE 1049B & \multirow{2}{*}{287 s $\times$ 56} & \multirow{2}{*}{HIP 45977} & 35 & 33 & 70 & 66 \\
 & WISE 1049A &  &  & 38 & 36 & 76 & 72 \\ \hline
\end{tabular}
\end{table*}

% B and A properties respectively
Many studies have reported that the B component is the main source of variability in the binary, with amplitudes of 5-11\% in optical to near-IR \citep{Gillon2013, Biller2013, Buenzli2015a}.
WISE 1049B's notable brightness and one of the largest variability amplitudes among T dwarfs make it the best possible (and nearly only) candidate for Doppler imaging with 8-m class telescopes.
On the other hand, WISE 1049A has a maximum measured variability of only $\sim$4\% at 0.8-1.15 $\mu$m \citep{Buenzli2015}. The polarization study by \cite{Millar-Blanchaer2020} and TESS long-term monitoring by \cite{Apai2021} reveals signs of band-like structure and zonal circulation in both A and B components.
Additionally, variability amplitudes for both components can vary notably from one rotational period to another (e.g. \citealt{Apai2021}), suggesting dramatic atmospheric evolution even in short terms.

Considering their brightness, high variability amplitude, and favorable rotational parameters, the WISE 1049AB system is an excellent test case for future Doppler imaging studies of more brown dwarfs with ELTs.
\cite{Crossfield2014} produced the first global TOA map of WISE 1049B through VLT/CRIRES high-resolution spectroscopic monitoring. The observation covered one rotation period of WISE 1049B ($\sim$5h) in a relatively narrow window in the K band centered on the 2.2 $\mu$m CO bandhead. The most prominent aspect of this map is the identification of large-scale bright and dark structures on WISE 1049B including a mid-latitude dark spot, indicating the existence of patchy clouds or chemical abundance variations in its atmosphere.
\cite{Luger2021} recently reproduced the WISE 1049B Doppler map with the stellar modeling code \textsc{starry} that features a spherical harmonics representation of the surface and a closed-form solution in a Bayesian posterior inference context. \cite{Plummer2022} also demonstrated a new surface mapping framework for ultracool objects that infers parametrized spots from high-resolution spectral time series and recovered the spot feature on WISE 1049B using the 2014 data set. 

Building upon previous studies, this work focuses on new Doppler imaging observations for WISE 1049 AB from the Gemini IGRINS spectrograph, which covers a much wider wavelength range than the 2014 study and with simultaneous coverage in the H ($\sim$1.6 $\mu$m) and K ($\sim$2.1 $\mu$m) band. This enables us to assess the longevity of the previously discovered atmospheric structure and to map different pressure levels in the atmosphere for the first time. The observation and data reduction are presented in Section \ref{sec:obs}. The spectral fitting is described in Section \ref{sec:mf}. The Doppler imaging procedures and resulting maps are described in Section \ref{sec:dime} and \ref{sec:maps}, followed by interpretations and discussion in Section \ref{sec:interp}. Finally, we simulated effects that may affect our result, and discuss the limitations and prospects of Doppler imaging in Section \ref{sec:disc}.

%%%%%%%%%%%%%%%%%%%%%%%%%%%%%%%%%%%%%%%%%%%%%%%%%%%%%%%%
%%%%%%%%%%%%%%%%%%%%%%%%%%%%%%%%%%%%%%%%%%%%%%%%%%%%%%%%
%%%%%%%%%%%%%%%%%%%%%%%%%%%%%%%%%%%%%%%%%%%%%%%%%%%%%%%%
\section{Observations and Data Reduction}
\label{sec:obs}

We conducted high-resolution spectroscopic observations of WISE 1049AB on the nights of 9 and 11 February 2020, using the Immersion GRating INfrared Spectrometer (IGRINS, \citealt{Park2014, Mace2018}) mounted on the Gemini South telescope located at Cerro Pach\'{o}n, Chile. The observations lasted from UTC 2020-02-10 04:00:27 to 09:01:48 and UTC 2020-02-12 04:20:08 to 09:32:35, each consisting of 56 exposures spanning 5 hours. Our spectra cover the H and K bands (1.45 to 2.48 $\mu$m) simultaneously with spectral resolving power R$\sim$45,000. The two components were spatially resolved, and the spectrograph slit was aligned to the position angle of the binary to simultaneously disperse both brown dwarfs with a separation of $\sim$17 pixels on the IGRINS detector or $\sim$1.5" on the sky. A telluric A0V star was observed at a similar airmass with the same telescope and instrument configuration for telluric line removal. The observation parameters are summarized in Table \ref{tab:obs}.

The data were reduced with the IGRINS Pipeline Package (PLP; \citet{Lee2016}). The PLP performs sky subtraction, flat-fielding, bad-pixel correction, aperture extraction, wavelength calibration, and telluric correction, yielding wavelength-calibrated, telluric-corrected fluxes and signal-to-noise ratios (SNR) for individual points in the spectrum. Since IGRINS has a fixed spectral format, the wavelength solution was first derived from an empirical template and then refined using sky OH emission in a 300s SKY exposure. It was then further refined using telluric absorption features in the A0V. Telluric lines were removed by dividing the target spectrum by the spectrum of the A0V standard star. 
To increase the number of individual spectra for each component of WISE 1049AB we used the 2-dimensional spectra from the PLP that were rectified, flat-fielded, and wavelength calibrated. We then used custom IDL routines to optimally extract the A and B components. The telluric standard was extracted with the same routine. The final data products for each night were 56 individual telluric corrected spectra for both WISE 1049 A and B, along with their SNR spectra. The H and K band mean SNRs of the reduced spectra are listed in Table \ref{tab:obs}.

We binned the 56 time-resolved spectra into 14 timesteps to enhance the signal-to-noise level. The total observation time per timestep is thus $\sim$20 minutes. The H and K band SNRs for the final binned spectra are listed in the last column of Table \ref{tab:obs}. The pixels located on the edges of spectral orders have high uncertainties, so they were removed for the analysis. During data analysis, we found that the telluric lines are not completely removed by the standard star in the reduction pipeline. We applied a customized filtering routine on each spectrum to mask out the remaining noisy spikes by replacing spikes or outliers with local median values. The flux was normalized so that the continuum level is unity. The binned, filtered, and normalized spectra of WISE 1049B and A on the night of Feb 11 are shown in Fig. \ref{fig:Bfit} and \ref{fig:Afit}.

%%%%%%%%%%%%%%%%%%%%%%%%%%%%%%%%%%%%%%%%%%%%%%%%%%%%%%%%
%%%%%%%%%%%%%%%%%%%%%%%%%%%%%%%%%%%%%%%%%%%%%%%%%%%%%%%%
\section{Model fitting and Line profile extraction}
\subsection{Model fitting}
\label{sec:mf}

\begin{figure*}
    \begin{centering}
    \begin{minipage}[b]{1\textwidth}
        \centering 
        \includegraphics[width=0.92\textwidth]{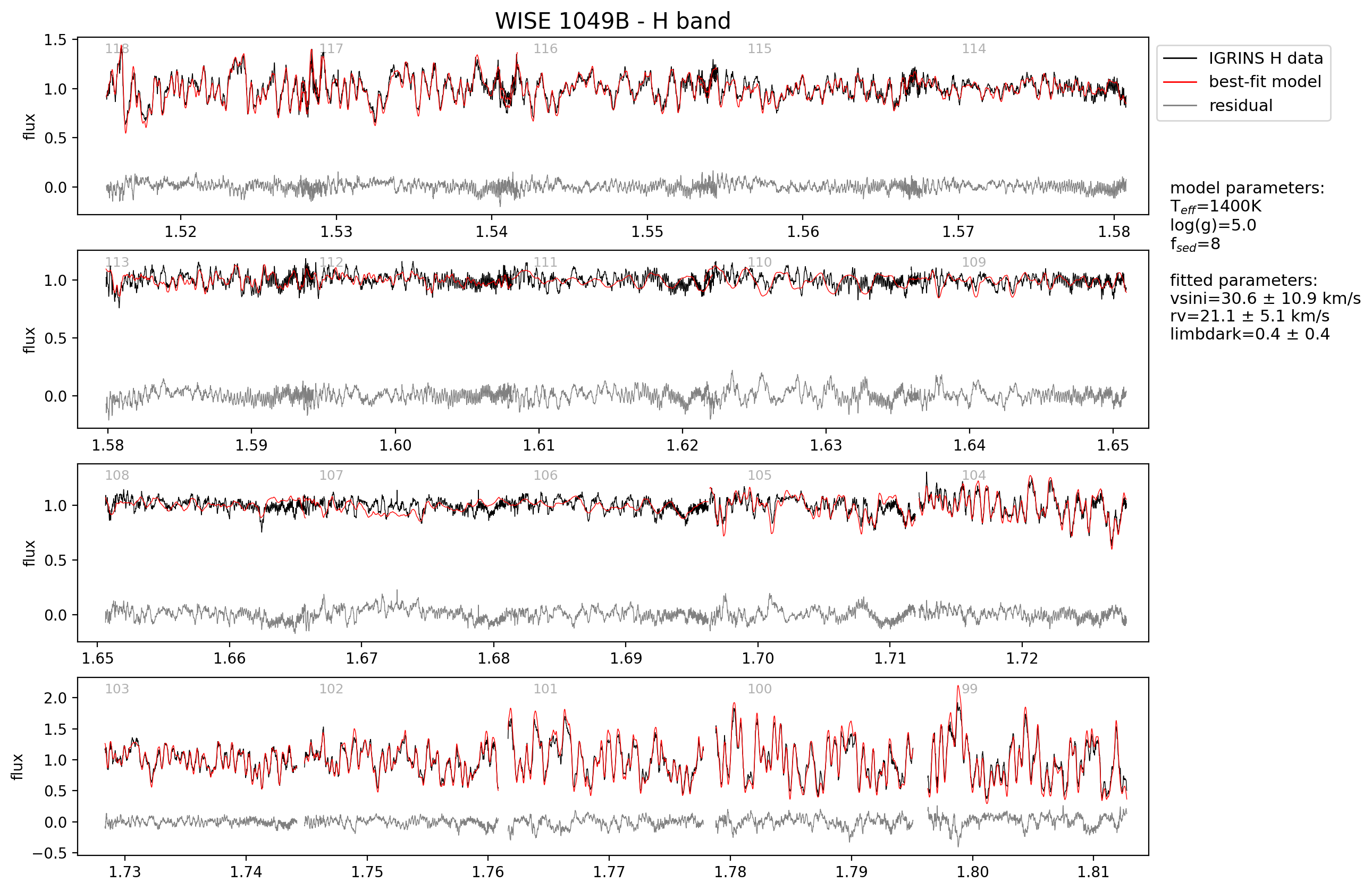}
    \end{minipage}
    \\[10pt]
    \begin{minipage}[b]{1\textwidth}
        \centering
        \includegraphics[width=0.92\textwidth]{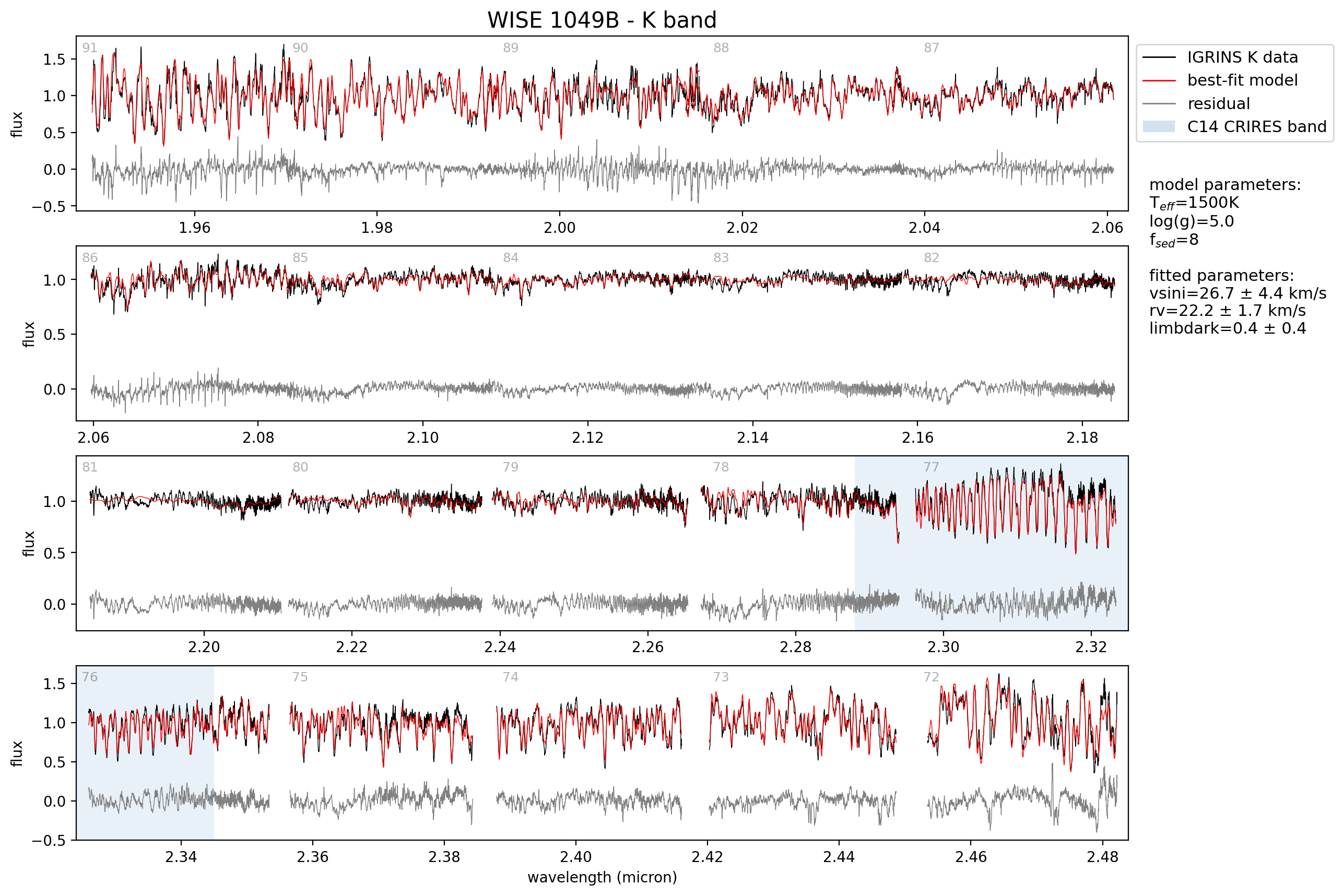}
    \end{minipage}
    \caption{IGRINS H and K band spectrum and best-fit Sonora Diamondback model for WISE 1049B on the night of Feb 11. The fitting is performed for each spectral order and each timestep of observation separately, but only data from one selected timestep is shown in this figure. The wavelength covered by the original CRIRES Doppler imaging observations in \protect\cite{Crossfield2014} are shaded in blue.}
    \label{fig:Bfit}
    \end{centering}
\end{figure*}

\begin{figure*}
    \begin{centering}
    \begin{minipage}[b]{1\textwidth}
        \centering 
        \includegraphics[width=0.92\textwidth]{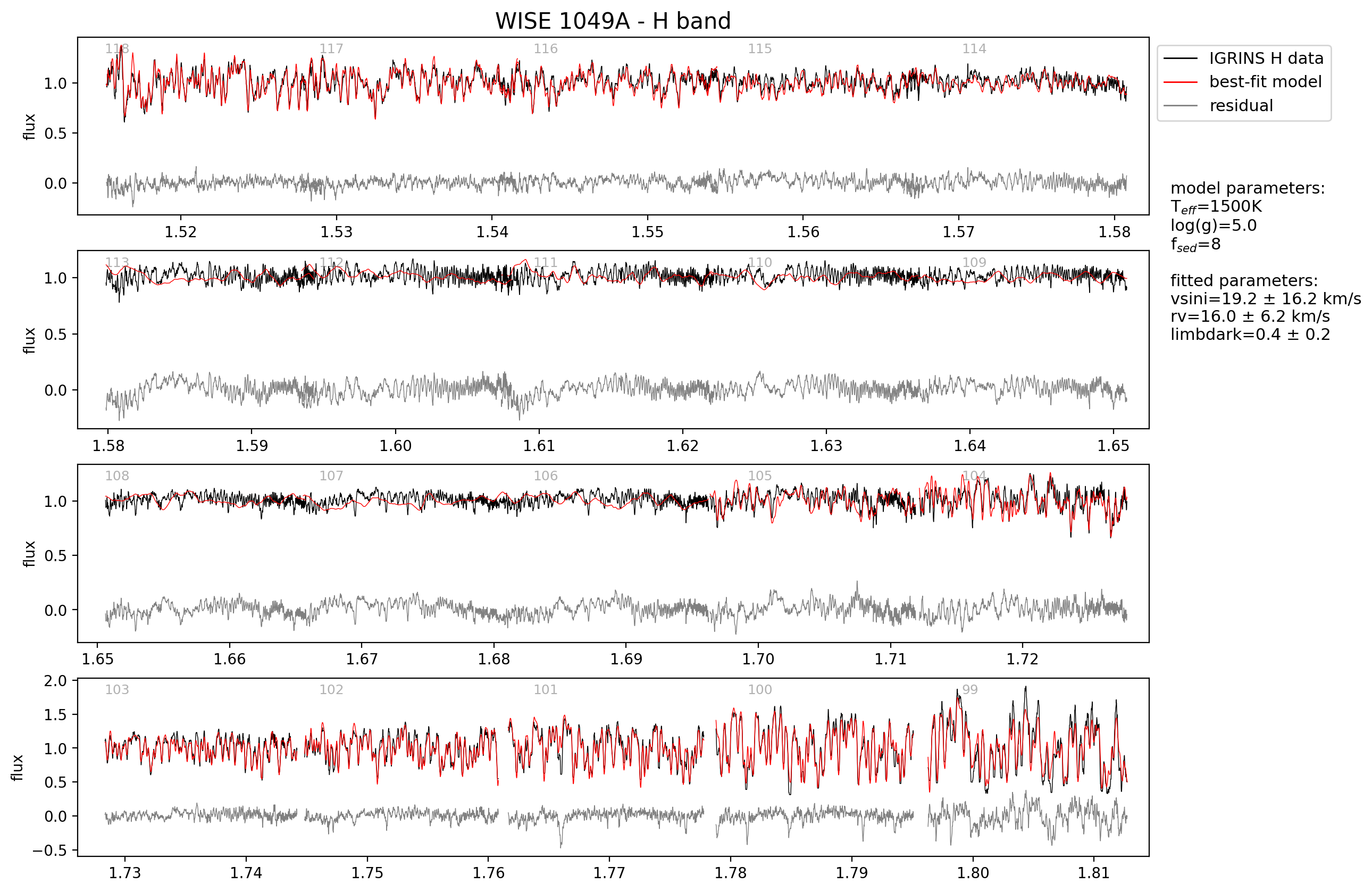}
    \end{minipage}
    \\[10pt]
    \begin{minipage}[b]{1\textwidth}
        \centering
        \includegraphics[width=0.92\textwidth]{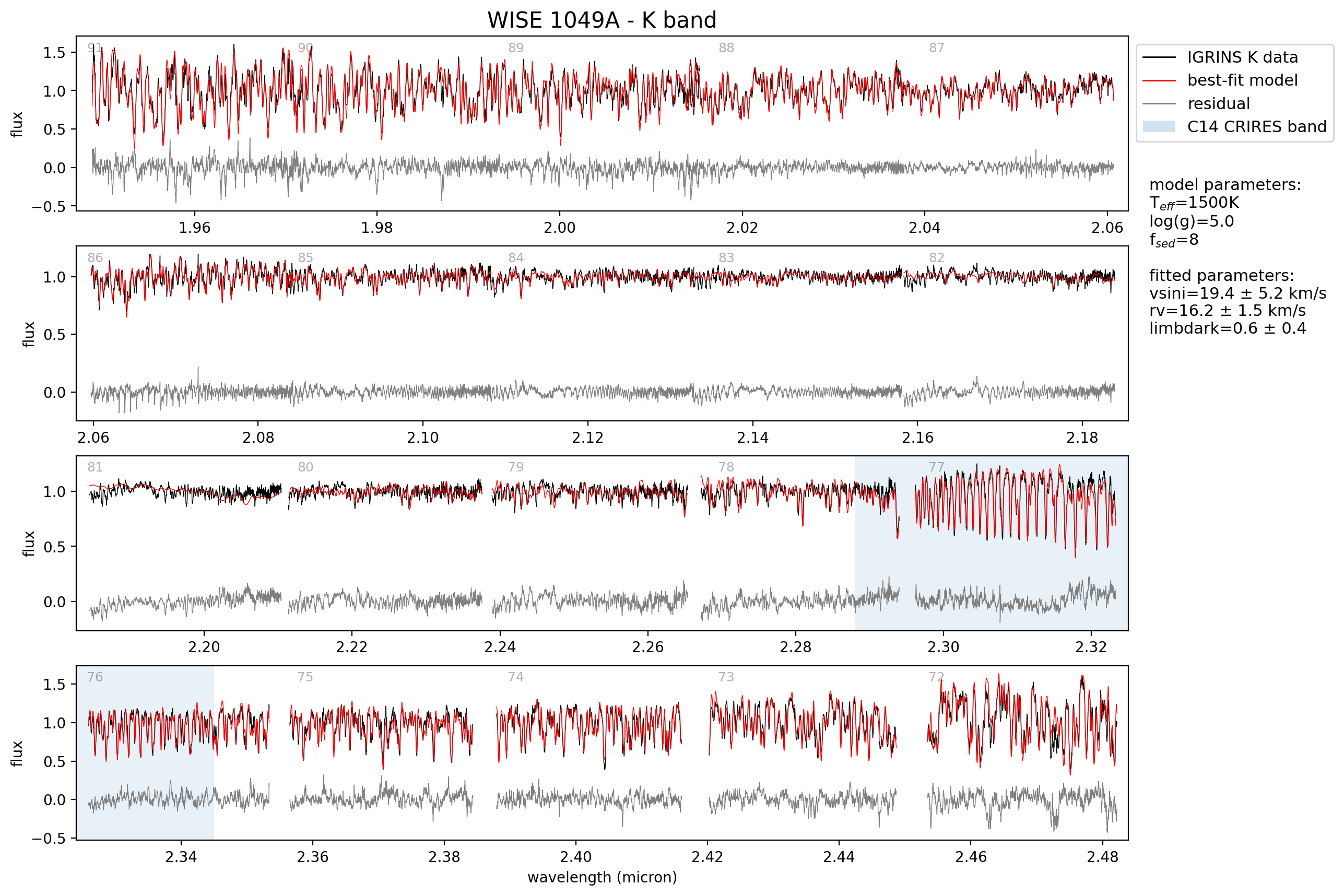}
    \end{minipage}
    \caption{Same as Fig. \ref{fig:Bfit} but for WISE 1049A, on the night of Feb 11.}
    \label{fig:Afit}
    \end{centering}
\end{figure*}

\begin{table*}
    \centering
    \caption{Best-fitting parameters for WISE 1049AB from Sonora Diamondback models ([Fe/H] =1.0, C/O=1.0) and comparison with values reported in C14.}
    \label{tab:fit}
    \begin{tabular}{ccccccccccccc}
    \hline
                   &&                   &         &                &&\multicolumn{3}{c}{\textit{Feb 11, 2020}} &&\multicolumn{3}{c}{\textit{Feb 9, 2020}}  \\
    \hline
    Target + band  && $T_\text{eff}$ (K)& $\log g$&$f_\mathrm{sed}$&& $v\sin i$ (km/s)& RV (km/s)  & limbdark  &&$v\sin i$ (km/s)& RV (km/s)   & limbdark  \\
    \hline
    WISE 1049B H   && 1400              & 5.0     & 8              && 30.6 ± 10.9     & 21.1 ± 5.1 & 0.4 ± 0.4 &&  30.9 ± 11.5   & 21.1 ± 3.2  & 0.5 ± 0.4 \\
    WISE 1049B K   && 1500              & 5.0     & 8              && 26.7 ± 4.4      & 22.2 ± 1.7 & 0.4 ± 0.4 &&  26.9 ± 4.6    & 21.6 ± 13.1 & 0.5 ± 0.4 \\
    WISE 1049B avg &&  /                &  /      & /              && 27.2 ± 4.1      & 22.1 ± 1.6 & 0.4 ± 0.3 &&  27.5 ± 4.3    & 21.1 ± 3.1  & 0.5 ± 0.3 \vspace{3pt}\\
    WISE 1049A H   && 1500              & 5.0     & 8              && 19.2 ± 16.2     & 16.0 ± 6.2 & 0.4 ± 0.3 &&  16.7 ± 11.1   & 16.0 ± 3.0  & 0.5 ± 0.2 \\
    WISE 1049A K   && 1500              & 5.0     & 8              && 19.4 ± 5.2      & 16.2 ± 1.5 & 0.6 ± 0.4 &&  18.9 ± 4.8    & 16.6 ± 6.8  & 0.8 ± 0.4 \\
    WISE 1049A avg &&  /                &  /      & /              && 19.4 ± 5.0      & 16.2 ± 1.5 & 0.5 ± 0.2 &&  18.6 ± 4.4    & 16.1 ± 2.7  & 0.6 ± 0.2 \\
    \hline
                   &&                   &         &                &&\multicolumn{3}{c}{\cite{Crossfield2014}} &&                &             &           \\ 
    \hline
    WISE 1049B K   &&  1450             & 5.0     & /              && 26.1 ± 0.2      & 17.4 ± 0.5 & /         &&                &             &           \\
    WISE 1049A K   &&  1500             & 5.0     & /              && 17.6 ± 0.1      & 20.1 ± 0.5 & /         &&                &             &           \\
    \hline
    \end{tabular}
\end{table*}

We fitted the spectra with atmospheric models to determine the bulk parameters and the best non-broadened model spectra for WISE 1049AB, which are required for the Doppler imaging procedure. The free parameters in our $\chi^2$ model fitting routine (and their uniform priors used in fitting) are the projected rotational velocity ($v$sin$i$, 0-100 km/s), radial velocity (0-100 km/s), linear limb-darkening coefficient (0-1), and two continuum-normalization coefficients (-10, 10). The error for $\chi^2$ fitting is determined from the SNR values of the reduced spectra. 

We first performed fitting with the BTSettl models \citep{Allard2013} that spans a grid of effective temperatures $T_\text{eff}$ = 1200K - 1800K and surface gravities $\log g$ = 4.0-5.5, with metallicity and C/O fixed to solar value. Based on the fitting results with BTSettl grids, we narrowed down the preferred $T_\text{eff}$ and gravity ranges to $T_\text{eff}$ = 1400K to 1500K and $\log g$ = 5.0.
We then ran new atmospheric models that use the temperature-pressure profiles, abundances, and cloud parameters from the Sonora Diamondback grid (Morley et al. 2024 in prep) with updated line lists. These models are an extension of the cloud-free Sonora Bobcat grid by \cite{Marley2021} with the addition of cloud opacity. The molecular opacities included in the models are H$_2$/He CIA, H$_2$O, CO, CH$_4$, NH$_3$, and FeH. The models span effective temperatures from $T_\text{eff}$ = 1400K to 1500K, $\log g$ = 5.0, and cloud sedimentation factors $f_\text{sed}$ = 1, 2, 4, 8 and no cloud. $f_\text{sed}$ describes the efficiency with which particles can settle out of the cloud, which tunes the cloud particle sizes and controls the vertical extent of clouds \citep{Ackerman2001}. Smaller $f_\mathrm{sed}$ leads to vertically extended clouds with smaller particles, while greater $f_\mathrm{sed}$ corresponds to thin clouds with large particles. 

We performed individual fitting to each of the 20 spectral orders of IGRINS H and K bands at each of the 14 timesteps. The spectra and best-fit Sonora Diamondback models are shown in Fig. \ref{fig:Bfit}, \ref{fig:Afit}. 
The best-fit Sonora Diamondback model has $T_\text{eff}$ = 1500 K, $\log g$ = 5.0, and $f_\mathrm{sed}$ = 8 for WISE 1049A (both H and K) and B (K band), and $T_\text{eff}$ = 1400 K for WISE 1049B in H band. The best-fit model remains the same for the data from two nights.
These fitted values are higher compared to effective temperatures derived from fitting evolutionary models to bolometric luminosity measurements (e.g., around 1200-1300K for both WISE 1049A and B; \citealt{Faherty2014, Biller2024}). It is a well-documented issue that atmospheric models tend to require higher $T_\text{eff}$ and unphysically small radii compared to evolutionary models to fit observed spectrum  (e.g. \citealt{Marois2008, Carter2023}), partly due to uncertainties in cloud parametrization in the models. Thus the best-fit effective temperatures reported in this work should be interpreted with caution and considered alongside values derived from evolutionary models.

For both the BTSettl and Sonora Diamondback models, we found that the same model at certain $T_\text{eff}$ and $\log g$ does not always give the best fit to the data from both H and K bands, and the best-fit parameters vary across each order and timestep considerably, as shown in Fig. \ref{fig:fitparamH_B}-\ref{fig:fitparamK_A} in the Appendix. This is in general due to the varied data quality at different orders and timesteps. The fit quality is also affected by the number and strength of lines in different orders. 
In general, the IGRINS H band exhibits poorer fits and yields larger error bars on the fitted parameters compared to the K band. This arises from multiple reasons: First, the brown dwarfs are fainter in the H band than in the K band; second, the H band contains fewer spectral lines to fit with; and third, the H band is more susceptible to telluric contamination. Orders containing dense spectral lines, such as the CO bandhead in the K band, typically yield well-fitted results. Another possible factor affecting the fit in the H band is that there could be less CH$_4$ at the atmospheric level probed by Doppler Imaging due to non-equilibrium chemistry driving vertical mixing \citep{Tremblin2020}.
Additionally, atmospheres in the L/T transition are strongly affected by heterogeneous clouds (e.g. \citealt{Marley2010, Radigan2012, Apai2013, Zhou2018, Vos2023}). It is thus common that a single 1-D model is insufficient to fit the entire spectrum, but a combination of multiple spectral components is needed to model the effect of these heterogeneous clouds.
Since a detailed spectral analysis goes beyond the scope of this paper, we chose to adopt the median value of fitted parameters over all the orders and timesteps after removing outliers beyond 3$\sigma$. We then combined the values from the H band and K band using a weighted average.

As a result, we measured a 
projected rotational velocity of 27.2 ± 4.1 km/s and radial velocity of 22.1 ± 1.6 km/s for WISE 1049B,
along with a $v\sin i$ of 19.4 ± 5.0 km/s and RV of 16.2 ± 1.5 km/s for WISE 1049A.
The fitted model grid parameters and free parameters for the two nights are summarized in Table \ref{tab:fit}. The fitted parameters from the two nights agree within their error bars. Compared to the values reported in \citet{Crossfield2014}, the $v\sin i$ measurement aligns within error bars, while the radial velocity measurement differs due to the brown dwarfs being in a different position.
We ran a simple Monte-Carlo simulation to estimate the inclination of WISE 1049AB from our measurements of $v\sin i$. Assuming 0.9-1.1 Jupiter radius \citep{Burrows2001} and a period of 5h for WISE 1049B and 7h for A, the measured $v\sin i$ translates to an inclination of 80$^{+10}_{-17}$ degrees for WISE 1049B and 69$^{+21}_{-28}$ degrees for WISE 1049A. The posterior distributions from the Monte-Carlo runs are shown in the Appendix (Fig. \ref{fig:incMC}).

%%%%%%%%%%%%%%%%%%%%%%%%%%%%%%%%%%%%%%%%%%%%%%%%%%%%%%%%
\subsection{Least-Square deconvolution}
\label{sec:LSD}

\begin{figure*}
\centering
\includegraphics[width=0.9\textwidth]{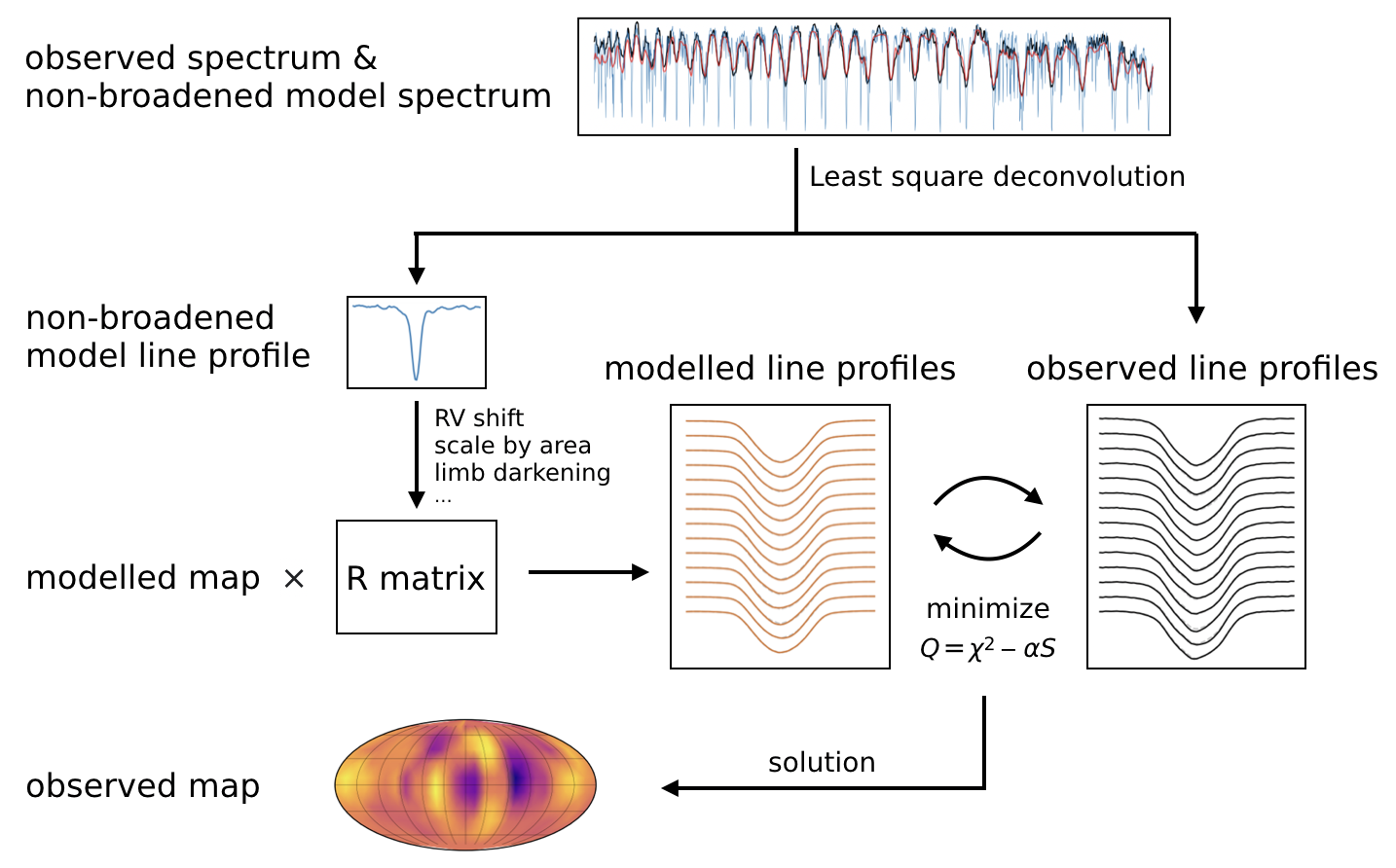}
\caption{A schematic of the Doppler imaging framework implemented in this work. The model fitting and least square deconvolution (LSD) process are described in Section \ref{sec:mf} and \ref{sec:LSD}. The formulation of the Doppler imaging problem and construction of the response $R$ matrix are described in Section \ref{sec:dime}. Implementation details can be found in the code repository on Github\protect\footnotemark.}
\label{fig:flowchart}
\end{figure*}

Most Doppler imaging studies to date use least-squares deconvolution (LSD; \citealt{Donati1997}) to enhance the signal-to-noise of spectral features (e.g. \citealt{Marsden2005, Cameron2010, Crossfield2014}). The LSD algorithm combines all available spectral lines in the observed spectra to derive an average line profile (LP) with significantly increased line SNR. An observed spectrum is the convolution of the rest-frame intrinsic spectrum with a broadening kernel which incorporates the rotational and instrumental broadening. LSD works by deconvolving the observed spectrum with a delta spectrum containing all the lines in the model spectra through a least-square fitting process, which results in an averaged line profile kernel in the radial velocity space. 

To generate a reference line list for the LSD procedure, we used \textsc{DAOspec} \citep{Stetson2008} to automatically detect lines from the best-fit Sonora Diamondback spectrum. We then generated a delta spectrum from this line list and applied the LSD algorithm as implemented in \cite{Crossfield2014} to each spectral order at each observed timestep, resulting in a time series of observed line profiles, as shown in Fig. \ref{fig:flowchart}. The same LSD procedure was repeated on the best-fit Sonora Diamondback non-broadened model spectrum to obtain an intrinsic non-broadened model line profile, which will be used in constructing the Doppler imaging response matrix in later steps. 

We then calculated the empirical SNR of each LSD profile following the method in \cite{Luger2021}. We estimated the noise using the median absolute deviation (MAD), i.e., $\mathit{noise} = 1.4826 \cdot \mathit{median}(|x_i - \Tilde{x}|)$, where $x_i$ is the difference between the LSD profile and a smoothed profile at each pixel and $\Tilde{x}$ is the median of $x$. The signal is estimated by the average line depth. Note that the definition of the SNR of an LSD line profile is the ratio between the line depth and the line profile noise, whereas SNRs reported in Table \ref{tab:obs} are defined as the ratio between the flux and error of an IGRINS spectral pixel. Due to this difference in definitions, these two SNRs are not directly comparable. Therefore, we make a distinction by referring to the SNR calculated for an LSD profile as the \textit{line SNR} hereafter. We found that the typical line SNR of an LSD profile is around 20-30. While it is difficult to determine the “line SNR” for the original spectrum due to the presence of numerous lines with varying depths that are often overlapping or blended, we attempted to estimate this value by applying the same formula to a few selected lines in the pre-LSD spectrum. We found that in regions with deep, isolated lines, the "line SNR" can reach $\sim$10, however in the majority of regions where lines are shallower and blended, this value is typically only around 2-4. This demonstrates the necessity of LSD for achieving sufficient signal for line inversion.
\footnotetext{https://github.com/alphalyncis/doppler-imaging-maxentropy}

We tested how the number of pixels used in the LSD kernel, $n_k$, affects the LSD result. The resolution of the resulting LSD line profile is determined by the instrument's spectral resolution. $R = \lambda/\Delta \lambda \sim$ 45,000 with IGRINS corresponds to a resolution of $\sim$ 1.5 km/s per pixel in the velocity space. Thus, the resolution (and therefore the width) of the LSD line profile is only determined by the $v$sin$i$ of the brown dwarf. The pixels outside the ±$v\sin i$ value of the brown dwarf are padded with the continuum level, and $n_k$ only controls the width of the flat paddings outside the line profiles. We looped over different values of $n_k$ and calculated the line SNR for the resulting LSD line profiles. The time-averaged line SNR for each spectral order versus the $n_k$ value used in LSD is shown in Fig. \ref{fig:nktest} in the Appendix.
We found that $n_k$ of 120$\sim$150 yields the best LSD result with the highest line SNR values. In the later image reconstruction step, we also found that the choice of $n_k$ also affects the latitude of the retrieved spot. This effect is also observed by \cite{Plummer2022}. Thus, finding a suitable value of $n_k$ is important for recovering trustworthy map details. We adopted $n_k=125$ for our LSD routine.

We observed that LSD profiles with low line SNR (< 20) usually correspond to the spectral orders that are poorly fitted in the previous step, and these low-SNR profiles tend to introduce more noise than valuable mapping information during the later stages of line inversion. We therefore choose to only include orders with line SNR > 20 in the subsequent Doppler imaging process. This includes orders 72-77 \& 87-90 for the IGRINS K band, and orders 99-104 \& 115-118 for the IGRINS H band (see Table 2. in \citealt{Tannock2022} for wavelengths of IGRINS orders and their major absorption lines).
Next, we average the line profiles over the selected orders at each timestep, resulting in an order-averaged time series of observed line profiles, as shown in Fig. \ref{fig:flowchart}. These line profiles are now ready for line inversion to reconstruct the surface brightness map.

%%%%%%%%%%%%%%%%%%%%%%%%%%%%%%%%%%%%%%%%%%%%%%%%%%%%%%%%
\section{Maximum entropy Doppler imaging}
\label{sec:dime}

\subsection{Method}

Reconstructing the surface brightness map of a brown dwarf from a time series of line profiles requires a line inversion technique. In Doppler imaging, the line inversion solution can be found by iteratively solving the much easier forward problem, searching among all possible maps until one is found that fits all the observed data within the known noise level of those data.
 
The forward problem in Doppler imaging is relatively straightforward, i.e., to compute the rotationally modulated spectral time series from a known surface brightness map and a non-broadened line profile. 
This can be formulated as a linear transformation between the map vector and the line profile vector through a response matrix. We follow the formulation in \cite{Vogt1987}, where we divide the spherical surface into rectangular cells and represent the surface brightness by the map vector $I$ with size $n_\mathrm{cell}$, each element representing the brightness value of the corresponding cell.
%cell division

\begin{figure}
\centering
\includegraphics[width=\columnwidth]{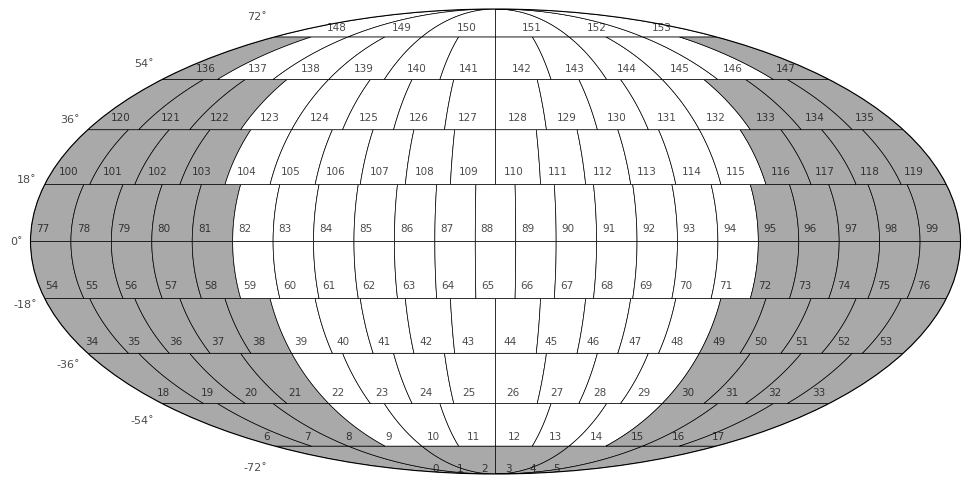}\\
\vspace{15pt}
\includegraphics[width=\columnwidth]{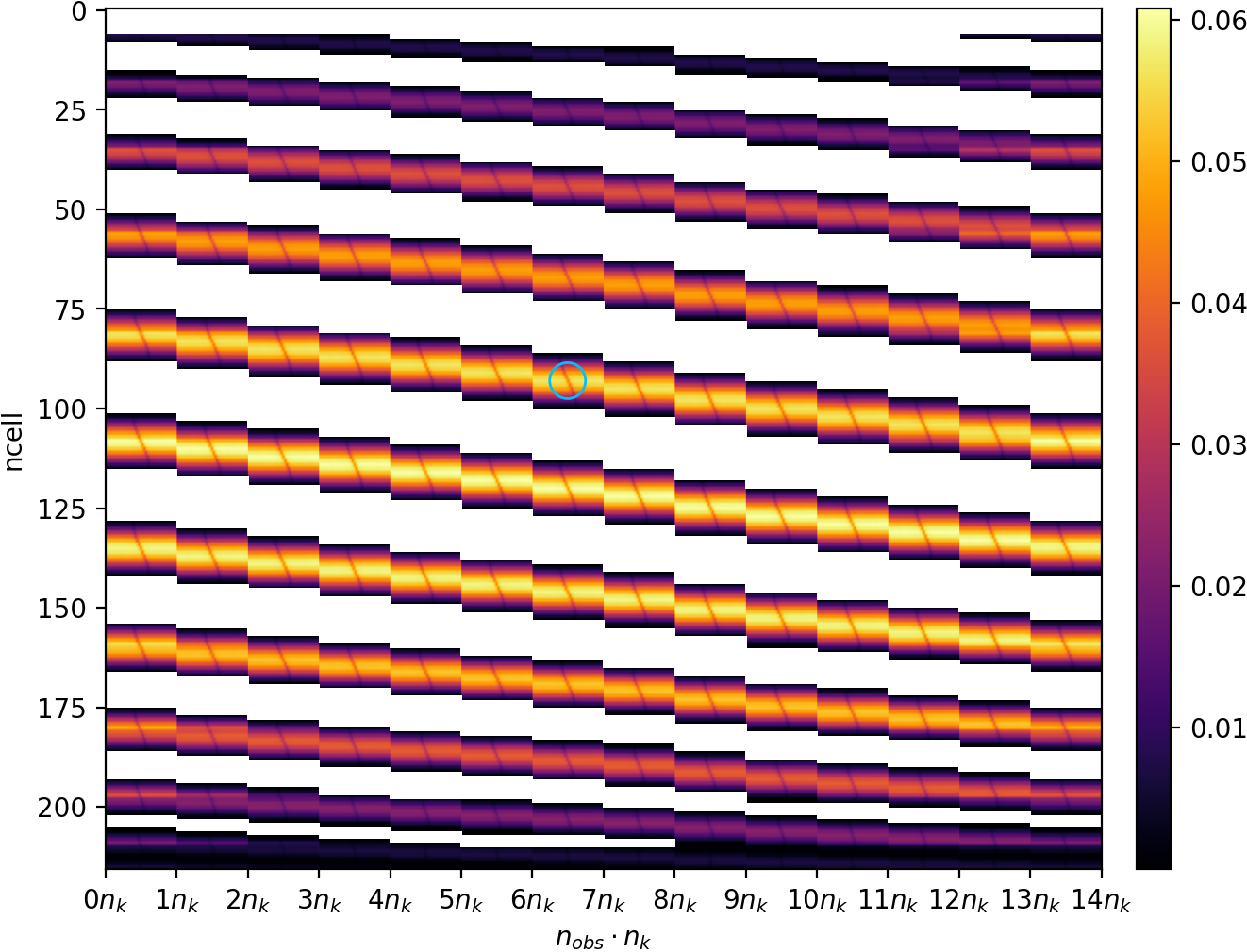}
\caption{\textit{upper}: A schematic of the cell map used in Doppler imaging with $\sim$150 cells. The cells marked in gray are the cells that are not visible at the beginning of observation. \textit{lower}: The Doppler imaging response matrix $R$ corresponding to the map above. The dark trace in the center of each block (e.g. circled in blue) illustrates the Doppler shift of the intrinsic line profile contributed by a row of adjacent cells mapped to different projected velocities across the visible disk.}
\label{fig:cells}
\end{figure}

%R matrix
The Doppler imaging response matrix, $R$, encodes how the disk-integrated rotationally broadened line profile at a particular rotational phase responds to changes in the surface brightness of a particular cell.
$R$ is a matrix with $n_\mathrm{cell}$ rows and $n_\mathrm{obs} \times n_k$ (number of observed phases $\times$ width of line profile) columns. Each row of $R$ consists of $n_\mathrm{obs}$ blocks of line profiles concatenated end-to-end, such that block $R_{ij}$ represents the flux contributed from the $i^\mathrm{th}$ cell at phase $j$. In other words, $R_{ij}$ is the non-broadened modelled line profile shifted to the radial velocity of the $i^\mathrm{th}$ cell and scaled by its projected area and limb-darkening factor at phase $j$.
% RV shift
The radial velocity of a particular cell at rotational phase $j$ is given by $rv_\mathit{cell} = v\sin i \cdot \overline{y}_\mathit{cell}$, where $\overline{y}$ is the mean of $y$ coordinates of the cell vertices.
%projected area
The projected area of a fully visible cell is given by the polygon area formula $A=\frac{1}{2} \sum_{k=1}^4\left(y_k z_{k+1}-y_{k+1} z_k\right)$\footnote{https://mathworld.wolfram.com/PolygonArea.html}, where $(y_k, z_k)$ is the coordinate on the plane perpendicular to the line of sight of the $k^\mathrm{th}$ vertex of the cell. For a partially visible cell, the vertices hidden behind the sphere are replaced by new vertices on the limb, from which the visible projected area is calculated as above. 
%limb darkening
The linear limb darkening term is given by $I(\theta)/I(0) = 1 - \epsilon + \epsilon \cdot \cos{\theta}$ where $\epsilon$ is the linear limb darkening coefficient and $\theta$ is the angle between the line of sight and the normal vector of the cell. 
The final $R$ matrix is populated by concatenating these $R_{ij}$ blocks for each rotational phase, resulting in a full representation of the spectral response as a function of location on the sphere and rotational phase. A visualization of the $R$ matrix for a map with $\sim$ 150 cells is shown in Fig. \ref{fig:cells}. 

Given a map vector $I$, the generated time series of spectral line profiles, $M$, is simply given by the product of the map vector $I$ and the response matrix $R$, i.e.,
\begin{equation}
    \label{eq:fwd}
    M = IR
\end{equation}

\begin{figure*}
\centering
\includegraphics[width=0.95\textwidth]{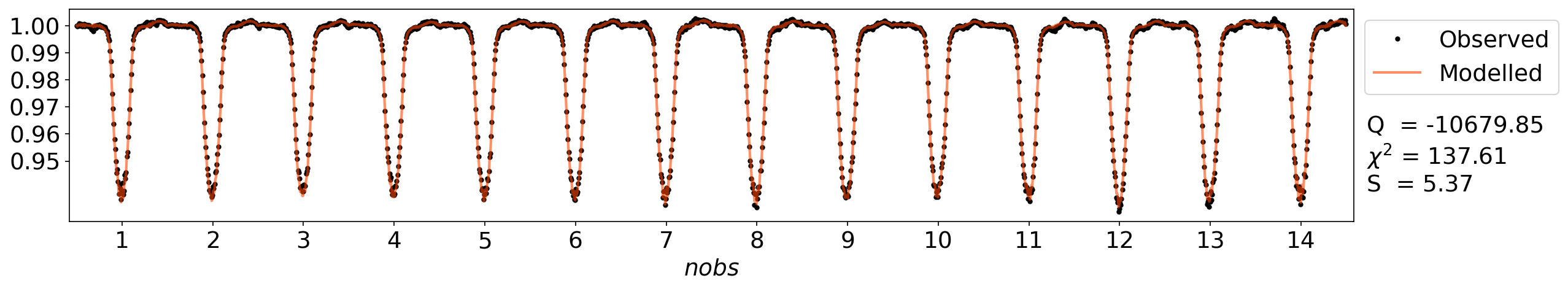}
\begin{minipage}{\textwidth}\centering
\vspace{10pt}
\includegraphics[width=0.93\textwidth]{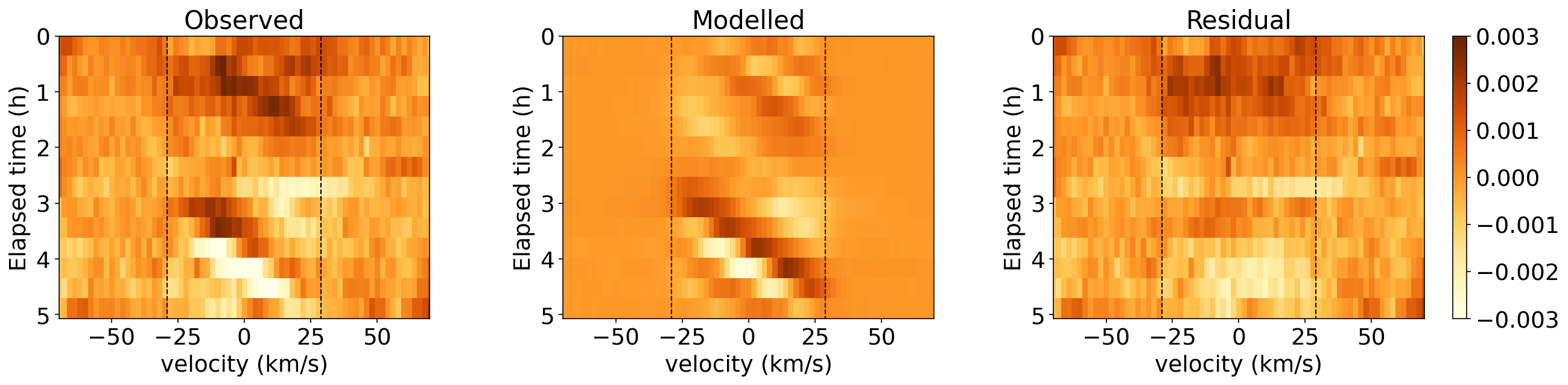}
\end{minipage}
\caption{{\it Upper}: The observed time series of LSD line profiles (black) and the modelled line profiles (red) which give the best-fit solution to the maximum entropy image reconstruction for the WISE 1049B IGRINS K band on the night of Feb 11. The values of $\chi^2$, map entropy $S$, and the metric $Q=\chi^2-\alpha S$ are shown alongside. {\it Lower}: Deviation from the time-averaged profile for the observed and modeled line profiles, as well as the residual between them. The black dashed lines in the deviation plots mark the edges of the visible brown dwarf disk at the equator where the projected velocity is maximum. The black dashed lines mark the edge of the brown dwarf disk where the projected velocity is maximum. The dark trace seen in the deviation plot is the signal of a dark patch rotating across the visible disk. } 
\label{fig:residual_fit}
\end{figure*}

$M$ is a vector of size $n_\mathrm{obs} \times n_k$ representing the rotationally broadened line profiles at every observed phase modelled from the given map. Finally, the observed data vector $D$ is formed by concatenating the time series of observed line profiles end-to-end,  so it is also of size $n_\mathrm{obs} \times n_k$. An example of the model vector $M$ and observed data vector $D$ is shown in Fig. \ref{fig:residual_fit}.

To reconstruct the surface brightness map $I$ from the set of observed data $D$ requires solving the inverse problem of Eq. \ref{eq:fwd}.
$I$ can be iteratively solved by minimizing the difference between the observed spectrum $D$ and forwardly-computed spectrum $M$, i.e., minimize
\begin{equation}
    \label{eq:chi2}
    \chi^2=\sum_{i=1}^{n_\mathrm{obs} \cdot n_k} (D_i - M_i )^2 / \sigma_i^2
\end{equation}

where $1 / \sigma_i^2$ are the weights of each data point, taken from the empirical error value for each spectral order computed in the LSD procedure in Section \ref{sec:LSD}. 

A well-known problem for Doppler imaging is that often more than one surface map can be consistent with the same observed line profile variation, and a criterion is needed to select the best map. One possible criterion is to choose the simplest or smoothest image, i.e., the one with the least amount of information (thus maximum entropy), which is still consistent with all the data within the known noise level of the data. \cite{Vogt1987} introduced maximum entropy image reconstruction for solving Doppler imaging problems, which involves finding the image with the largest configuration entropy $S$, which is defined as: 
\begin{equation}
    \label{eq:entropy}
    S=-\sum_j p_j \log (p_j)
\end{equation}

where $p_j$ is the brightness value of the $j^\mathrm{th}$ image pixel.
Thus, the problem becomes finding the image that produces spectra that fit all the real data subject to the additional constraint of maintaining the highest possible entropy. This is implemented by minimizing the metric
\begin{equation}
    \label{eq:metric}
    Q=\chi^2-\alpha S
\end{equation}
where $\chi^2$ is the goodness of fit from Eq. \ref{eq:chi2}, $S$ is the entropy of the image, and $\alpha$ is a regularization parameter that balances the two.

%%%%%%%%%%%%%%%%%%%%%%%%%%%%%%%%%%%%%%%%%%%%%%%%%%%%%%%%
\subsection{Implementation}
\label{sec:pipeline}

We adopted the same implementation of maximum entropy image reconstruction as in \cite{Crossfield2014}. In this code, the brown dwarf surface is represented by a grid of rectangular cells whose edges lie along lines of latitude and longitude. We implemented a new partition of the spherical surface such that each cell has approximately equal area. This approach allows the cells in polar and equatorial regions to be treated equally. Given the total desired number of cells, the code automatically finds the number of cells per row and the number of rows needed. An illustration of the cell division is shown at the top of Fig. \ref{fig:cells}.
In our Doppler imaging reconstruction, we found that a total number of $\sim$200 cells on the map is sufficient to capture the main features. Further increasing the number of cells does not enhance the final image quality. While implementing the chi-square function in Equation \ref{eq:chi2}, we take only the deviation from the time-averaged line profile to construct the observed spectra $D_i$, i.e., $D_i = D_\text{time-avg} + D_{\text{diff},i}$. This way, we removed the effect of the instrumentation profile on the line profile shapes.

The next question is to determine the value of $\alpha$, the regularization parameter that balances the fit to data and the entropy of the reconstructed image (i.e. the smoothness of the map).
Using an $\alpha$ value that is too small may lead to overfitting to noise, creating spurious features on the reconstructed map. Using an $\alpha$ value that is too large would avoid overfitting but may lose information in the true map.
To determine the optimal value of $\alpha$, we conducted a test using cross-validation, comparing fitted maps with different $\alpha$ values.
In cross-validation, the dataset is divided into a training set and a validation set. The fitting is performed only on the training set, and the fit quality is evaluated on the validation set by a cross-validation score (CV score), which is computed based on the discrepancy between the prediction of the model and the data points in the validation set. 
In our cross-validation test, for a specific choice of $\alpha$, we repeated our maximum entropy routine $n_\mathrm{obs}$ times, each time omitting data points from one observation timestep (the validation set) and fit only the rest data points in the time series (the training set). This is implemented by putting the weights of the omitted timestep to zero in the fitting. We then computed the CV score as the residual sum of squares (RSS) between the fitted model and the validation data points. We take the average CV score over the $n_\mathrm{obs}$ runs as the final CV score for a given $\alpha$ value. 
We tested $\alpha$ values from 0 to 10000 and found that the best CV score is achieved around $\alpha$ = 2000-5000, as shown in Fig. \ref{fig:cv}. This value matches with the empirical choice by checking the map quality by eye and is also consistent with the value chosen in \cite{Crossfield2014}. We thus adopt an $\alpha$ value of 2000 in our reconstruction of WISE 1049AB maps.

Finally, we set the physical properties of WISE 1049AB in the Doppler imaging routine. 
The linear limb darkening coefficient is set to 0.4 based on fitting results. Tests showed that varying the linear limb darkening coefficient within $\pm$0.3 (typical uncertainty level from fitting) in the Doppler code does not significantly affect the resulting Doppler map structure. The $v\sin i$ and RV values are set to the best-fit value in Section \ref{sec:mf}. We set a period of 5 hr and an inclination of 80$^\circ$ for WISE 1049B, and a period of 7 hr and an inclination of 70$^\circ$ for A. During testing, we found that uncertainty in the period and inclination can affect the size, latitude, and longitude of the mapped features. These effects can be degenerate with other factors that influence the shape and location of the features, making the interpretation of the map more complex. We discuss some of these effects in Section \ref{sec:disc}.

After setting the parameters, the code loads the observed and model data, performs least square deconvolution and computes the $R$ matrix based on the model LSD profile and physical parameters. Model maps are generated and their $\chi^2$, entropy $S$ and metric $Q$ are computed and minimized iteratively using the 
\texttt{scipy.optimize.minimize} function until an optimal Doppler imaging solution is found. A representative plot showing the best-fit Doppler imaging solution $M$ to the data vector $D$ is shown in the upper panel of Fig. \ref{fig:residual_fit}, along with the metric, entropy of the map, and $\chi^2$ value of the fit. 
% dev plots
To better visualize the subtle changes in the line profile shapes over time, we plot only the deviation of each line profile from the time-averaged line profile using a colormap, as shown in the lower panel of \ref{fig:residual_fit}. We present this deviation plot for the observed line profiles, the modelled line profiles from the best-fit solution, and the residual between the two. The black dashed lines mark the edge of the brown dwarf disk where the projected velocity is maximum. The dark trace seen in the deviation plot is the signal of a dark patch rotating across the visible disk. It can be seen from the observed and modeled deviations in Fig. \ref{fig:residual_fit} that the Doppler imaging solver provides a good fit in this case, capturing the main deviation signals.
We further discuss the features of the deviation plots for each observation in Section 5 along with the Doppler map solutions.

%%%%%%%%%%%%%%%%%%%%%%%%%%%%%%%%%%%%%%%%%%%%%%%%%%%%%%%%
\section{Doppler maps of WISE 1049AB}
\label{sec:maps}

\begin{figure*}
\begin{center}
\includegraphics[width=0.95\textwidth]{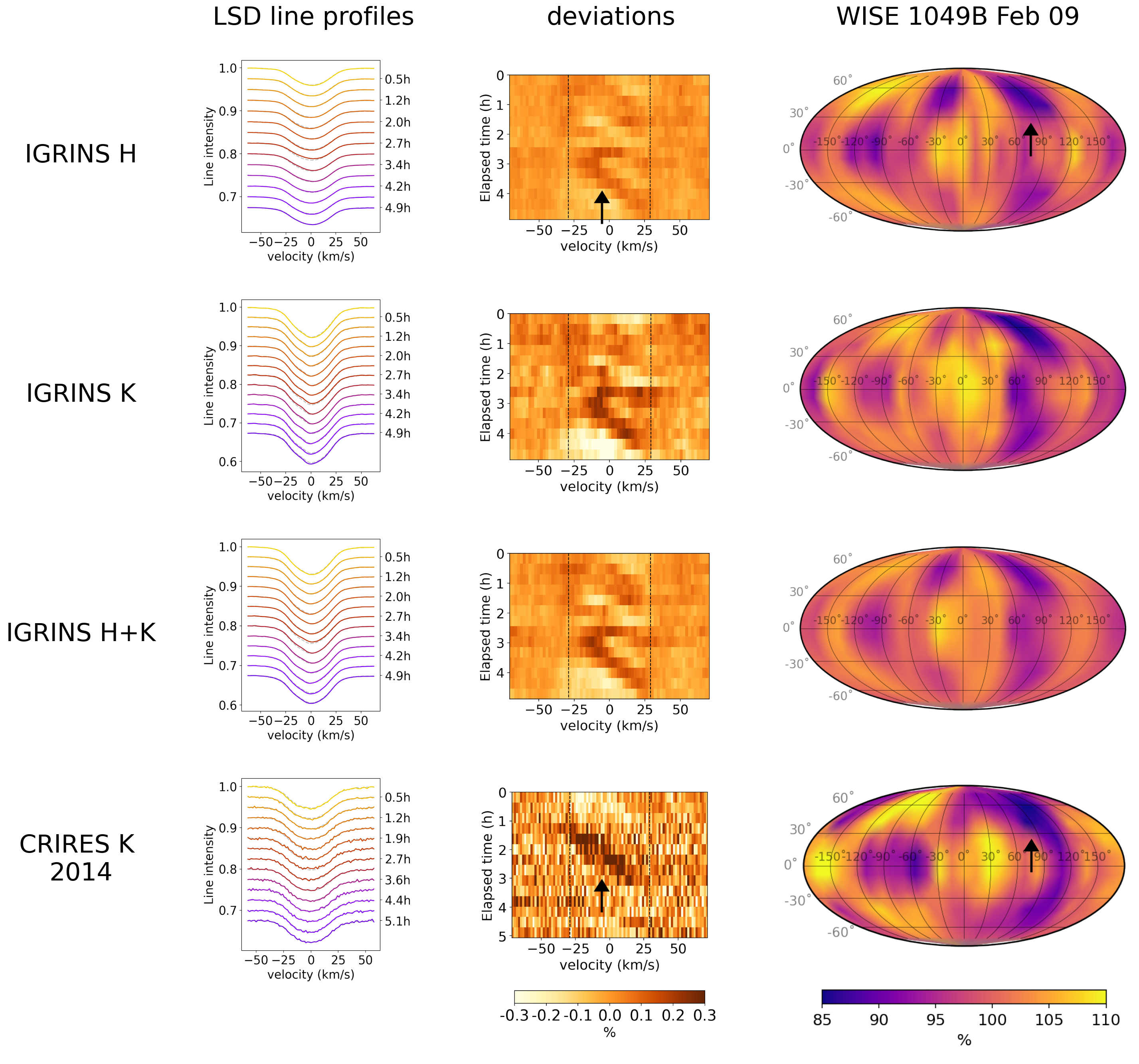}
\caption{The observed line profiles (\textit{left}), deviation from the time-averaged profile (\textit{middle}), and resulting Doppler map (\textit{right}) of WISE 1049B on the night of Feb 9, 2020. The first 3 rows show the maps from the IGRINS H band, K band, and H+K combined respectively. The black dashed lines in the middle column mark the edges of the visible brown dwarf disk at the equator where the projected velocity is maximum. All maps are shown on the same color scale. Unmappable areas below 80$^\circ$ south are marked in grey. The main feature of the Doppler maps is a dark spot-like pattern spanning from mid-to-high latitude regions in the northern hemisphere around 90$^\circ$ longitude, indicated by a black arrow. The reproduction of the Doppler map using the CRIRES K band data from \citet{Crossfield2014} is shown on the bottom row, with the resulting map rotated in phase such that the main dark feature aligns with the feature discovered on the IGRINS maps.}
\label{fig:mapsB1}
\end{center}
\end{figure*}

\begin{figure*}
\begin{center}
\includegraphics[width=0.95\textwidth]{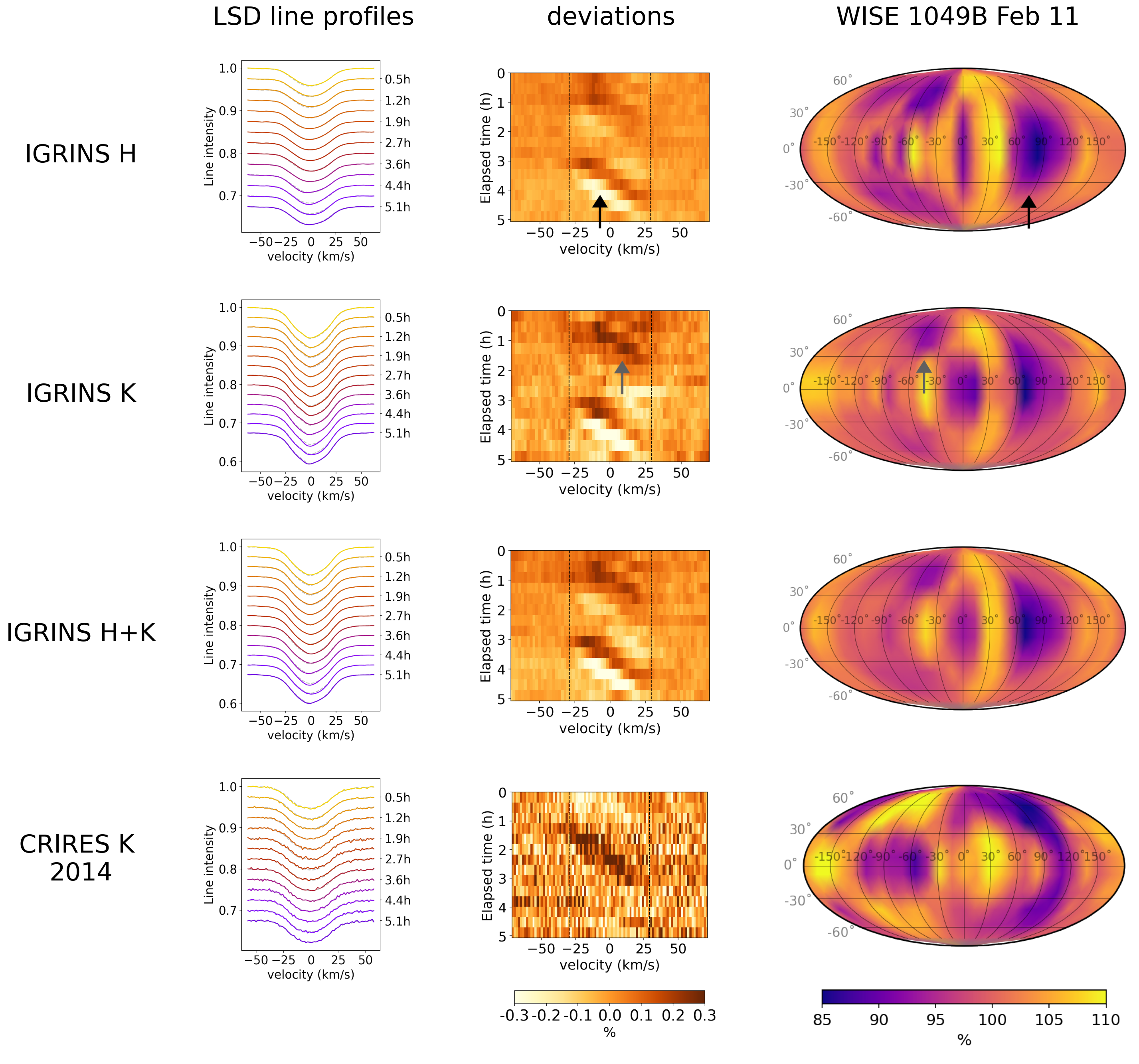}
\caption{The observed line profiles, deviations, and Doppler maps of WISE 1049B as in Figure \ref{fig:mapsB1}, but on the night of Feb 11, 2020. All maps are shown on the same color scale as in Fig. \ref{fig:mapsB1}. The main dark feature and possible secondary feature are marked with black and gray arrows respectively. The reproduction of the CRIRES 2014 map is rotated in phase such that the dark feature aligns with the feature discovered on the IGRINS maps.}
\label{fig:mapsB}
\end{center}
\end{figure*}

\begin{figure*}
\begin{center}
\includegraphics[width=0.95\textwidth]{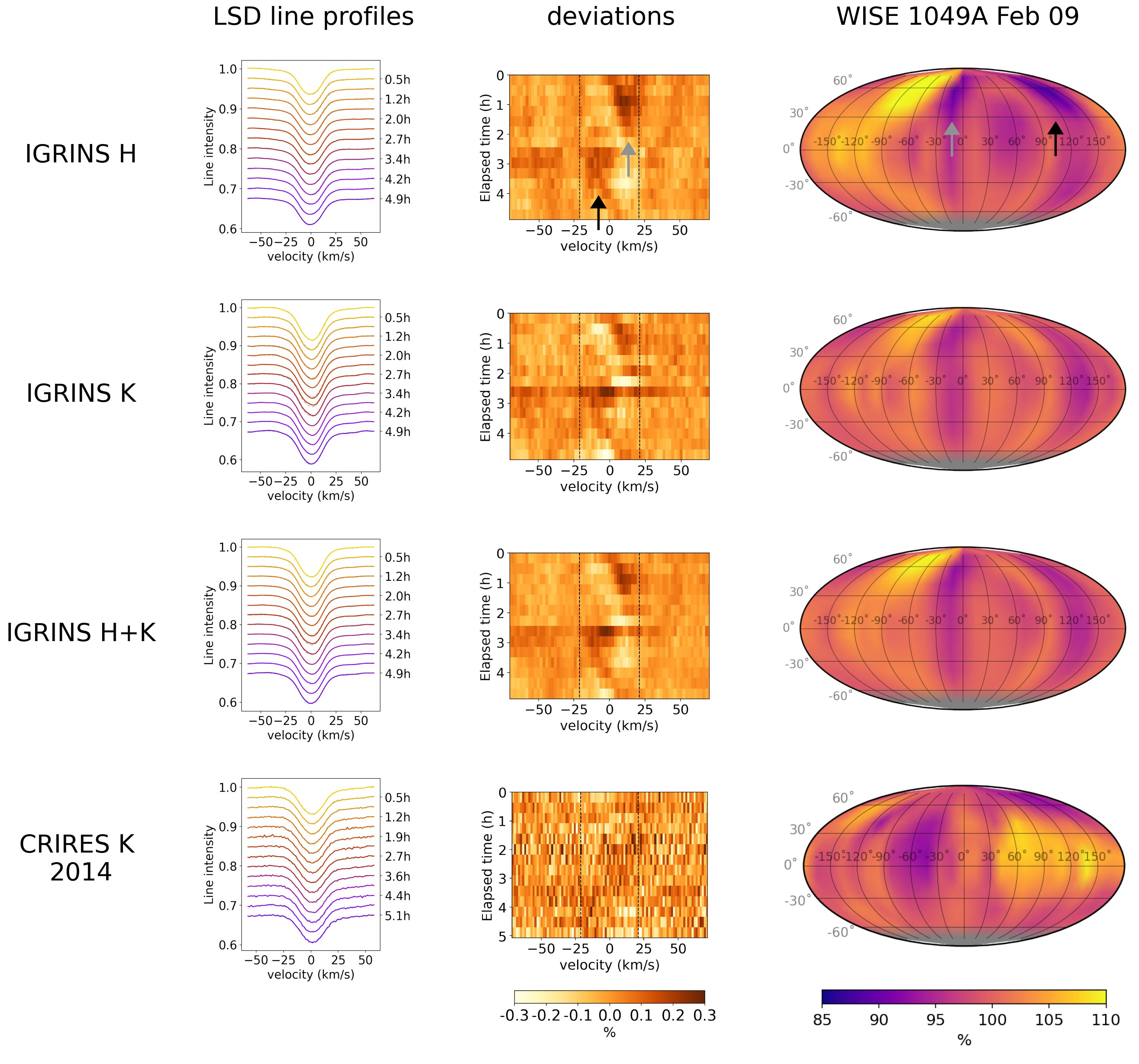}
\caption{Same as Figure \ref{fig:mapsB1}, but for WISE 1049A, on the night of Feb 9, 2020. All maps are displayed on the same color scale as the WISE 1049B maps to facilitate a direct comparison of signal strengths between the A and B components. Unmappable areas below 70$^\circ$ South are marked in grey. The identified polar features on the map and their corresponding traces in the deviation plots are marked with arrows. The reproduction of the CRIRES WISE 1049A map from \citet{Crossfield2014} is shown on the bottom row and is not rotated due to the lack of features.}
\label{fig:mapsA1}
\end{center}
\end{figure*}

\begin{figure*}
\begin{center}
\includegraphics[width=0.95\textwidth]{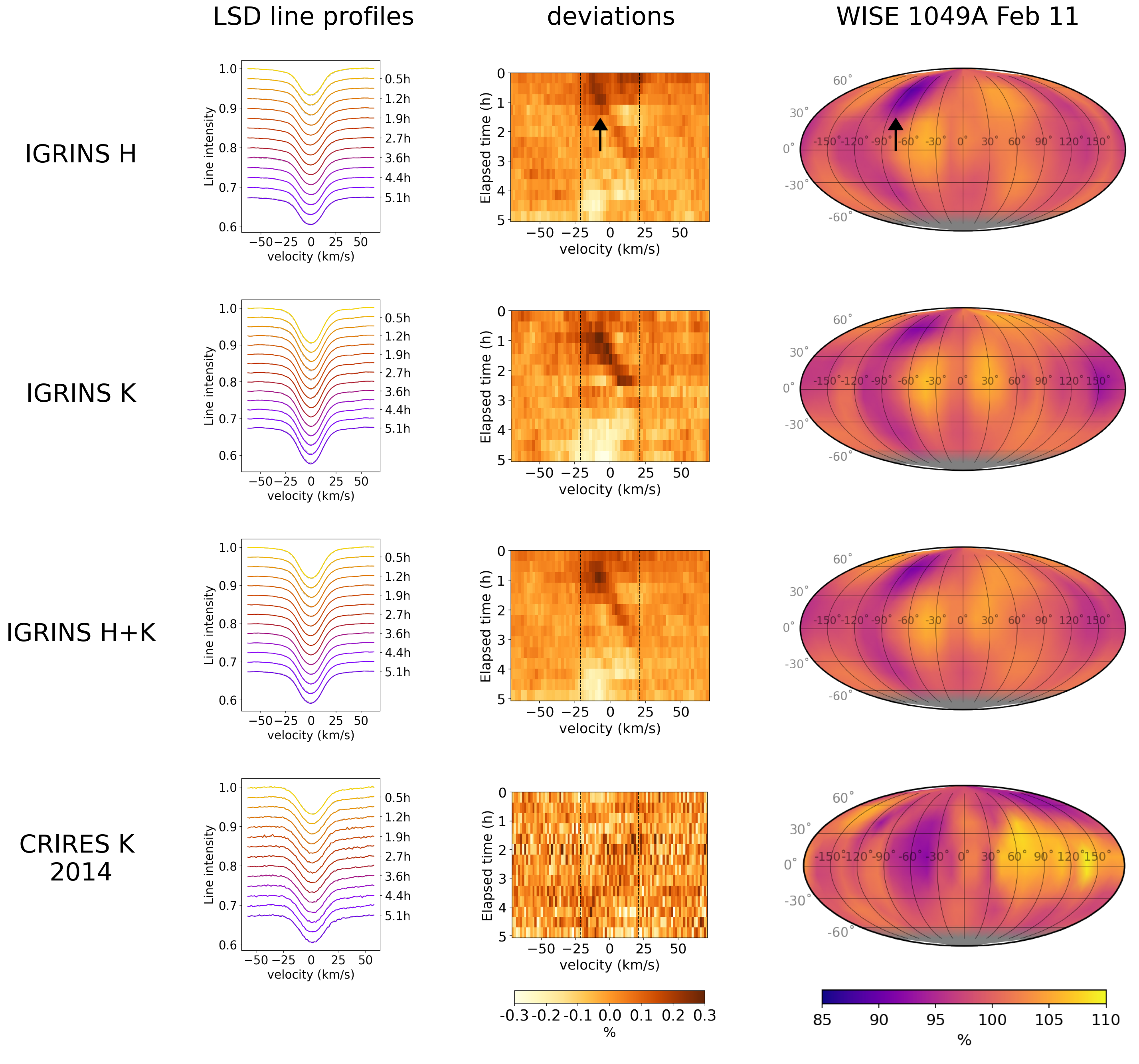}
\caption{Same as Figure \ref{fig:mapsB}, but for WISE 1049A, on the night of Feb 11, 2020. All maps are displayed on the same color scale as the WISE 1049B maps. The identified polar feature is marked with black arrows on the map and the deviation plot. The reproduction of the CRIRES WISE 1049A map is shown on the bottom row and not rotated due to the lack of features.}
\label{fig:mapsA}
\end{center}
\end{figure*}

%%%%%%%%%%%%%%%%%%%%%%%%%%%%%%%%%%%%%%%%%%%%%%%%%%%%%%%%
\subsection{WISE 1049B maps}

% defining the map coords
The retrieved Doppler maps for WISE 1049B in both H and K bands using the maximum entropy method are shown in the third column of Fig. \ref{fig:mapsB1} and \ref{fig:mapsB} for Feb 9 and Feb 11, with the corresponding deviation plots shown in the second column. The $\chi^2$, entropy, and metric along with the modelled and residual deviation profiles as in Fig. \ref{fig:residual_fit} for the rest Doppler maps are included in Appendix \ref{app:residual}. We defined the longitude directly facing the observer at the beginning of observation as 0$^\circ$. Due to the 80$^\circ$ inclination of WISE 1049B, the regions below 80$^\circ$ south are not visible to Doppler mapping, thus the fluxes shown in those regions are flat initial guesses. All maps are shown on the same color scale which represents the brightness temperature of the TOA in percentage compared to a uniform background. 

\subsubsection{First night map}
% main feature on Feb 9 W1049B map
On the Feb 9 Doppler map in Fig. \ref{fig:mapsB1}, a prominent dark spot-like feature is observed, spanning from mid-to-high latitude regions in the northern hemisphere around 90$^\circ$ longitude, marked with a black arrow in the H band map. The presence of a dark trace in the deviation plot in the second column of Fig. \ref{fig:mapsB1} from 3h to $\sim$4.5h (also marked with an arrow) suggests that the dark spot is a true signal from the data, rather than artifacts introduced by the image reconstruction routine. This feature is detected in both IGRINS H and K bands at similar longitudes. 
We also observe a fainter mirror image of this feature in the southern hemisphere, nearly connecting with that in the northern hemisphere and creating the appearance of an elongated pattern. This is a known limitation of the Doppler imaging technique when applied to objects with nearly equator-on inclinations (\citealt{Vogt1987}, see also simulations in Section \ref{sec:sims}). 
While some other potential secondary features are discernible in the H and K maps, such as fainter spots in equatorial regions, their correspondence to features in the deviation plot cannot be confirmed, as all other signals in the modelled deviation plot (see Fig. \ref{fig:res_B1H} and \ref{fig:res_B1K}) are many times fainter and blended together, making them indistinguishable from noise.

Because we are essentially probing a map of brightness temperature with Doppler imaging, the darker regions in the map correspond to high-altitude, colder regions in the atmosphere, whereas the brighter regions correspond to a view into the hotter, deeper layers of the atmosphere. To confirm the nature of these identified features, simulated maps with injected TOA models are presented and discussed in Section \ref{sec:interp}, taking into account the emission contributions from the H and K bands.

\subsubsection{Second night map}

% main feature on Feb 11 W1049B map
The main feature seen on the Feb 11 map in Fig. \ref{fig:mapsB} is a dark spot extending between the mid-latitudes of both hemispheres, located around 60$^\circ$-90$^\circ$ longitudes by our definition (marked with a black arrow). The corresponding trace can be seen in the deviation plots between 3h to 4.5h in Fig. \ref{fig:mapsB} as indicated by the black arrow. The feature is also detected in both the H band and K band. A possible secondary feature also emerges in the Feb 11 deviation plots and is stronger than that in the first night. This is indicated by a dark trace between 0.5h to 1.5h, particularly noticeable in the K band, as pointed out with a gray arrow. This feature potentially appeared in the reconstructed maps as a higher-latitude spot located around -60$^\circ$ longitude, also marked with a gray arrow on the map. 

% Compare the two nights
Comparing the map from the two nights, they both feature a dominant spot-like structure in their TOA maps. The traces left by the spots in the deviation plot are similar in terms of shape, extent, and slope.
In the deviation plot, spots near the equator are expected to induce deviations across the entire line profile and should move across the full span of projected velocities (between the vertical dashed lines), while spots at higher latitudes should move more slowly, affecting a narrower range of velocities and resulting in a larger slope of the trace pattern. 
Although the reconstructed maps indicate that the spot is at a higher latitude on the first night compared to the second, we do not observe a significant difference in the slope or extent of the traces between the two nights. This suggests that the two features we detected on the two nights could be the same evolving structure in the atmosphere of WISE 1049B.
In addition, we do notice a slight difference in the trace pattern between the two nights: on the first night, the signal was mainly dark (i.e., spot fainter than the background level), while on the second night, it is a dark trace directly followed by a bright trace, which results in a map with alternating dark and bright regions. The cause of this difference remains unclear within our dataset's noise level.

Nevertheless, the fact that the two nights' traces appeared at almost the same rotational phase is a striking coincidence, if not caused by a systematic error. Assuming that what we detected is the same TOA feature returning to the exact same phase angle after a separation of 48.3 hours between the two nights, this would imply an alternative way of measuring the true rotational period for WISE 1049B: Considering that its period is around 5h (which is the assumed period in constructing the Doppler maps), the brown dwarf would have a period of 4.83h if it has gone through 10 rotations, or a period of 5.37h if it has gone through 9 rotations.
The fact that WISE 1049B showed a similar TOA structure on two nearby nights provides important insights into the timescale of the atmosphere dynamics of similar L/T transition objects, suggesting that these structures persist for at least two days.

\subsubsection{H vs. K band}
% Compare H and K map for Feb 11 - pressures

Spectra in different wavelengths probe different pressure levels and thus different vertical layers in the brown dwarf atmosphere. 
% similarity in H and K
Comparing the map in H and K bands, we did not find a significant difference in the shape, size, and location of the dark feature between the two bands in general. Considering the similarity in the H and K maps, we also produced an H+K band map by running a reconstruction using all available spectral orders from both the H and K bands, shown in the third row of Fig. \ref{fig:mapsB}. 
The combined map shows a similar structure to the separate maps, with a cleaner trace pattern in the deviation plot which translates to a slightly enhanced spot contrast in the maps. This confirms that there are no significant phase shifts between the H and K bands. 
%Previously, phase shifts between the sinusoidally varying light curves have been found between different wavelengths for WISE 1049B \citep{Biller2013}, indicating that the TOA structures causing the variability may differ in position in the vertical layers of the atmosphere.
% slight differences between H and K
Meanwhile, we do notice some slight differences between the H and K maps: In the Feb 9 map, a potential spot is seen in H  around 30$^\circ$, but it is not significant in K; In the Feb 11 map, a careful comparison of the latitude of the retrieved spots in the H and K bands potentially shows that the spot in the K band is smaller in size and slightly shifted to the left (i.e. smaller longitude) than that in the H band map. However, given the noise level of our results, we refrain from interpreting these small differences.
We further discuss the implication of the H and K maps on the vertical structure of WISE 1049AB atmospheres in Section \ref{sec:pres}.

\subsubsection{Comparison with the 2014 map}
% Compare with CRIRES map - long-term trends

Our findings are directly comparable to those of \cite{Crossfield2014} as we employed a similar method but with data obtained from distinct epochs separated by 7 years.
Therefore, we applied our modified Doppler imaging routine to the original CRIRES dataset from \cite{Crossfield2014}, which was observed in May 2013. The resulting maps are presented in the last row of Fig. \ref{fig:mapsB}. Our re-analysis successfully reproduced the main spot feature as found in the 2014 map. 
The reproduced CRIRES map is shown on the bottom rows of Fig. \ref{fig:mapsB1} and \ref{fig:mapsB}, and is rotated such that the dark spot feature (marked with a black arrow) lies roughly at the same longitude as the new IGRINS maps.
%Compared to the CRIRES map, the new IGRINS maps have a wider wavelength coverage and thus contain information from many more lines. However, the new IGRINS maps are noisier than the 2014 maps due to the lower spectral resolution (only half that of CRIRES).

The IGRINS maps from Feb 9 and 11, 2020 both resemble the 2014 map, as they all reveal a prominent dark spot feature. The Feb 11 map more closely matches the 2014 map in terms of the spot’s location at higher latitudes compared to the Feb 9 map.
Although it is not possible to compare the longitude of the spots found in the 2014 vs. 2020 maps nor to confirm if the spots found are the same evolving structure due to the long time interval between the two observation epochs, this still provides evidence that large patches of TOA structure may be common and long-lasting in brown dwarfs. 
%This is supported by the general circulation models of \cite{Tan2021b}. 
The potential presence of a persistent cloud feature on WISE 1049B is proposed in \cite{Karalidi2016} by examining light curves collected from various epochs of observations separated by months. 
Our dataset contributes two additional epochs to this ongoing discussion and again identifies a possible stable cloud structure.
The interpretation of the longevity of this structure remains uncertain, especially considering the rapidly evolving light curves from recent TESS \citep{Apai2021, Fuda2024} and JWST observations \citep{Biller2024}.
It could suggest the presence of a stable formation like the Great Red Spot on Jupiter, or recurring structures that form and dissipate over different periods, such as the evolving patchy storms in globally propagating waves predicted by the general circulation models (e.g. \citealt{Tan2021b}). In the context of the latter case, our results could imply a preferred size of TOA structure on brown dwarfs in the L/T transition. We further discuss the physical interpretation of the Doppler maps in Section \ref{sec:interp}.

General circulation models of similar objects predict that east-west traveling global-scale equatorial waves can form hot or cold patches. Such structures can produce wave-like light curves in single rotations, and their formation, propagation, and dissipation can cause irregular long-term light curve evolution (\citealt{Tan2021b}, Tan et al. in prep). This scenario is supported by the two sets of Doppler imaging observations from 2014 and this work, while also being able to explain the multiple planetary-scale waves observed in long-term photometry (e.g. \citealt{Apai2021, Fuda2024}).

%Longer multi-wavelength Doppler imaging observations of WISE 1049B over continuous rotations could help to clarify the nature of this structure.

%%%%%%%%%%%%%%%%%%%%%%%%%%%%%%%%%%%%%%%%%%%%%%%%%%%%%%%%
\subsection{WISE 1049A maps}

The reconstructed maps for WISE 1049A using the maximum entropy method are shown in Fig. \ref{fig:mapsA1} and \ref{fig:mapsA}. Similar to earlier sections, this figure includes maps in the H, K, combined H+K bands, and those from the 2014 CRIRES data for comparative analysis.

Contrary to findings from the CRIRES data by \cite{Crossfield2014}, which suggested WISE 1049A to be likely featureless, our new dataset has tentatively revealed a spot-like feature in A. 
% 1st night
We identified a polar spot on the Feb 9 map at $\sim$140$^\circ$ longitude, which corresponds to the dark trace from 3h to 5h in the deviation plots, as pointed out with black arrows in the first row of Fig. \ref{fig:mapsA1}. We identified another trace from 0h to around 1.5h, which corresponds to a second dark patch at polar regions at around -10$^\circ$ longitude, marked with gray arrows.
% 2nd night
For the second night of Feb 11,
we identified a dark trace spanning shortly after 0h to $\sim$2.5h in the deviation plots, matched with a high-latitude spot appearing around -90$^\circ$ longitude in the Doppler map. This feature is again marked with black arrows in Fig. \ref{fig:mapsA}.
For all these identified polar features, we can see their image in the northern hemisphere and a weak mirror image in the southern hemisphere. These features are retrieved in both the H band and the K band, as well as the combined map. 
Like the case of WISE 1049B, the traces from two nights of WISE 1049A also show similar shape, extent, and slope. The feature marked with black arrows on Feb 9 could be the same TOA feature as the one found on Feb 11, but shifted in phase by $\sim$230$^\circ$ after 48.3 hours between the two observations.
The trace of this feature appeared $\sim$3h earlier in the deviation plots on Feb 11 compared to Feb 9, which means that if we are probing the same TOA feature, it would return to the initial phase angle after a total of 51.3 hours. This allows a measurement of WISE 1049A's rotational period. Considering a period of around 7h for A, this would imply a period of 6.4h if A has completed 8 rotations, or a period of 7.3h if it has completed 7 rotations.

% comparison to 2014
In comparison to the new WISE 1049A maps from the 2020 dataset, our reproduction of the 2014 map from CRIRES did not reveal a similar polar spot feature. This confirms that the spot is a distinct feature from the new IGRINS data set and not an artifact introduced by our Doppler imaging routine. The appearance of a similar pattern on both nights but at different rotational phases further supports that this is an actual TOA feature on the brown dwarf seen at different phase angles. Compared with the WISE 1049B map on the same color scale, the TOA feature on WISE 1049A is fainter, which could imply either it has less temperature contrast with the background or that it is smaller in size. This observation aligns with many previous studies that suggest WISE 1049A is less variable than the B component (e.g., \citealt{Biller2013, Buenzli2015}). The TOA feature on WISE 1049A is closer to the polar region, whereas the spot found on the B component is at lower latitudes. This could imply different formation mechanisms for the observed structures on WISE 1049A and B under different dynamical regimes. This is supported by 3D global circulation models that have shown equatorial regions to be dominated by zonal jets, while polar regions host more vortices and storms (e.g., \citealt{Tan2021b}).

% caveats
It should also be noted that the map of A should be interpreted with caution due to its low amplitude of variability, less constrained period, and incomplete phase coverage, which can lead to spurious map features like dark spots/stripes. We provide a simulated map with a spot model in Section \ref{sec:interp} to confirm the nature of the discovered structure, and we further discuss the effect of incomplete phase coverage in Section \ref{sec:disc}. 

%%%%%%%%%%%%%%%%%%%%%%%%%%%%%%%%%%%%%%%%%%%%%%%%%%%%%%%%
\subsection{Comparing with other Doppler imaging solvers}

To test the reliability of the Doppler mapping results with our default reconstruction routine, we ran our data through another pipeline built on the Doppler imaging module of the \textsc{starry} package. This code uses a linear combination of spherical harmonics to represent the brightness distribution on a spherical surface, and employs a variety of solving strategies for the inversion \citep{Luger2021}. For specific details regarding the implementation of Doppler imaging solvers in \textsc{starry}, we direct readers to \cite{Luger2021}. 

We conducted tests with the Feb 11 WISE 1049B data set using both the linear and optimization solvers provided by \textsc{starry}, both with and without the least-square deconvolution step. We include the results in Fig. \ref{fig:solvers} in the Appendix for interested readers.
We found that all solvers successfully captured the primary dark spot in the maps, except for the direct optimization without LSD pre-processing. 
We also found that in general, solvers with LSD pre-processing perform better than those without it. 
This is expected, as achieving a good Doppler imaging solution without LSD at the SNR level of our data is challenging due to the inherent faintness of brown dwarfs for this technique. 
The fact that \textsc{starry} linear solver without LSD retrieved a broadly similar feature compared to maps including LSD rules out the possibility that the feature is an artifact of the LSD process. 
The agreement between maximum entropy maps and \textsc{starry} LSD maps confirms that the retrieved map features are not artifacts arising from either of the mapping codes.
Considering all, it is safe to conclude that the observed feature is a true signal derived from the data and not an artifact of the data processing methods. 
However, to interpret the structures found in Doppler maps, further tests using simulations are necessary, as detailed in the next section.

%%%%%%%%%%%%%%%%%%%%%%%%%%%%%%%%%%%%%%%%%%%%%%%%%%%%%%%%
%%%%%%%%%%%%%%%%%%%%%%%%%%%%%%%%%%%%%%%%%%%%%%%%%%%%%%%%
%%%%%%%%%%%%%%%%%%%%%%%%%%%%%%%%%%%%%%%%%%%%%%%%%%%%%%%%
\section{Interpretation of the maps}
\label{sec:interp}

%%%%%%%%%%%%%%%%%%%%%%%%%%%%%%%%%%%%%%%%%%%%%%%%%%%%%%%%
\subsection{Comparison with simulated maps from models}
\label{sec:sims}

\begin{figure*}
\begin{center}
\includegraphics[width=\textwidth]{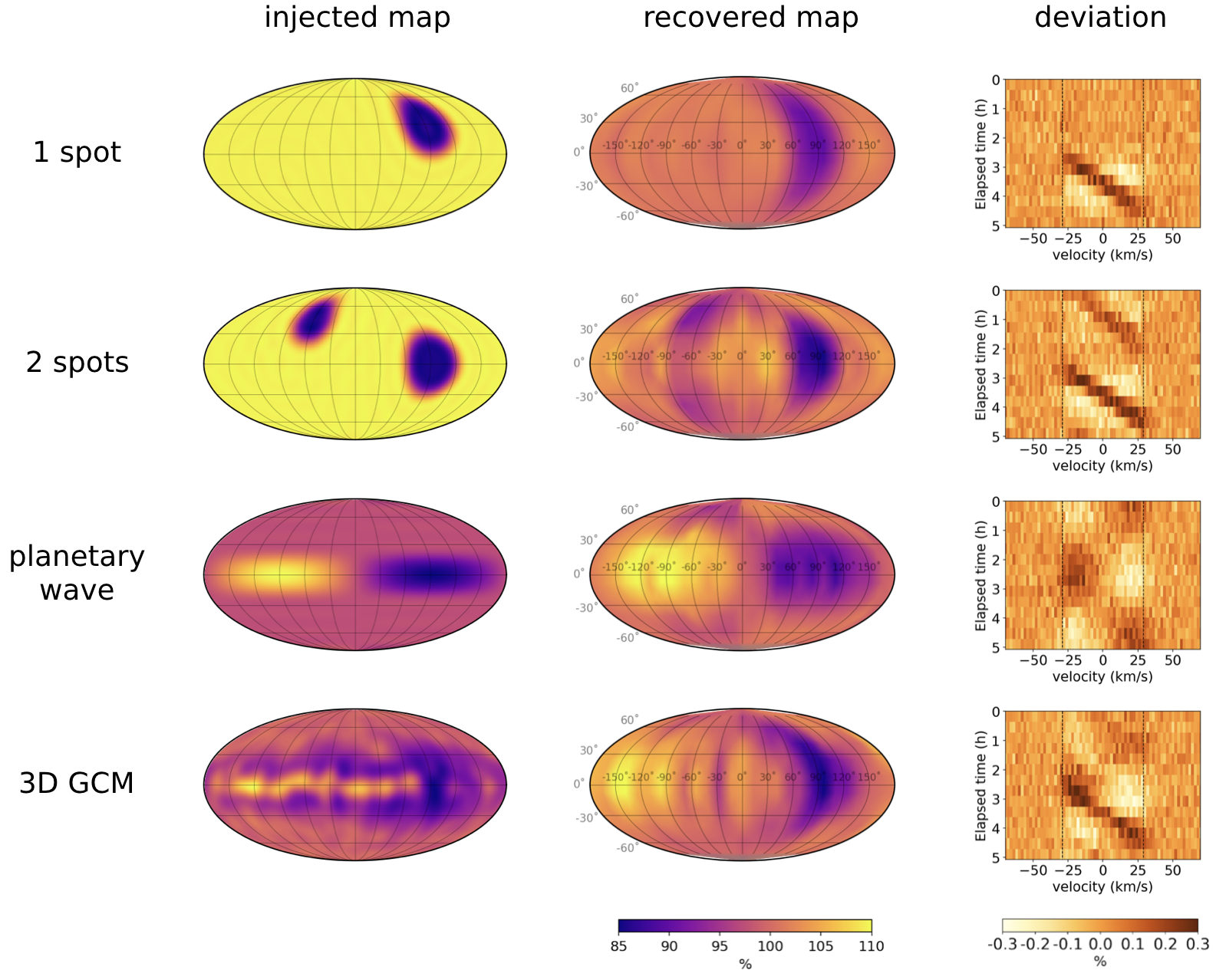}
\caption{Simulated Doppler maps and the corresponding deviation plots using different injected models. The spot, planetary-wave, and GCM models showed distinct structures in the recovered Doppler maps and deviation plots, demonstrating that Doppler imaging can to some extent distinguish between these scenarios. We found that the observed Doppler maps more closely match models with GCM or spot than models with only a global wave. In the GCM model, global-scale equatorial waves with hot and cold spots are present, but only the dominant patchy structure is recovered in the simulated map.}
\label{fig:simsB}
\end{center}
\end{figure*}

To investigate the nature of the discovered structures, we simulated Doppler maps retrieved from brightness maps with injected TOA features. To achieve this, we first generated a surface map with a modelled surface brightness distribution (e.g. flat background with a spot or band). Next, we generated a series of rotationally broadened spectra at different rotational phases from the best-fit model spectra using the \textsc{starry} package and added random noise to each spectrum. We kept the noise at the same SNR level as the observed spectrum.
We then performed LSD and applied the same Doppler image reconstruction routine described in Section \ref{sec:dime} to the noise-added mock spectral data set. We first test three different models (spots, planetary-scale waves, and general circulation model) with WISE 1049B, and then discuss the case of WISE 1049A. 

\subsubsection{Spots}

We first explore a one-spot model for WISE 1049B. By fitting spot parameters to photometric lightcurves, \cite{Karalidi2016} recovered 4 spots on WISE 1049B, including a mid-latitude dark spot consistent with \cite{Crossfield2014}. Our analysis of the IGRINS data revealed a similar dark spot feature.

We injected spot on a latitude and longitude such that the dark trace in the deviation plot can be reproduced at the same location in the time-velocity space. 
For WISE 1049B, we first injected a mid-latitude spot with parameters similar to those found in \cite{Crossfield2014}, with a radius of $\sim$30$^\circ$ and positioned at 30$^\circ$ latitude, 115$^\circ$ longitude (defining 0$^\circ$ as the longitude directly facing the observer at t=0). 
The spot brightness was set to 80\% of the background. In our tests with different spot parameters, we observed a degeneracy between spot size and spot brightness contrast in the resulting map. Also, a smaller brightness contrast has a similar effect to a lower SNR in the spectra.

The first row of Fig.~\ref{fig:simsB} displays the injected map, retrieved map, and the corresponding deviation plot for this one-spot model.
The trace of the spot in the deviation plot resembles the one observed in the IGRINS H and K band for WISE 1049B. 
Our Doppler imaging routine successfully recovered the spot in the simulated map, capturing both its size and longitude reasonably well. 
The most noticeable distortion is that even though the spot was initially placed in the mid-latitude region of the northern hemisphere, both the original spot and a weak mirror image in the southern hemisphere are retrieved. They appear almost connected, forming an elongated dark strip that is nearly symmetric in both hemispheres. 
This limitation is well-known in Doppler imaging, arising from the degeneracy in Doppler shift in the latitudinal direction, particularly at nearly edge-on inclinations. Patterns in one hemisphere can induce an artificial mirror image in the opposite hemisphere, resulting in longitudinally elongated patterns (e.g. \citealt{Vogt1987}).
This explains why the dark features in almost all the observed IGRINS maps are elongated and suggests that these patterns should originally be spot-like. 
Despite the mirroring issue, a closer inspection of the recovered map shows that the spot still appears darker in the northern hemisphere than in the southern part. This brightness asymmetry provides clues about where the spot is originally located.

From our simulation, it can be seen that a single-spot model can already capture the most dominant feature in the observed maps. Since we also observed a secondary trace-like structure between 0-2h, we also explored the possibility of a second spot. 
We tested a more realistic two-spot model. This time we kept the primary spot the same size as in the 1-spot map but placed it near the equator, which matches better with the Feb 11 maps. 
The second spot was injected at a higher latitude with a smaller radius, positioned at -90$^\circ$ longitude to reproduce the dark trace seen between 0-2h. 
As a result, we retrieved two traces in the simulated deviation plot, with the second trace much fainter than the first one. 
As seen in the second row of Fig. \ref{fig:simsB}, the smaller polar spot is also recovered, although with faint contrast. However, the longitude of the recovered spot at $\sim$-70$^\circ$ does not precisely align with that of the observed secondary spot at approximately -50$^\circ$ longitude. Thus, we refrain from definitively attributing this feature to an authentic structure, acknowledging the possibility of it being a result of noise.

\subsubsection{Planetary wave in bands}

Apart from elliptical spots, both WISE 1049A and B have shown evidence of zonal circulation in previous studies. Models involving band-like structures in brown dwarf atmospheres have been proposed based on long-term monitoring \citep{Apai2017, Apai2021, Fuda2024} and polarization measurements \citep{Millar-Blanchaer2020}. \citet{Apai2017, Apai2021} proposed that planetary-scale waves formed in bands with various wind speeds based on several observed peaks in the periodogram of WISE 1049B. \cite{Millar-Blanchaer2020} proposed the existence of cloud bands on WISE 1049A to explain the observed polarization level. Modeling work by \citet{Mukherjee2021} demonstrated that enhanced cloud bands at low latitudes help to explain the observed polarization observations of WISE 1049B. \citet{Vos2017} and \citet{Suarez2023} reported correlations between inclination angle, color, and silicate absorption that support enhanced equatorial clouds or zonal banding of clouds on brown dwarfs. Zonal bands are also observed on Solar system gas giants. Probing these zonal structures is important to constrain the atmospheric dynamics of brown dwarfs and giant exoplanets. Thus, we assess the capability of Doppler imaging to detect such structures. 

Since Doppler imaging relies on the time-variation of line profile shapes, which is only sensitive to longitudinally varying structures, it is practically incompetent in detecting globally uniform bands. 
The effect of a uniform band-like structure can only be discerned from deviations in the time-mean line shape from the modelled profile. This requires careful consideration of additional line-broadening and line-shifting effects within the model spectrum, which is beyond the scope of this study.
We ran a simulation involving a non-varying band positioned at the equator and another one at a higher latitude. In both cases, no significant variation was detected in the resulting maps.

Therefore, we only present here the simulation of a sinusoidally varying planetary-scale wave confined in a band, which is inspired by the model in \cite{Apai2017, Apai2021}.
We modelled a band with a width of 30$^\circ$, and a brightness variation amplitude of 30\%, such that the brightness at the trough is 70\% of the background level. We positioned the band at the equator, aligning the trough roughly with the observed dark feature at 90$^\circ$. The resulting maps are shown in the third row of Fig. \ref{fig:simsB}. 

The band with a planetary-scale wave is recovered as an equatorial large-scale, extended dark and bright patches centered on the trough of the wave in the simulated map. When comparing the deviation plots, the dark traces in the planetary-scale wave model are more spread out in time and display an alternate bright and dark pattern, which we do not observe for WISE 1049B.

The simulation qualitatively suggests that a spot-like model more closely matches the observed WISE 1049B Doppler map than the model with one sinusoidally-varying global wave. With our detection sensitivity, we can rule out waves with amplitude greater than 30\%. 
Although we cannot exclude the possibility of band-like structures with smaller brightness variations or multiple bands with shifted phases that cancel each other out, it is safe to say that the dominant features detected in the Doppler maps are more consistent with a spot-like morphology.
Nevertheless, it is important to recognize that both spots and bands are simplified representations and do not fully capture the complex atmospheric dynamics of a brown dwarf. The true scenario likely involves an interplay of both types of structures, co-existing and co-evolving within the atmosphere.

\subsubsection{General circulation models}
\label{sec:gcm}

Since brown dwarf atmospheres are 3D in nature, 3D models are necessary to explain the observed features. To address this, we simulated a Doppler map using a brightness temperature map derived from a general circulation model (GCM). The GCM in this work has been updated from that in \cite{Tan2021a, Tan2021b} by using a more realistic non-grey radiative transfer and assuming equilibrium chemistry.  We model a typical L7 type planetary-mass companion with $T_\text{eff}$ = 1000K, $\log g$ = 3.2, and solar composition (Tan et al. in prep). 
The global GCM map shows a global-scale equatorial wave driven by cloud radiative feedback as the dominant surface inhomogeneous feature, accompanied by relatively small-scale turbulent features at mid-to-high latitudes. We reduced the spatial resolution of the GCM map before inputting it to \textsc{starry} to produce synthetic spectra. Thus small-scale vortices were filtered out in our simulations but the dominant structure was fully preserved. The contrast between the darkest areas and the mean background level was normalized to roughly the same as the spot simulations ($\sim$ 80\%).

The recovered maps and corresponding deviation plots are shown in the fourth row of Fig. \ref{fig:simsB}.
Only the main patchy structure located around 90$^\circ$ longitude is recovered in the Doppler map. The equatorial band is recovered as a faint background with planetary wave-like brightness variation, but the latitudinal information is lost completely. This indicates that Doppler maps are dominated by the hot spots and cold spots in the TOA, and only the largest-scale structures are detectable in Doppler maps. 
%The actual scenario is likely a mixture of spots and bands, similar to the results of GCM simulations.
It is also noteworthy that the Doppler imaging reconstruction drastically smooths out TOA structures. Therefore we recovered the most dominant global-scale wave feature that significantly affects the light curve variability, but leaving smaller-scale spots and turbulences unresolved, which is no surprise.
This tells us that we should be cautious when making conclusions about the size scale of TOA structures from Doppler maps alone.

% discussion on spot vs. wave
The simulated map from GCM showed good consistency with the observed WISE 1049B Doppler map presented in this work. 
The bright and dark patches identified in the WISE 1049B map can be attributed to the hot and cold patches formed by the traveling global-scale equatorial waves predicted by the GCM.
These hot and cold patchy spots can produce wave-like light curves in single rotations, and their formation, propagation and dissipation can drive irregular long-term light curve evolution (\citealt{Tan2021b}, Tan et al. in prep). This scenario, while consistent with Doppler imaging observations, also explains well the multiple planetary-scale waves observed in long-term photometry (e.g. \citealt{Apai2021, Fuda2024}), which is often considered in tension with spot-like models.
The fact that GCM aligns well with both Doppler imaging and photometry supports the idea that these two methods could be both probing the same complex atmospheric circulation scenario from different perspectives, due to their sensitivity to different feature morphologies. Both types of studies will contribute to a comprehensive understanding of the complex mechanisms governing weather on brown dwarfs and giant exoplanets.

%%%%%%%%%%%%%%%%%%%%%%%%%%%%%%%%%%%%%%%%%%%%%%%%%%%%%%%%%%%%%%%%

\begin{figure*}
\begin{center}
\includegraphics[width=\textwidth]{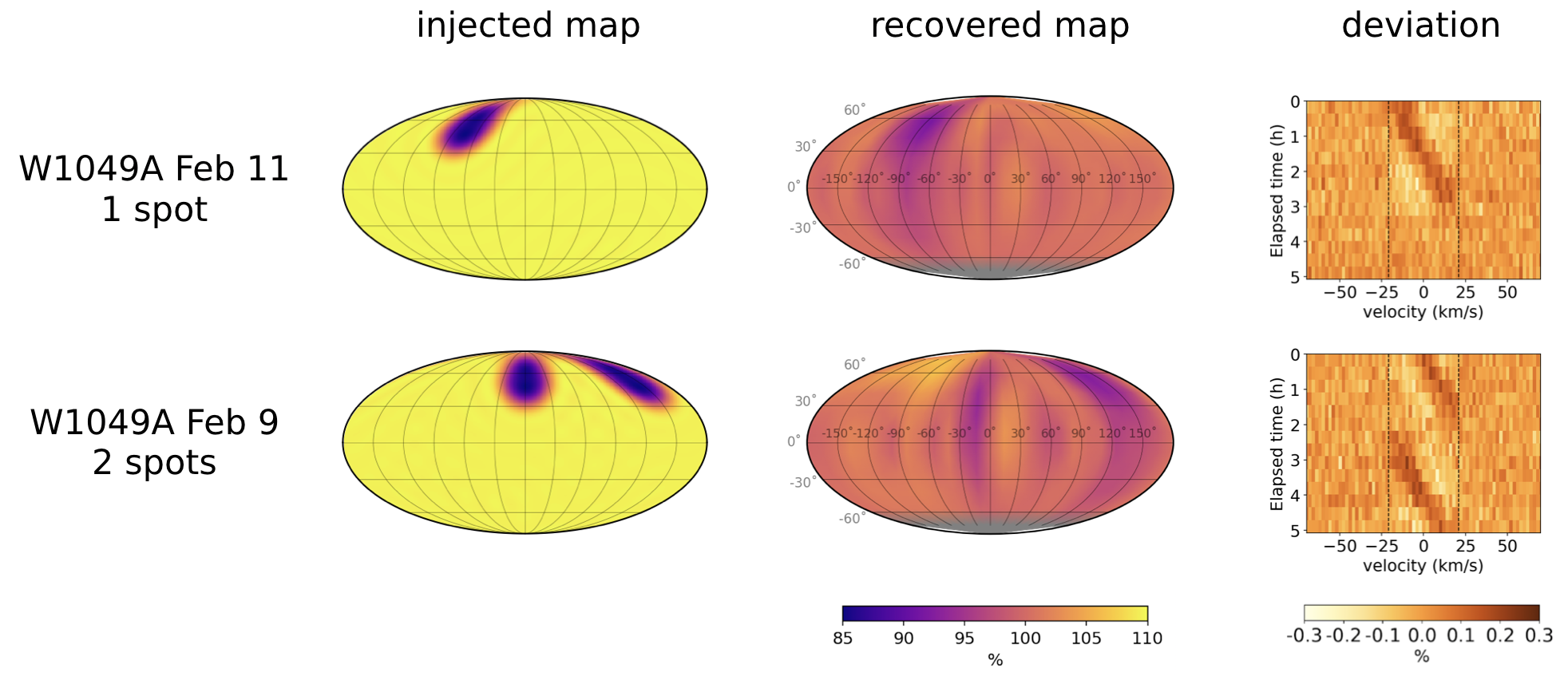}
\caption{Simulated Doppler maps for WISE 1049A with injected spot models. On the night of Feb 11, a single polar spot model reproduces the observed map, while on Feb 9, a model with two polar spots is needed to reproduce the observed map.}
\label{fig:simsA}
\end{center}
\end{figure*}

\subsubsection{WISE 1049A models}

For WISE 1049A, we simulated a model with a high-latitude spot with 30$^\circ$ radius at -75$^\circ$ longitude for the Feb 11 map, assuming a true period of 7h. We then modelled the Feb 9 map by shifting the spot to 140$^\circ$ longitude and adding another fainter spot at 0$^\circ$ longitude. We recovered both the Feb 11 spot and the two Feb 9 spots in our simulated maps, shown in Fig. \ref{fig:simsA}. This supports that the dark spot found in the IGRINS Doppler map is likely a real structure in the brown dwarf atmosphere. 

Photometric monitoring and previous mapping efforts have indicated that the measured variability for WISE 1049AB changes over time, with a general observation that A is less variable than the B component. Given the maximum observed variability amplitude of 4\%, any dominant cloud feature on A would exhibit a much smaller brightness contrast or radius compared to the B component. Consequently, the A map is more susceptible to being influenced by noise.

Another caveat is that our data covers only 5 hours in total, which is not enough to cover a full rotation of WISE 1049A assuming a 7-hour period. The missing phase coverage can appear as dark spot-like features in the retrieved map (see Section \ref{sec:disc}). The spot on WISE 1049A is located at a longitude that crosses the visible hemisphere in the first half of the observation, so it should be less affected by the missing phase coverage. In addition, due to the smaller variability amplitude, the rotational period of WISE 1049A is not as well-constrained as WISE 1049B. To see how incomplete phase coverage and unknown periods are affecting the Doppler mapping, we conducted a set of simulations with varied assumed rotation periods and observed phase coverage in Section \ref{sec:disc}.

%%%%%%%%%%%%%%%%%%%%%%%%%%%%%%%%%%%%%%%%%%%%%%%%%%%%%%%%
\subsection{Pressure-dependent features}
\label{sec:pres}

\begin{figure*}
  \centering
  \begin{minipage}[b]{0.49\textwidth}
    \includegraphics[width=\textwidth]{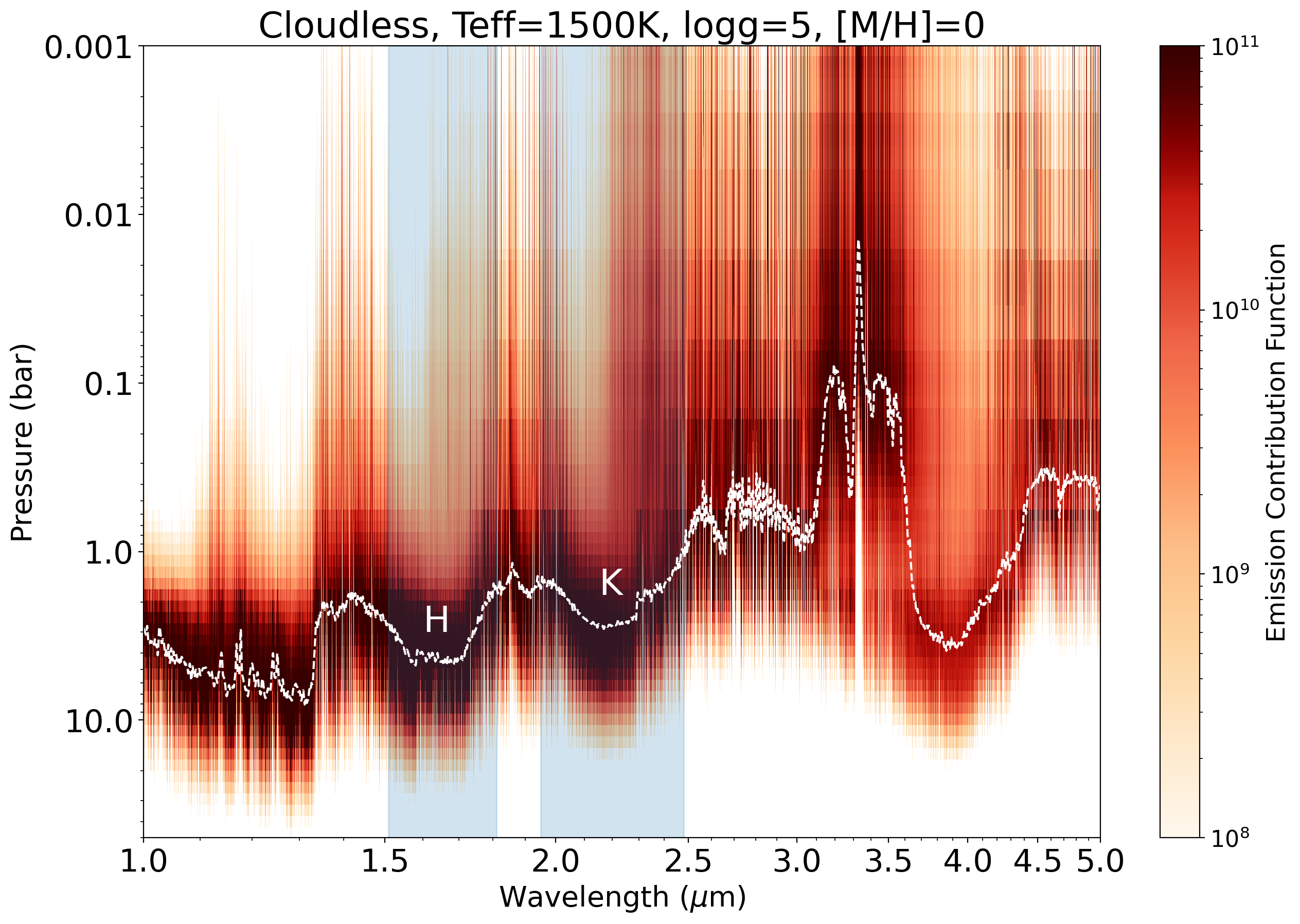}
  \end{minipage}
  \hfill
  \begin{minipage}[b]{0.49\textwidth}
    \includegraphics[width=\textwidth]{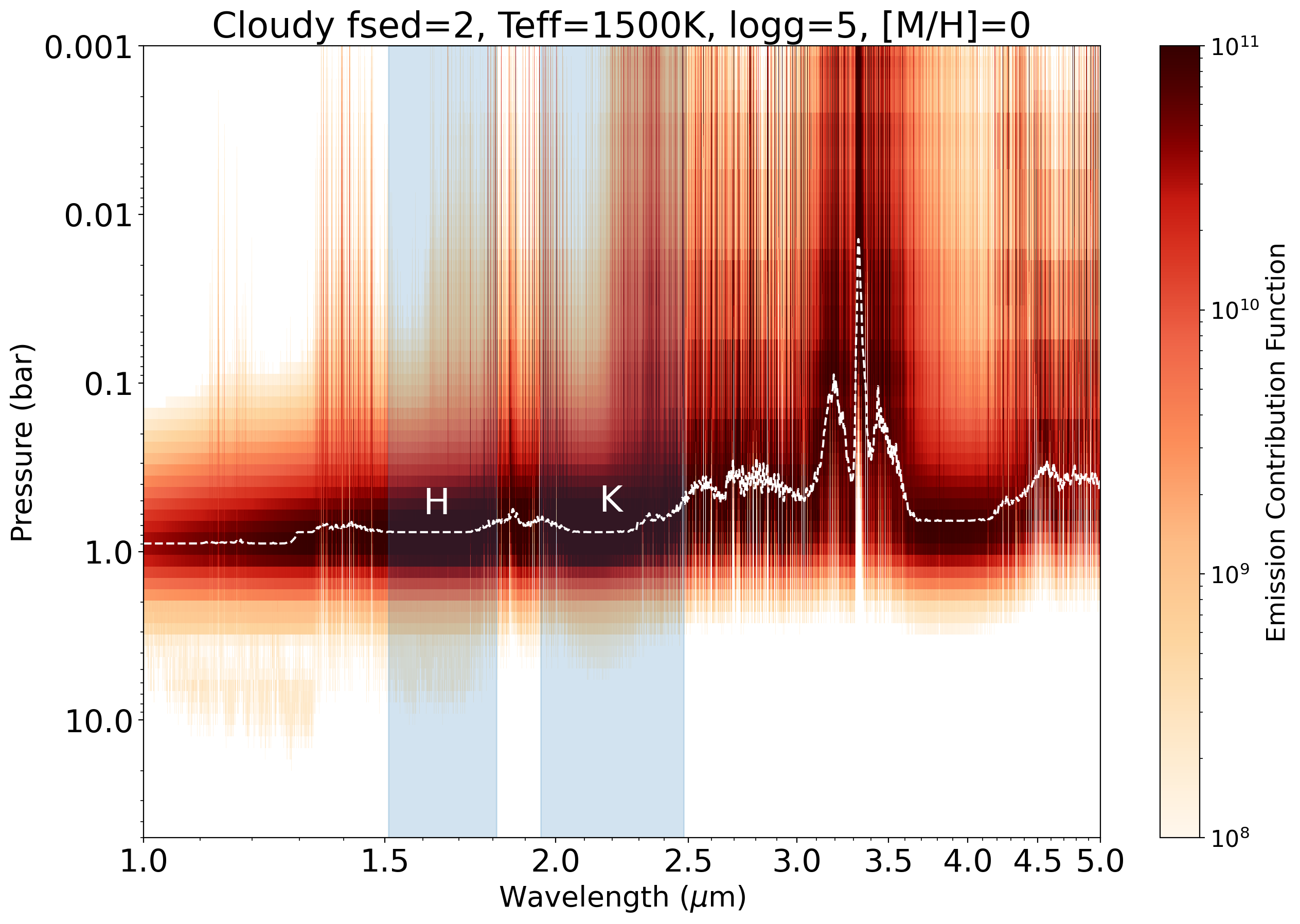}
  \end{minipage}
  \caption{\textit{Left}: The flux contribution from different pressure levels as a function of wavelength, calculated using a cloudy T-P profile from the best-fit Sonora Diamondback model with $T_\text{eff}$ = 1500K, $\log g$ = 5 and $f_\text{sed}$ = 8. The wavelength range of IGRINS H and K bands are marked in grey shaded areas. \textit{Right}: }
  \label{fig:cf}
\end{figure*}

Different wavelengths probe different pressures in the atmosphere because the opacity of the atmosphere varies with wavelength. Multi-wavelength and spectroscopic monitoring opens up the possibility of probing different depths into brown dwarf and planetary-mass companion atmospheres, revealing their vertical structures. The correspondence between wavelength and the pressure level probed can be established by emission contribution functions calculated from radiative-transfer atmospheric models. 

Emission contribution functions illustrate the pressures at which the flux is emitted for specific wavelengths (e.g. \citealt{Lothringer2018}). We calculated the contribution function for WISE 1049B using the \textsc{picaso} package \citep{Batalha2019}. 
Because the patchiness of the 2-D TOA maps of WISE 1049B suggests that its atmosphere likely consists of both cloudy and clear columns, a single contribution function computed from a 1-D model cannot adequately represent this atmosphere. Therefore, we present both a cloudless and a cloudy version of the contribution function in Fig.~\ref{fig:cf}, which would represent a clear and cloudy column in the WISE 1049B atmosphere. These were generated using a Sonora T-P profile with the best-fit parameters from our spectrum ($T_\text{eff}$ = 1500K, $\log g$ = 5). For the cloudy case, we included a cloud layer composed of Fe, MgSiO$_3$ and Mg$_2$SiO$_4$ with $f_\text{sed}$=2 using the \textsc{virga} module. The pressures at which the maximum flux is emitted for each wavelength are marked with a white dashed line.

The thermal contribution function in the left panel of Fig.~\ref{fig:cf} shows that the emission in the H band primarily originates between pressure levels of 2-10 bar under cloudless conditions. The K band probes slightly higher pressure levels in the atmosphere, mainly at 1-5 bar, with some overlaps with the H band. The right panel of Fig.~\ref{fig:cf} shows that the emission peaks are almost at the same altitude for H and K in the cloudy case, with flux originating from a lower pressure around 1 bar. The addition of clouds smooths out the contribution function and shifts the emission continuum to higher altitudes.
% effect of CO opacities in the K narrow band
The contribution function is also shaped by molecular opacities, particularly at high spectral resolution where individual molecular lines are resolved. The line cores probe levels much higher than the clouds in the atmosphere, reaching up to around 1-0.001 bar in both the cloudless and cloudy cases. The molecular opacity primarily affects the K band, especially towards the end of the K band around 2.3-2.5 $\mu$m where a bandhead of dense CO lines lies (e.g. \citealt{Crossfield2014}).

% implication on Doppler map
It can be learned from the contribution functions that the observed brightness temperature in Doppler maps is affected by both gas molecular opacity and cloud opacity:
A brighter area on the Doppler map indicates we are looking at a clear column where emission comes from deeper in the atmosphere. A darker region, on the other hand, suggests that we are probing either the top of a cloud deck which is higher and thus cooler in the atmosphere, or a localized CO patch where the emission also originates from higher up. Such CO patchiness can be formed due to convective plumes driving vertical mixing, which creates regions of localized chemical disequilibrium, a process believed to be common on L/T transition objects \citep{Tremblin2020}. 
% caveats
Distinguishing between cloud-related and chemistry-related mechanisms for the probed TOA structure is challenging because both factors play a role in the complex atmospheric processes in these objects. Additionally, it is hard to pinpoint the exact altitude probed by K and H band Doppler maps, since Doppler imaging uses information from the entire line width which is formed across a wide range of pressures, and gaining sufficient SNR requires combining multiple lines from different spectral orders across the H or K band. This means that we are inevitably probing an altitude-averaged quantity through these measurements. Therefore, the contribution functions should be viewed as indicative guides rather than precise mappings.
 
% possible scenarios
With these considerations in mind, we now discuss possible scenarios for the observed Doppler maps. If there are no clouds in the atmosphere and the darker regions in the Doppler map are solely due to CO patchiness, one would expect these features to be prominent in the K band, but not significant in the H band, as CO opacity primarily affects the K band. Conversely, if the dark features are only caused by clouds and there is no CO patchiness (i.e. CO is uniformly distributed throughout the atmosphere), one would expect the map in H and K band to show similar structures as the clouds block the view into the deeper atmosphere. 
% observed
In our data, since we observed similar TOA structures at non-CO wavelengths (e.g. the IGRINS H band) as in the K band, we can infer that the cooler spots are not solely due to CO absorption and must involve cloud formations. Whether these structures are also affected by CO patchiness can not be definitively concluded from our dataset. It’s also important to note that the scenarios discussed based on the cloudy and cloud-free cases are still greatly simplified and do not account for varying cloud thicknesses or the presence of multi-layered clouds at different altitudes.
Simultaneous variability monitoring in a longer wavelength range covering the 10 $\mu$m silicate feature which directly probes the cloud particles could provide further constraints on this problem \citep{Luna2021}.

% implication of no phase shift
%The fact that no significant differences such as phase shifts are found between the H and the K bands in our maps indicates that they probe pressure levels that are close and partly overlapping in the atmosphere, and that cloud structures are vertically extended across these pressures. 

% Comparison with literature
The finding of similar TOA structure in H and K bands is consistent with previous multi-wavelength variability studies where large phase shifts are often only found between widely separated wavelength windows such as between near-IR (1-2 $\mu$m) and mid-IR (3-4$\mu$m Spitzer channels). One exception to this is the claim in \citet{Biller2013} where a $\sim$100$^\circ$ phase shift between H and K band is found for WISE 1049B, indicating that the clouds causing the variability may differ in position in their vertical layers of the atmosphere. 
A possible explanation for this is that the light curves in \citet{Biller2013} were very noisy and the detected phase shift might not have been seen in higher quality light curves or with a longer time baseline of monitoring.

%%%%%%%%%%%%%%%%%%%%%%%%%%%%%%%%%%%%%%%%%%%%%%%%%%%%%%%%
%%%%%%%%%%%%%%%%%%%%%%%%%%%%%%%%%%%%%%%%%%%%%%%%%%%%%%%%
%%%%%%%%%%%%%%%%%%%%%%%%%%%%%%%%%%%%%%%%%%%%%%%%%%%%%%%%

\section{Discussion}
\label{sec:disc}

In this final section, we present a few tests to demonstrate to what extent uncertainties in the target's physical parameters affect the result of Doppler imaging. We also summarize the factors influencing and limiting the performance of Doppler imaging.

\begin{figure*}
\begin{center}
\includegraphics[width=1\textwidth]{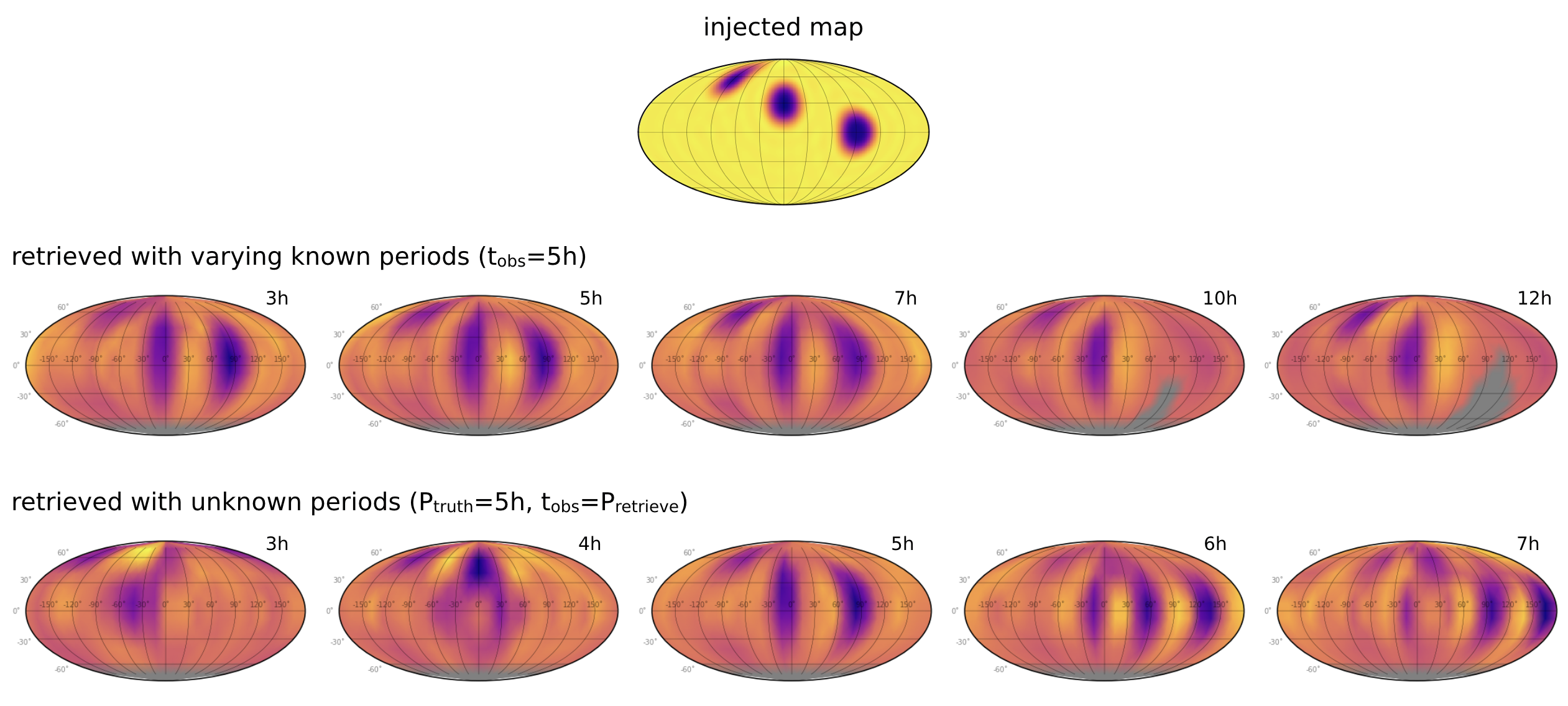}
\caption{The effects of rotation period assumptions on simulated Doppler maps with multiple spots. The upper row shows the effect of repeated or incomplete phase coverage. Completely uncovered areas during the observed phases are marked in gray. Spots rotating across the visible hemisphere during the covered phases are mostly unaffected, while spots that do not complete a full crossing of the visible side during the covered time are lost in the Doppler map. The lower row shows the effect of using wrong period assumptions in Doppler map reconstruction. Retrieving the map with a period too long or too short shifts the longitude of recovered spots causing the features to mix up.}
\label{fig:sim_per}
\end{center}
\end{figure*}
%%%%%%%%%%%%%%%%%%%%%%%%%%%%%%%%%%%%%%%%%%%%%%%%%%%%%%%%
\subsection{Effects of incomplete phase coverage and unknown period}
\label{sec:period}

Doppler imaging relies on the rotation of inhomogeneous top-of-atmosphere structures across the visible hemisphere to extract information about their sizes, shapes, and locations. Thus, successfully retrieving information depends on obtaining a complete set of spectra covering an entire rotation cycle of the brown dwarf and is quite sensitive to the period assumption in the reconstruction process. However, in practice, brown dwarf period measurements have large uncertainties, since the light curves of variable objects often show non-periodic changes and undergo dramatic evolution between rotations. Also, it is often difficult to obtain high-resolution spectra of a brown dwarf for a complete rotation due to practical limitations. Specifically, our data set for WISE 1049A only covers roughly 5/7 of its rotational period based on current best measurements. Therefore, it is important to understand the effect of incomplete phase coverage and uncertainties in the prior knowledge of their rotational period on the performance of Doppler imaging.

To fully capture a TOA feature, it is required that the spectrum is taken as the feature rotates to a full set of positions in the projected radial velocity space. If the sampled phase coverage is smaller than 360$^\circ$, some information loss will occur.
If $t_\text{obs}$ is smaller than half of the rotational period, there will be a portion of the object's surface that is completely unseen. If the observation covered $\phi_\text{cover}$ out of the full 360$^\circ$ rotation, in the case of edge-on inclination, the angular size of the completely uncovered area will be (180$^\circ$ - $\phi_\text{cover}$). All the rest area will be partially covered, and no area on the surface will be fully covered (a fully covered area means that every line of longitude in this area has the chance to cross the entire visible hemisphere, and a partially covered area means that lines of longitude in this area will only cross a limited part of the visible hemisphere). 

If $t_\text{obs}$ is larger than half of the rotational period but smaller than one full period, all areas on the surface will be at least partially covered, and some regions will be fully covered. The angular size of the fully covered area will be ($\phi_\text{cover}$ - 180$^\circ$). The longitudes opposite to the fully covered area will always lack (360$^\circ$ - $\phi_\text{cover}$) of coverage. For the rest of the spherical surface, the longitudes ahead of the fully covered area will have decreasing coverage from 360$^\circ$ to (360$^\circ$ - $\phi_\text{cover}$), while the longitudes behind the fully covered area will have increasing coverage from (360$^\circ$ - $\phi_\text{cover}$) to 360$^\circ$.

This then implies that, if a spot is only partially covered, the image reconstruction routine will be fitting an incomplete trace in the deviation plot.
If the spot is located at a longitude that is entirely uncovered, the feature will be lost. However, as long as the phase coverage exceeds half of the period, we will obtain at least partial information on all features, regardless of their longitudinal location. We conducted a series of simulations to see how well the pipeline can recover only partially covered features.

We used a test map that includes 3 spots placed at different latitudes and longitudes. We placed the spot such that at t=0, the leftmost spot began to emerge on the left limb while the rightmost spot started to exit the view from the right limb. The inclination is fixed at 70$^\circ$.
Synthetic spectra are generated with true rotation periods from 3h to 12h, but only the phases corresponding to the first $t_\text{obs}$=5h of data are used in the map reconstruction. We assume that the period is known exactly in the reconstruction. The results are shown in the first row of Fig. \ref{fig:sim_per}. We see that retrieval is unaffected when $t_\text{obs}$ is larger than or equal to the period. When the true period goes to 7h, all three spots are still faithfully recovered. This is because the two spots on the left lie in an area that is still completely mapped, and the spot on the right is also partially covered. 
With a true period of 10h, which is twice the observed time, completely unmapped areas begin to show up (indicated in grey on the maps).  The rightmost spot is situated in this area, resulting in its loss in the reconstructed map. The same happens when the period is 12h, which is more than twice the observation time. While the rightmost spot is missing, the two spots on the left appear to be well-recovered, even though they are only partially mapped. This test shows that as long as the phase coverage is not significantly less than the rotational period, the chance of missing out on a feature is low.

Next, we tested the effect of poorly-constrained rotational periods on the retrieved map. This time we fixed the true rotational period to 5h and adopted various period assumptions for Doppler imaging (i.e., generating synthetic spectra with a 5h period, but using 3-7h of observation to perform imaging reconstruction respectively). The results are shown in the second row of Fig. \ref{fig:sim_per}. 
We found that using a period different from the true period in the Doppler imaging routine leads to changes in both longitude and latitude for the recovered feature. The features become incomplete, shifted, and even mixed up.
If the period assumed is shorter than the true period (e.g. 3h and 4h case), the recovered feature would be shifted opposite of the rotational direction and higher in latitude, and if the period used in the mapping routine is larger than the true period, the recovered feature would be shifted towards the rotational direction and lower in latitude.
Combined with the impact of incomplete phase coverage, this means that assuming a period significantly shorter or longer than the true period (e.g., in the cases of 3h and 7h) can distort the features to the extent that they no longer accurately represent the true maps.
For targets whose period is largely unconstrained, this may raise a problem. In the case of WISE 1049A, previous variability monitoring suggests a period ranging from 3-4h \citep{Biller2013} up to $\sim$8h \citep{Mancini2015}. Luckily, the phase coverage is at least more than half of the period, and the detected spot lies in the fully covered region. Assuming that the true period is not far from 7h, the recovered map should be a faithful representation of its true TOA structure.

\begin{figure*}
\begin{center}
\includegraphics[width=1\textwidth]{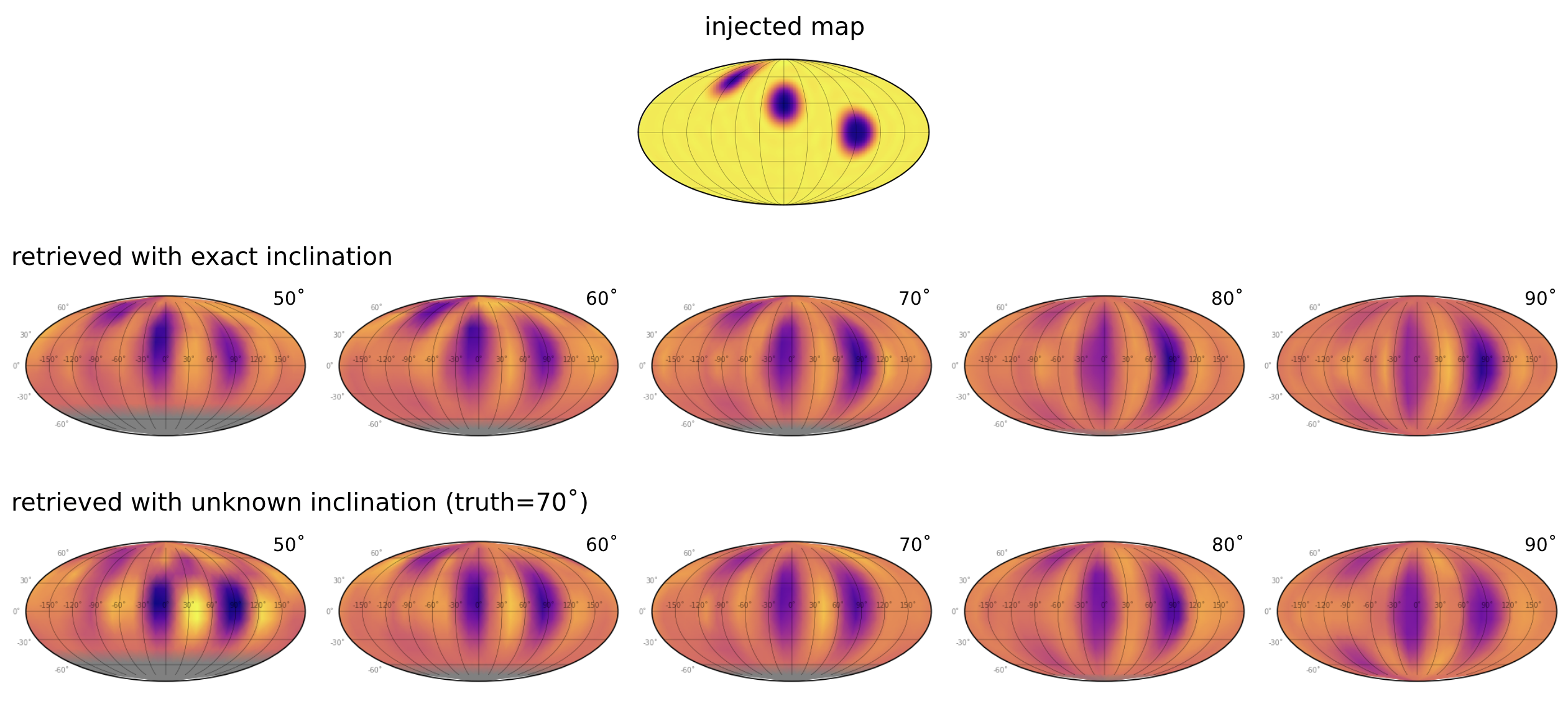}
\caption{The effects of inclination assumptions on simulated Doppler maps with multiple spots. Completely unobservable areas are marked in gray. The upper row shows the effect of the true inclination of the object. Objects with nearly edge-on geometry suffer from mirroring between the northern and southern hemispheres. The lower row shows the effect of using wrong inclination assumptions in Doppler map reconstruction. Retrieving the map with a smaller inclination causes the retrieved spot to be more concentrated and shifted to the equatorial region.}
\label{fig:sim_inc}
\end{center}
\end{figure*}

%%%%%%%%%%%%%%%%%%%%%%%%%%%%%%%%%%%%%%%%%%%%%%%%%%%%%%%%
\subsection{Effects of unknown inclination}
\label{sec:inc}

The inclination of brown dwarfs is usually poorly constrained because the measurement of inclination depends on obtaining the measurements of $v\sin i$, period, and radius (e.g. \citealt{Vos2017}) -- each of which will likely have large uncertainties.
We test the impact of unknown inclinations on our Doppler imaging routine. 

First, we demonstrate the performance of Doppler imaging on objects with varying inclinations. We set the inclination of the brown dwarf to 50-90$^\circ$ to generate a set of synthetic spectra and use that precise inclination for image reconstruction. The recovered map is shown in the first row of Fig. \ref{fig:sim_inc}, with the unseen areas marked in grey. The most obvious effect is that the visible area available for mapping increases when the inclination goes to 90$^\circ$ (aka more edge-on). 
Another important effect is that nearly edge-on objects suffer from more severe mirroring in the southern hemisphere because rotational Doppler broadening is symmetric at the same latitude on both hemispheres. Because of this degeneracy, the recovered surface features appear more elongated than the actual TOA feature injected. This effect is also demonstrated in various Doppler imaging papers using different code implementations (e.g. \citealt{Vogt1987, Luger2021}) and seen in our WISE 1049AB map and simulations as well.

With the impact of varying intrinsic inclinations in mind, we further test how sensitive the Doppler imaging routine is to uncertainties in the inclination assumed. 
We generated synthetic spectra with a true inclination of 70$^\circ$ and attempted to recover the map assuming an inclination from 50-90$^\circ$. We again used a test map with 3 spots at different latitudes and longitudes. The results are shown in the second row of Fig. \ref{fig:sim_inc}.

When the inclination used in the reconstruction is smaller than the true inclination (e.g. in the case of 50$^\circ$ and 60$^\circ$), the retrieved spot becomes more concentrated and shifts toward the equatorial region. The spots are also recovered with a higher brightness contrast due to the compressed size of the spot. Given a smaller size, the spot must be made darker to match the fixed amount of deviations of the line profiles. Conversely, if the inclination used is larger than the true value (e.g. the case of 80$^\circ$ and 90$^\circ$), the Doppler imaging reconstruction results in more elongated features, causing the spots to be more spread out and less dark. The two hemispheres also suffer from more mirroring effects and become indistinguishable. 

This tells us that if the inclination is poorly constrained beforehand, an incorrect assumption about inclination can lead to features shifting in latitude and stretching in longitude. This can result in a loss of information regarding spot size, location, and brightness. Therefore, the choice of a target with a well-constrained inclination is crucial for obtaining reliable information from Doppler maps.

%%%%%%%%%%%%%%%%%%%%%%%%%%%%%%%%%%%%%%%%%%%%%%%%%%%%%%%%

\subsection{Capability of Doppler imaging}

We summarize the limiting factors that might affect our Doppler imaging results.
Factors that affect the latitude of the retrieved features include the true inclination of the object (Section \ref{sec:inc}), the uncertainty in the inclination and period measurements (Section \ref{sec:inc}, \ref{sec:period}), the number of pixels $n_k$ used in the least-square deconvolution (Section \ref{sec:LSD}), and the $v\sin i$ of the mapped object.
Factors that affect the longitude of the retrieved features include the phase coverage of observation and the uncertainty in the period measurement (Section \ref{sec:period}).
Factors that affect the shape and size of the recovered feature include the inclination of the object (Section \ref{sec:inc}), uncertainties in the measured parameters (Section \ref{sec:inc}, \ref{sec:period}), the noise level of the spectra (e.g. Section \ref{sec:sims}), as well as cell size and smoothing parameter used in the image reconstruction pipeline (Section \ref{sec:pipeline}).

In addition to uncertainties arising from prior knowledge about the brown dwarf parameters, other sources of uncertainties in our Doppler maps, listed in order of importance, include the quality of LSD pre-processing,  the signal-to-noise ratio (SNR) of the spectroscopic data, the instrumental line profile of the spectrograph, and the influence of telluric lines. Improved LSD algorithms would significantly enhance the precision of Doppler imaging techniques by extracting precise line profile shapes from the observed spectra. The effectiveness of Doppler imaging is crucially dependent on the spectral resolution and SNR of spectroscopic observations, emphasizing the importance of high-quality data.

Many factors have degenerate effects in shaping the resulting Doppler map.
This tells us that only the largest-scale structures discovered by Doppler maps should be trusted. One should be especially cautious while interpreting the latitudinal information in Doppler maps.
While Doppler imaging is proficient at identifying longitudinal variations, it is less effective at retrieving latitudinal information, making the detection of axis-symmetric features (such as zonal banding) more challenging compared to non-symmetric features like spots or patches. 
Other observational methods capable of discerning latitudinal variations can complement the information derived from Doppler imaging. For example, observations of brown dwarfs with different viewing geometries suggest increased cloudiness at the equator compared to the poles \citep{Vos2017, Suarez2023}; short-term monitoring (single rotations) versus long-term monitoring (hundreds of days) can probe modulations originating from different latitudes (i.e., equatorial vs. poles; \citealt{Apai2021, Fuda2024}). These methods provide additional insights into the latitudinal distribution of clouds and atmospheric structures and help to build a more comprehensive picture.

The advent of upcoming extremely large telescopes (ELTs) will revolutionize the capability of Doppler imaging. Based on simulations by \cite{Crossfield2014a}, over 50 brown dwarfs and planetary-mass companions with spectral types >M9 will become available for Doppler imaging, and a smaller number of exoplanets will be potentially mappable. Notably, the L2 type planet $\beta$ Pic b \citep{Lagrange2009, Lagrange2010} stands out as a promising candidate to be the first exoplanet to obtain Doppler maps, while the L0 planet AB Pic b \citep{Chauvin2005} is another possible target. Instruments like MODHIS on the Thirty Meter Telescope (TMT) and METIS on the E-ELT have the potential of mapping planetary-mass objects like VHS 1256 b (L7 type planetary-mass companion; \citealt{Gauza2015}), SIMP 0136 (T2.5 dwarf on the planet-dwarf boundary; \citealt{Artigau2009, Vos2023}), and Beta Pic b in one rotation \citep{Plummer2023}. It's worth noting that the HR8799 planets \citep{Marois2008, Wang2022} might not be optimal targets due to their nearly face-on inclination.

%%%%%%%%%%%%%%%%%%%%%%%%%%%%%%%%%%%%%%%%%%%%%%%%%%%%%%%%
%%%%%%%%%%%%%%%%%%%%%%%%%%%%%%%%%%%%%%%%%%%%%%%%%%%%%%%%
%%%%%%%%%%%%%%%%%%%%%%%%%%%%%%%%%%%%%%%%%%%%%%%%%%%%%%%%
\section{Conclusions}

We obtained time-resolved high-resolution IGRINS spectroscopy of WISE 1049AB for over 5h each on the night of Feb 9 and Feb 11, 2020. We fitted the spatially-resolved spectra with atmospheric models, extracted the mean line profile with least-square deconvolution (LSD), and produced Doppler maps for both WISE 1049AB in the K and H bands, which revealed persistent top-of-atmosphere (TOA) structures on both components. We also provided simulated Doppler maps to explore possible models underlying the observed features, and discussed factors affecting the accuracy of Doppler maps. Our main findings are summarized below:

\begin{enumerate}
    \item For WISE 1049B, we discovered a prominent dark feature in both the Feb 9 and Feb 11 maps. 
    The presence of a similar TOA structure on both nights suggests the stability of atmospheric features over the timescale of days.
    The size and shape of the newly discovered TOA structure are similar to those observed in the 2014 map by \cite{Crossfield2014}, indicating that such structures may be stable or recurring on WISE 1049B. 
    
    \item For WISE 1049A, unlike the featureless 2014 map, we discovered polar spots on both nights of observation. Modeling with simulated maps suggests that the features are likely real and not affected by the missing phase coverage.

    \item We tested our Doppler mapping technique using models of spots, planetary-scale waves, and 3D general circulation (GCM) models, and showed this technique is to some extent effective in distinguishing these scenarios. We found that the observed Doppler maps more closely match models with GCM or spot than models with only a global wave. In the GCM model, a global scale equatorial wave with hot and cold spots is present, but only the dominant patchy structure is recovered in the simulated map. This suggests that Doppler imaging is more sensitive to detecting patchy and latitudinally-varying structures, which may explain the tension with findings from photometric monitoring where planetary-scale waves are found to explain variability on L/T transition objects like WISE 1049B.

    \item We did not find significant differences between the H and K band maps for both WISE 1049B and A. The thermal contribution function derived from the Sonora Diamondback model suggests that the K band probes higher up than the H band in a cloudless atmosphere, whereas the addition of clouds smooths out this difference and results in both bands probing similar pressure levels higher up in the atmosphere. A portion of the K band is also affected by CO opacities which shift the probed pressure level even higher than that of the clouds. However, because we observed similar TOA structures both in and out of CO wavelengths, the cooler spots are not solely due to CO patchiness and must involve cloud formations.
    
    \item We demonstrated the capability and limitations of Doppler imaging through simulations. The uncertainty in the target's physical parameters such as period and inclination affect the location of retrieved features, and these effects are often degenerate. While missing phase coverage is not a major concern, a poorly constrained period or inclination can distort the map significantly. 
    
    \item The signal-to-noise ratio (SNR) of the spectral data is a critical limiting factor in this study and it will remain the primary challenge for future Doppler imaging observations. Achieving sensible imaging reconstruction results requires a line SNR (line depth over line profile noise) over $\sim$20 in the LSD line profiles, which translates to a per-pixel SNR of around 100 in the IGRINS spectra. Upcoming extremely large telescopes (ELTs) offer the potential to map a much broader range of brown dwarfs and even a few directly-imaged giant exoplanets.
    
\end{enumerate}

\section*{Acknowledgements}

This paper contains data based on observations obtained at the Gemini Observatory (Program ID GS-2020A-Q-204),
a program of NSF’s NOIRLab, which is managed by the Association of Universities for Research in Astronomy (AURA) under a cooperative agreement with the National Science Foundation on behalf of the Gemini Observatory partnership: the National Science Foundation (United States), National Research Council (Canada), Agencia Nacional de Investigaci\'{o}n y Desarrollo (Chile), Ministerio de Ciencia, Tecnolog\'{i}a e Innovaci\'{o}n (Argentina), Minist\'{e}rio da Ci\^{e}ncia, Tecnologia, Inova\c{c}\~{o}es e Comunica\c{c}\~{o}es (Brazil), and Korea Astronomy and Space Science Institute (Republic of Korea).
This work used the Immersion Grating Infrared Spectrometer (IGRINS) that was developed under a collaboration between the University of Texas at Austin and the Korea Astronomy and Space Science Institute (KASI) with the financial support of the Mt. Cuba Astronomical Foundation, of the US National Science Foundation under grants AST-1229522 and AST-1702267, of the McDonald Observatory of the University of Texas at Austin, of the Korean GMT Project of KASI, and Gemini Observatory. This paper also includes observations collected at the European Organisation for Astronomical Research in the Southern Hemisphere under ESO program 093.C-0335(A).
Xueqing thanks the support from the China Scholarship Council (CSC) under Grant CSC No. 202208170018. B.B. acknowledges funding by the UK Science and Technology Facilities Council (STFC) grant no. ST/V000594/1. J. M. V. acknowledges support from a Royal Society - Science Foundation Ireland University Research Fellowship (URF$\backslash$1$\backslash$221932). 
Xueqing thanks Genaro Suarez for the helpful discussion.

%%%%%%%%%%%%%%%%%%%%%%%%%%%%%%%%%%%%%%%%%%%%%%%%%%
\section*{Data Availability}

The raw IGRINS data for WISE J104915.57-531906.1AB are available in the IGRINS Archive (RRISA; \citealt{Sawczynec2023}) and the Gemini Archive under Program ID GS-2020A-Q-204. The reduced spectrum and best-fitting Sonora Diamondback models are available on Zenodo (DOI: 10.5281/zenodo.13340163). The Doppler imaging code used for generating the results and figures of this paper can be found on Github: \url{https://github.com/alphalyncis/doppler-imaging-maxentropy} or Zenodo (DOI: 10.5281/zenodo.12599801).



\bibliographystyle{mnras}
\bibliography{references} % if your bibtex file is called example.bib

\begin{thebibliography}{}
\makeatletter
\relax
\def\mn@urlcharsother{\let\do\@makeother \do\$\do\&\do\#\do\^\do\_\do\%\do\~}
\def\mn@doi{\begingroup\mn@urlcharsother \@ifnextchar [ {\mn@doi@}
  {\mn@doi@[]}}
\def\mn@doi@[#1]#2{\def\@tempa{#1}\ifx\@tempa\@empty \href
  {http://dx.doi.org/#2} {doi:#2}\else \href {http://dx.doi.org/#2} {#1}\fi
  \endgroup}
\def\mn@eprint#1#2{\mn@eprint@#1:#2::\@nil}
\def\mn@eprint@arXiv#1{\href {http://arxiv.org/abs/#1} {{\tt arXiv:#1}}}
\def\mn@eprint@dblp#1{\href {http://dblp.uni-trier.de/rec/bibtex/#1.xml}
  {dblp:#1}}
\def\mn@eprint@#1:#2:#3:#4\@nil{\def\@tempa {#1}\def\@tempb {#2}\def\@tempc
  {#3}\ifx \@tempc \@empty \let \@tempc \@tempb \let \@tempb \@tempa \fi \ifx
  \@tempb \@empty \def\@tempb {arXiv}\fi \@ifundefined
  {mn@eprint@\@tempb}{\@tempb:\@tempc}{\expandafter \expandafter \csname
  mn@eprint@\@tempb\endcsname \expandafter{\@tempc}}}

\bibitem[\protect\citeauthoryear{Ackerman \& Marley}{Ackerman \&
  Marley}{2001}]{Ackerman2001}
Ackerman A.~S.,  Marley M.~S.,  2001, \mn@doi [The Astrophysical Journal]
  {10.1086/321540}, 556, 872

\bibitem[\protect\citeauthoryear{Allard, Homeier, Freytag, Schaffenberger  \&
  Rajpurohit}{Allard et~al.}{2013}]{Allard2013}
Allard F.,  Homeier D.,  Freytag B.,  Schaffenberger W.,   Rajpurohit A.~S.,
  2013, Memorie della Societa Astronomica Italiana Supplementi, 24, 128

\bibitem[\protect\citeauthoryear{Apai, Radigan, Buenzli, Burrows, Reid  \&
  Jayawardhana}{Apai et~al.}{2013}]{Apai2013}
Apai D.,  Radigan J.,  Buenzli E.,  Burrows A.,  Reid I.~N.,   Jayawardhana R.,
   2013, \mn@doi [The Astrophysical Journal] {10.1088/0004-637X/768/2/121},
  768, 121

\bibitem[\protect\citeauthoryear{Apai et~al.,}{Apai et~al.}{2017}]{Apai2017}
Apai D.,  et~al., 2017, \mn@doi [Science] {10.1126/science.aam9848}, 357, 683

\bibitem[\protect\citeauthoryear{Apai, Nardiello  \& Bedin}{Apai
  et~al.}{2021}]{Apai2021}
Apai D.,  Nardiello D.,   Bedin L.~R.,  2021, \mn@doi [The Astrophysical
  Journal] {10.3847/1538-4357/abcb97}, 906, 64

\bibitem[\protect\citeauthoryear{Artigau, Bouchard, Doyon  \&
  Lafreni{\`e}re}{Artigau et~al.}{2009}]{Artigau2009}
Artigau {\'E}.,  Bouchard S.,  Doyon R.,   Lafreni{\`e}re D.,  2009, \mn@doi
  [The Astrophysical Journal] {10.1088/0004-637X/701/2/1534}, 701, 1534

\bibitem[\protect\citeauthoryear{{Batalha}, {Marley}, {Lewis}  \&
  {Fortney}}{{Batalha} et~al.}{2019}]{Batalha2019}
{Batalha} N.~E.,  {Marley} M.~S.,  {Lewis} N.~K.,   {Fortney} J.~J.,  2019,
  \mn@doi [\apj] {10.3847/1538-4357/ab1b51}, \href
  {https://ui.adsabs.harvard.edu/abs/2019ApJ...878...70B} {878, 70}

\bibitem[\protect\citeauthoryear{Bedin, Pourbaix, Apai, Burgasser, Buenzli,
  Boffin  \& Libralato}{Bedin et~al.}{2017}]{Bedin2017}
Bedin L.~R.,  Pourbaix D.,  Apai D.,  Burgasser A.~J.,  Buenzli E.,  Boffin H.
  M.~J.,   Libralato M.,  2017, \mn@doi [Monthly Notices of the Royal
  Astronomical Society] {10.1093/mnras/stx1177}, 470, 1140

\bibitem[\protect\citeauthoryear{Biller et~al.,}{Biller
  et~al.}{2013}]{Biller2013}
Biller B.~A.,  et~al., 2013, \mn@doi [The Astrophysical Journal]
  {10.1088/2041-8205/778/1/L10}, 778, L10

\bibitem[\protect\citeauthoryear{Biller et~al.,}{Biller
  et~al.}{2018}]{Biller2018}
Biller B.~A.,  et~al., 2018, \mn@doi [The Astronomical Journal]
  {10.3847/1538-3881/aaa5a6}, 155, 95

\bibitem[\protect\citeauthoryear{Biller et~al.,}{Biller
  et~al.}{2024}]{Biller2024}
Biller B.~A.,  et~al., 2024, \mn@doi [Monthly Notices of the Royal Astronomical
  Society] {10.1093/mnras/stae1602}, 532, 2207

\bibitem[\protect\citeauthoryear{Buenzli, Apai, Radigan, Reid  \&
  Flateau}{Buenzli et~al.}{2014}]{Buenzli2014}
Buenzli E.,  Apai D.,  Radigan J.,  Reid I.~N.,   Flateau D.,  2014, \mn@doi
  [The Astrophysical Journal] {10.1088/0004-637X/782/2/77}, 782, 77

\bibitem[\protect\citeauthoryear{Buenzli, Saumon, Marley, Apai, Radigan, Bedin,
  Reid  \& Morley}{Buenzli et~al.}{2015a}]{Buenzli2015}
Buenzli E.,  Saumon D.,  Marley M.~S.,  Apai D.,  Radigan J.,  Bedin L.~R.,
  Reid I.~N.,   Morley C.~V.,  2015a, \mn@doi [The Astrophysical Journal]
  {10.1088/0004-637X/798/2/127}, 798, 127

\bibitem[\protect\citeauthoryear{Buenzli, Marley, Apai, Saumon, Biller,
  Crossfield  \& Radigan}{Buenzli et~al.}{2015b}]{Buenzli2015a}
Buenzli E.,  Marley {\relax Mark}.~S.,  Apai D.,  Saumon D.,  Biller B.~A.,
  Crossfield I. J.~M.,   Radigan J.,  2015b, \mn@doi [The Astrophysical
  Journal] {10.1088/0004-637X/812/2/163}, 812, 163

\bibitem[\protect\citeauthoryear{Burgasser, Sheppard  \& Luhman}{Burgasser
  et~al.}{2013}]{Burgasser2013}
Burgasser A.~J.,  Sheppard S.~S.,   Luhman K.~L.,  2013, \mn@doi [The
  Astrophysical Journal] {10.1088/0004-637X/772/2/129}, 772, 129

\bibitem[\protect\citeauthoryear{{Burrows}, {Hubbard}, {Lunine}  \&
  {Liebert}}{{Burrows} et~al.}{2001}]{Burrows2001}
{Burrows} A.,  {Hubbard} W.~B.,  {Lunine} J.~I.,   {Liebert} J.,  2001, \mn@doi
  [Reviews of Modern Physics] {10.1103/RevModPhys.73.719}, \href
  {https://ui.adsabs.harvard.edu/abs/2001RvMP...73..719B} {73, 719}

\bibitem[\protect\citeauthoryear{Cameron et~al.,}{Cameron
  et~al.}{2010}]{Cameron2010}
Cameron A.~C.,  et~al., 2010, \mn@doi [Monthly Notices of the Royal
  Astronomical Society] {10.1111/j.1365-2966.2010.16922.x}, 407, 507

\bibitem[\protect\citeauthoryear{Carter et~al.,}{Carter
  et~al.}{2023}]{Carter2023}
Carter A.~L.,  et~al., 2023, The Astrophysical Journal Letters

\bibitem[\protect\citeauthoryear{Chauvin et~al.,}{Chauvin
  et~al.}{2005}]{Chauvin2005}
Chauvin G.,  et~al., 2005, \mn@doi [Astronomy \& Astrophysics]
  {10.1051/0004-6361:200500111}, 438, L29

\bibitem[\protect\citeauthoryear{Collier~Cameron}{Collier~Cameron}{1995}]{CollierCameron1995}
Collier~Cameron A.,  1995, \mn@doi [Monthly Notices of the Royal Astronomical
  Society] {10.1093/mnras/275.2.534}, 275, 534

\bibitem[\protect\citeauthoryear{Crossfield}{Crossfield}{2014}]{Crossfield2014a}
Crossfield I. J.~M.,  2014, \mn@doi [Astronomy \& Astrophysics]
  {10.1051/0004-6361/201423750}, 566, A130

\bibitem[\protect\citeauthoryear{Crossfield et~al.,}{Crossfield
  et~al.}{2014}]{Crossfield2014}
Crossfield I. J.~M.,  et~al., 2014, \mn@doi [Nature] {10.1038/nature12955},
  505, 654

\bibitem[\protect\citeauthoryear{Donati, Semel, Carter, Rees  \&
  Cameron}{Donati et~al.}{1997}]{Donati1997}
Donati J.-F.,  Semel M.,  Carter B.~D.,  Rees D.~E.,   Cameron A.~C.,  1997,
  \mn@doi [Monthly Notices of the Royal Astronomical Society]
  {10.1093/mnras/291.4.658}, 291, 658

\bibitem[\protect\citeauthoryear{{Dupuy} \& {Liu}}{{Dupuy} \&
  {Liu}}{2012}]{Dupuy2012}
{Dupuy} T.~J.,  {Liu} M.~C.,  2012, \mn@doi [\apjs]
  {10.1088/0067-0049/201/2/19}, \href
  {https://ui.adsabs.harvard.edu/abs/2012ApJS..201...19D} {201, 19}

\bibitem[\protect\citeauthoryear{Faherty, Beletsky, Burgasser, Tinney, Osip,
  Filippazzo  \& Simcoe}{Faherty et~al.}{2014}]{Faherty2014}
Faherty J.~K.,  Beletsky Y.,  Burgasser A.~J.,  Tinney C.,  Osip D.~J.,
  Filippazzo J.~C.,   Simcoe R.~A.,  2014, \mn@doi [The Astrophysical Journal]
  {10.1088/0004-637X/790/2/90}, 790, 90

\bibitem[\protect\citeauthoryear{Fuda, Apai, Nardiello, Tan, Karalidi  \&
  Bedin}{Fuda et~al.}{2024}]{Fuda2024}
Fuda N.,  Apai D.,  Nardiello D.,  Tan X.,  Karalidi T.,   Bedin L.~R.,  2024,
  \mn@doi [The Astrophysical Journal] {10.3847/1538-4357/ad2c84}, 965, 182

\bibitem[\protect\citeauthoryear{Garcia et~al.,}{Garcia
  et~al.}{2017}]{Garcia2017}
Garcia E.~V.,  et~al., 2017, \mn@doi [The Astrophysical Journal]
  {10.3847/1538-4357/aa844f}, 846, 97

\bibitem[\protect\citeauthoryear{Gauza, B{\'e}jar, {P{\'e}rez-Garrido}, Osorio,
  Lodieu, Rebolo, Pall{\'e}  \& Nowak}{Gauza et~al.}{2015}]{Gauza2015}
Gauza B.,  B{\'e}jar V. J.~S.,  {P{\'e}rez-Garrido} A.,  Osorio M. R.~Z.,
  Lodieu N.,  Rebolo R.,  Pall{\'e} E.,   Nowak G.,  2015, \mn@doi [The
  Astrophysical Journal] {10.1088/0004-637X/804/2/96}, 804, 96

\bibitem[\protect\citeauthoryear{Gillon, Triaud, Jehin, Delrez, Opitom, Magain,
  Lendl  \& Queloz}{Gillon et~al.}{2013}]{Gillon2013}
Gillon M.,  Triaud A. H. M.~J.,  Jehin E.,  Delrez L.,  Opitom C.,  Magain P.,
  Lendl M.,   Queloz D.,  2013, \mn@doi [Astronomy \& Astrophysics]
  {10.1051/0004-6361/201321620}, 555, L5

\bibitem[\protect\citeauthoryear{{Hammond}, {Mayne}, {Seviour}, {Lewis}, {Tan}
  \& {Mitchell}}{{Hammond} et~al.}{2023}]{hammond2023}
{Hammond} M.,  {Mayne} N.~J.,  {Seviour} W. J.~M.,  {Lewis} N.~T.,  {Tan} X.,
  {Mitchell} D.,  2023, \mn@doi [\mnras] {10.1093/mnras/stad2265}, \href
  {https://ui.adsabs.harvard.edu/abs/2023MNRAS.525..150H} {525, 150}

\bibitem[\protect\citeauthoryear{Hatzes}{Hatzes}{1998}]{Hatzes1998}
Hatzes A.~P.,  1998, Astronomy and Astrophysics, 330, 541

\bibitem[\protect\citeauthoryear{Heinze et~al.,}{Heinze
  et~al.}{2013}]{Heinze2013}
Heinze A.~N.,  et~al., 2013, \mn@doi [The Astrophysical Journal]
  {10.1088/0004-637X/767/2/173}, 767, 173

\bibitem[\protect\citeauthoryear{Karalidi, Apai, Schneider, Hanson  \&
  Pasachoff}{Karalidi et~al.}{2015}]{Karalidi2015}
Karalidi T.,  Apai D.,  Schneider G.,  Hanson J.~R.,   Pasachoff J.~M.,  2015,
  \mn@doi [The Astrophysical Journal] {10.1088/0004-637X/814/1/65}, 814, 65

\bibitem[\protect\citeauthoryear{Karalidi, Apai, Marley  \& Buenzli}{Karalidi
  et~al.}{2016}]{Karalidi2016}
Karalidi T.,  Apai D.,  Marley M.~S.,   Buenzli E.,  2016, \mn@doi [The
  Astrophysical Journal] {10.3847/0004-637X/825/2/90}, 825, 90

\bibitem[\protect\citeauthoryear{Kirkpatrick}{Kirkpatrick}{2005}]{Kirkpatrick2005}
Kirkpatrick J.~D.,  2005, \mn@doi [Annual Review of Astronomy and Astrophysics]
  {10.1146/annurev.astro.42.053102.134017}, 43, 195

\bibitem[\protect\citeauthoryear{Lagrange et~al.,}{Lagrange
  et~al.}{2009}]{Lagrange2009}
Lagrange A.-M.,  et~al., 2009, \mn@doi [Astronomy \& Astrophysics]
  {10.1051/0004-6361:200811325}, 493, L21

\bibitem[\protect\citeauthoryear{Lagrange et~al.,}{Lagrange
  et~al.}{2010}]{Lagrange2010}
Lagrange A.-M.,  et~al., 2010, \mn@doi [Science] {10.1126/science.1187187},
  329, 57

\bibitem[\protect\citeauthoryear{Lazorenko \& Sahlmann}{Lazorenko \&
  Sahlmann}{2018}]{Lazorenko2018}
Lazorenko P.~F.,  Sahlmann J.,  2018, \mn@doi [Astronomy \& Astrophysics]
  {10.1051/0004-6361/201833626}, 618, A111

\bibitem[\protect\citeauthoryear{Lee \& Gullikson}{Lee \&
  Gullikson}{2016}]{Lee2016}
Lee J.-J.,  Gullikson K.,  2016, plp: v2.1 alpha 3,
  \mn@doi{10.5281/zenodo.56067}, \url {https://doi.org/10.5281/zenodo.56067}

\bibitem[\protect\citeauthoryear{Lew et~al.,}{Lew et~al.}{2016}]{Lew2016}
Lew B. W.~P.,  et~al., 2016, \mn@doi [The Astrophysical Journal]
  {10.3847/2041-8205/829/2/L32}, 829, L32

\bibitem[\protect\citeauthoryear{Liu et~al.,}{Liu et~al.}{2023}]{Liu2023}
Liu P.,  et~al., 2023, \mn@doi [Monthly Notices of the Royal Astronomical
  Society] {10.1093/mnras/stad3502}, 527, 6624

\bibitem[\protect\citeauthoryear{Lothringer, Barman  \& Koskinen}{Lothringer
  et~al.}{2018}]{Lothringer2018}
Lothringer J.~D.,  Barman T.,   Koskinen T.,  2018, \mn@doi [The Astrophysical
  Journal] {10.3847/1538-4357/aadd9e}, 866, 27

\bibitem[\protect\citeauthoryear{Luger, Bedell, {Foreman-Mackey}, Crossfield,
  Zhao  \& Hogg}{Luger et~al.}{2021}]{Luger2021}
Luger R.,  Bedell M.,  {Foreman-Mackey} D.,  Crossfield I. J.~M.,  Zhao L.~L.,
   Hogg D.~W.,  2021, Mapping Stellar Surfaces {{III}}: {{An Efficient}},
  {{Scalable}}, and {{Open-Source Doppler Imaging Model}} (\mn@eprint {arxiv}
  {arXiv:2110.06271})

\bibitem[\protect\citeauthoryear{Luhman}{Luhman}{2013}]{Luhman_2013}
Luhman K.~L.,  2013, \mn@doi [The Astrophysical Journal Letters]
  {10.1088/2041-8205/767/1/L1}, 767, L1

\bibitem[\protect\citeauthoryear{Luna \& Morley}{Luna \&
  Morley}{2021}]{Luna2021}
Luna J.~L.,  Morley C.~V.,  2021, \mn@doi [The Astrophysical Journal]
  {10.3847/1538-4357/ac1865}, 920, 146

\bibitem[\protect\citeauthoryear{{Mace} et~al.,}{{Mace}
  et~al.}{2018}]{Mace2018}
{Mace} G.,  et~al., 2018, in {Evans} C.~J.,  {Simard} L.,   {Takami} H.,  eds,
  Society of Photo-Optical Instrumentation Engineers (SPIE) Conference Series
  Vol. 10702, Ground-based and Airborne Instrumentation for Astronomy VII. p.
  107020Q, \mn@doi{10.1117/12.2312345}

\bibitem[\protect\citeauthoryear{Mancini et~al.,}{Mancini
  et~al.}{2015}]{Mancini2015}
Mancini L.,  et~al., 2015, \mn@doi [Astronomy \& Astrophysics]
  {10.1051/0004-6361/201526899}, 584, A104

\bibitem[\protect\citeauthoryear{Marley, Saumon  \& Goldblatt}{Marley
  et~al.}{2010}]{Marley2010}
Marley M.~S.,  Saumon D.,   Goldblatt C.,  2010, \mn@doi [The Astrophysical
  Journal] {10.1088/2041-8205/723/1/L117}, 723, L117

\bibitem[\protect\citeauthoryear{Marley et~al.,}{Marley
  et~al.}{2021}]{Marley2021}
Marley M.~S.,  et~al., 2021, \mn@doi [The Astrophysical Journal]
  {10.3847/1538-4357/ac141d}, 920, 85

\bibitem[\protect\citeauthoryear{Marois, Macintosh, Barman, Zuckerman, Song,
  Patience, Lafreni{\`e}re  \& Doyon}{Marois et~al.}{2008}]{Marois2008}
Marois C.,  Macintosh B.,  Barman T.,  Zuckerman B.,  Song I.,  Patience J.,
  Lafreni{\`e}re D.,   Doyon R.,  2008, \mn@doi [Science]
  {10.1126/science.1166585}, 322, 1348

\bibitem[\protect\citeauthoryear{Marsden, Waite, Carter  \& Donati}{Marsden
  et~al.}{2005}]{Marsden2005}
Marsden S.~C.,  Waite I.~A.,  Carter B.~D.,   Donati J.-F.,  2005, \mn@doi
  [Monthly Notices of the Royal Astronomical Society]
  {10.1111/j.1365-2966.2005.08946.x}, 359, 711

\bibitem[\protect\citeauthoryear{{Millar-Blanchaer} et~al.,}{{Millar-Blanchaer}
  et~al.}{2020}]{Millar-Blanchaer2020}
{Millar-Blanchaer} M.~A.,  et~al., 2020, \mn@doi [The Astrophysical Journal]
  {10.3847/1538-4357/ab6ef2}, 894, 42

\bibitem[\protect\citeauthoryear{Mukherjee, Fortney, {Jensen-Clem}, Tan, Marley
   \& Batalha}{Mukherjee et~al.}{2021}]{Mukherjee2021}
Mukherjee S.,  Fortney J.~J.,  {Jensen-Clem} R.,  Tan X.,  Marley M.~S.,
  Batalha N.~E.,  2021, \mn@doi [The Astrophysical Journal]
  {10.3847/1538-4357/ac2d92}, 923, 113

\bibitem[\protect\citeauthoryear{Park et~al.,}{Park et~al.}{2014}]{Park2014}
Park C.,  et~al., 2014, in Ramsay S.~K.,  McLean I.~S.,   Takami H.,  eds,
  {{SPIE Astronomical Telescopes}} + {{Instrumentation}}. {Montr{\'e}al,
  Quebec, Canada}, p. 91471D, \mn@doi{10.1117/12.2056431}

\bibitem[\protect\citeauthoryear{Plummer \& Wang}{Plummer \&
  Wang}{2022}]{Plummer2022}
Plummer M.~K.,  Wang J.,  2022, \mn@doi [The Astrophysical Journal]
  {10.3847/1538-4357/ac75b9}, 933, 163

\bibitem[\protect\citeauthoryear{Plummer \& Wang}{Plummer \&
  Wang}{2023}]{Plummer2023}
Plummer M.~K.,  Wang J.,  2023, \mn@doi [The Astrophysical Journal]
  {10.3847/1538-4357/accd5d}, 951, 101

\bibitem[\protect\citeauthoryear{Radigan, Jayawardhana, Lafreni{\`e}re,
  Artigau, Marley  \& Saumon}{Radigan et~al.}{2012}]{Radigan2012}
Radigan J.,  Jayawardhana R.,  Lafreni{\`e}re D.,  Artigau {\'E}.,  Marley M.,
   Saumon D.,  2012, \mn@doi [The Astrophysical Journal]
  {10.1088/0004-637X/750/2/105}, 750, 105

\bibitem[\protect\citeauthoryear{Radigan, Lafreni{\`e}re, Jayawardhana  \&
  Artigau}{Radigan et~al.}{2014}]{Radigan2014}
Radigan J.,  Lafreni{\`e}re D.,  Jayawardhana R.,   Artigau E.,  2014, \mn@doi
  [The Astrophysical Journal] {10.1088/0004-637X/793/2/75}, 793, 75

\bibitem[\protect\citeauthoryear{{Roettenbacher} et~al.,}{{Roettenbacher}
  et~al.}{2017}]{Roettenbacher2017}
{Roettenbacher} R.~M.,  et~al., 2017, \mn@doi [\apj]
  {10.3847/1538-4357/aa8ef7}, \href
  {https://ui.adsabs.harvard.edu/abs/2017ApJ...849..120R} {849, 120}

\bibitem[\protect\citeauthoryear{Sahlmann \& Lazorenko}{Sahlmann \&
  Lazorenko}{2015}]{Sahlmann2015}
Sahlmann J.,  Lazorenko P.~F.,  2015, \mn@doi [Monthly Notices of the Royal
  Astronomical Society: Letters] {10.1093/mnrasl/slv113}, 453, L103

\bibitem[\protect\citeauthoryear{{Sawczynec}, {Mace}, {Gully-Santiago}  \&
  {Jaffe}}{{Sawczynec} et~al.}{2023}]{Sawczynec2023}
{Sawczynec} E.,  {Mace} G.,  {Gully-Santiago} M.,   {Jaffe} D.,  2023, in
  American Astronomical Society Meeting Abstracts. p. 207.14

\bibitem[\protect\citeauthoryear{{Showman}, {Tan}  \& {Zhang}}{{Showman}
  et~al.}{2019}]{showman2019}
{Showman} A.~P.,  {Tan} X.,   {Zhang} X.,  2019, \mn@doi [\apj]
  {10.3847/1538-4357/ab384a}, \href
  {https://ui.adsabs.harvard.edu/abs/2019ApJ...883....4S} {883, 4}

\bibitem[\protect\citeauthoryear{Snellen, Brandl, De~Kok, Brogi, Birkby  \&
  Schwarz}{Snellen et~al.}{2014}]{Snellen2014}
Snellen I. A.~G.,  Brandl B.~R.,  De~Kok R.~J.,  Brogi M.,  Birkby J.,
  Schwarz H.,  2014, \mn@doi [Nature] {10.1038/nature13253}, 509, 63

\bibitem[\protect\citeauthoryear{Stetson \& Pancino}{Stetson \&
  Pancino}{2008}]{Stetson2008}
Stetson P.~B.,  Pancino E.,  2008, \mn@doi [Publications of the Astronomical
  Society of the Pacific] {10.1086/596126}, 120, 1332

\bibitem[\protect\citeauthoryear{Strassmeier}{Strassmeier}{2009}]{Strassmeier2009}
Strassmeier K.~G.,  2009, \mn@doi [The Astronomy and Astrophysics Review]
  {10.1007/s00159-009-0020-6}, 17, 251

\bibitem[\protect\citeauthoryear{Su{\'a}rez, Vos, Metchev, Faherty  \&
  Cruz}{Su{\'a}rez et~al.}{2023}]{Suarez2023}
Su{\'a}rez G.,  Vos J.~M.,  Metchev S.,  Faherty J.~K.,   Cruz K.,  2023,
  \mn@doi [The Astrophysical Journal Letters] {10.3847/2041-8213/acec4b}, 954,
  L6

\bibitem[\protect\citeauthoryear{{Tan}}{{Tan}}{2022}]{tan2022}
{Tan} X.,  2022, \mn@doi [\mnras] {10.1093/mnras/stac344}, \href
  {https://ui.adsabs.harvard.edu/abs/2022MNRAS.511.4861T} {511, 4861}

\bibitem[\protect\citeauthoryear{Tan \& Showman}{Tan \&
  Showman}{2021a}]{Tan2021a}
Tan X.,  Showman A.~P.,  2021a, \mn@doi [Monthly Notices of the Royal
  Astronomical Society] {10.1093/mnras/stab060}, 502, 678

\bibitem[\protect\citeauthoryear{Tan \& Showman}{Tan \&
  Showman}{2021b}]{Tan2021b}
Tan X.,  Showman A.~P.,  2021b, \mn@doi [Monthly Notices of the Royal
  Astronomical Society] {10.1093/mnras/stab097}, 502, 2198

\bibitem[\protect\citeauthoryear{Tannock, Metchev, Hood, Mace, Fortney, Morley,
  Jaffe  \& Lupu}{Tannock et~al.}{2022}]{Tannock2022}
Tannock M.~E.,  Metchev S.,  Hood C.~E.,  Mace G.~N.,  Fortney J.~J.,  Morley
  C.~V.,  Jaffe D.~T.,   Lupu R.,  2022, \mn@doi [Monthly Notices of the Royal
  Astronomical Society] {10.1093/mnras/stac1412}, 514, 3160

\bibitem[\protect\citeauthoryear{Tremblin, Amundsen, Chabrier, Baraffe,
  Drummond, Hinkley, Mourier  \& Venot}{Tremblin et~al.}{2016}]{Tremblin2016}
Tremblin P.,  Amundsen D.~S.,  Chabrier G.,  Baraffe I.,  Drummond B.,  Hinkley
  S.,  Mourier P.,   Venot O.,  2016, \mn@doi [The Astrophysical Journal]
  {10.3847/2041-8205/817/2/L19}, 817, L19

\bibitem[\protect\citeauthoryear{Tremblin, Phillips, Emery, Baraffe, Lew, Apai,
  Biller  \& Bonnefoy}{Tremblin et~al.}{2020}]{Tremblin2020}
Tremblin P.,  Phillips M.~W.,  Emery A.,  Baraffe I.,  Lew B. W.~P.,  Apai D.,
  Biller B.~A.,   Bonnefoy M.,  2020, \mn@doi [Astronomy \& Astrophysics]
  {10.1051/0004-6361/202038771}, 643, A23

\bibitem[\protect\citeauthoryear{Vogt, Penrod  \& Hatzes}{Vogt
  et~al.}{1987}]{Vogt1987}
Vogt S.~S.,  Penrod G.~D.,   Hatzes A.~P.,  1987, \mn@doi [The Astrophysical
  Journal] {10.1086/165647}, 321, 496

\bibitem[\protect\citeauthoryear{Vos, Allers  \& Biller}{Vos
  et~al.}{2017}]{Vos2017}
Vos J.~M.,  Allers K.~N.,   Biller B.~A.,  2017, \mn@doi [The Astrophysical
  Journal] {10.3847/1538-4357/aa73cf}, 842, 78

\bibitem[\protect\citeauthoryear{Vos et~al.,}{Vos et~al.}{2019}]{Vos2019}
Vos J.~M.,  et~al., 2019, \mn@doi [Monthly Notices of the Royal Astronomical
  Society] {10.1093/mnras/sty3123}

\bibitem[\protect\citeauthoryear{Vos et~al.,}{Vos et~al.}{2023}]{Vos2023}
Vos J.~M.,  et~al., 2023, \mn@doi [The Astrophysical Journal]
  {10.3847/1538-4357/acab58}, 944, 138

\bibitem[\protect\citeauthoryear{Wang, Prato  \& Mawet}{Wang
  et~al.}{2017}]{Wang2017}
Wang J.,  Prato L.,   Mawet D.,  2017, \mn@doi [The Astrophysical Journal]
  {10.3847/1538-4357/aa6345}, 838, 35

\bibitem[\protect\citeauthoryear{Wang et~al.,}{Wang et~al.}{2022}]{Wang2022}
Wang J.~J.,  et~al., 2022, \mn@doi [The Astronomical Journal]
  {10.3847/1538-3881/ac8984}, 164, 143

\bibitem[\protect\citeauthoryear{{Zhang} \& {Showman}}{{Zhang} \&
  {Showman}}{2014}]{zhang2014}
{Zhang} X.,  {Showman} A.~P.,  2014, \mn@doi [\apjl]
  {10.1088/2041-8205/788/1/L6}, \href
  {https://ui.adsabs.harvard.edu/abs/2014ApJ...788L...6Z} {788, L6}

\bibitem[\protect\citeauthoryear{Zhou et~al.,}{Zhou et~al.}{2018}]{Zhou2018}
Zhou Y.,  et~al., 2018, \mn@doi [The Astronomical Journal]
  {10.3847/1538-3881/aaabbd}, 155, 132

\makeatother
\end{thebibliography}

%\begin{thebibliography}{99}
%\end{thebibliography}

%%%%%%%%%%%%%%%%%%%%%%%%%%%%%%%%%%%%%%%%%%%%%%%%%%
\clearpage

%%%%%%%%%%%%%%%%% APPENDICES %%%%%%%%%%%%%%%%%%%%%

\appendix

%%%%%%%%%%%%%%%%%%%%%%%%%%%%%%%%%%%%%%%%%%%%%%%%%%
\section{Model fitting results}
\begin{figure}
    \centering 
    \includegraphics[width=1\columnwidth]{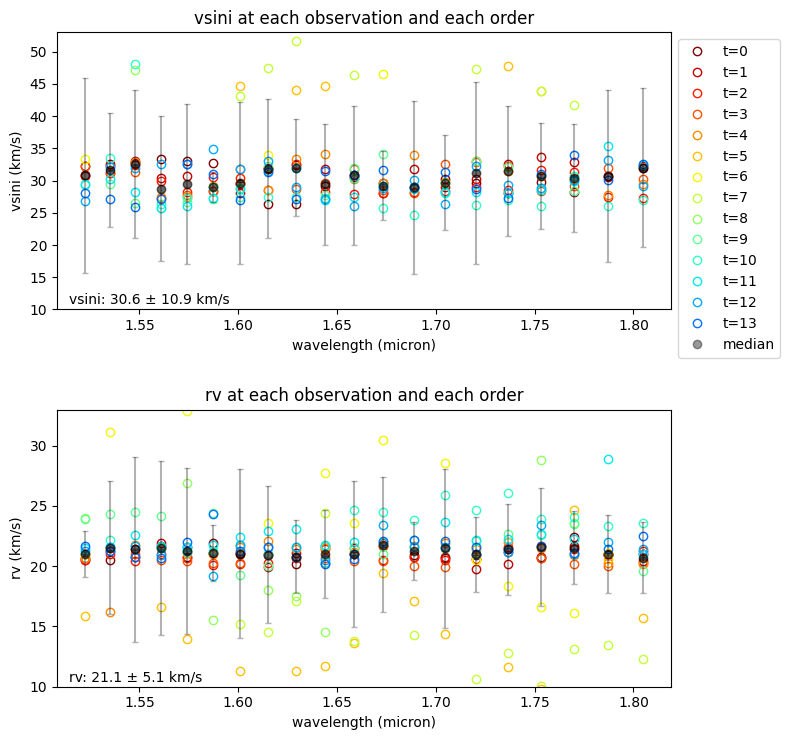}
    \caption{Fitted $v\sin i$ and rv for each observation and each spectral order for WISE 1049B, IGRINS H band.}
    \label{fig:fitparamH_B}
\end{figure}

\begin{figure}
    \centering
    \includegraphics[width=1\columnwidth]{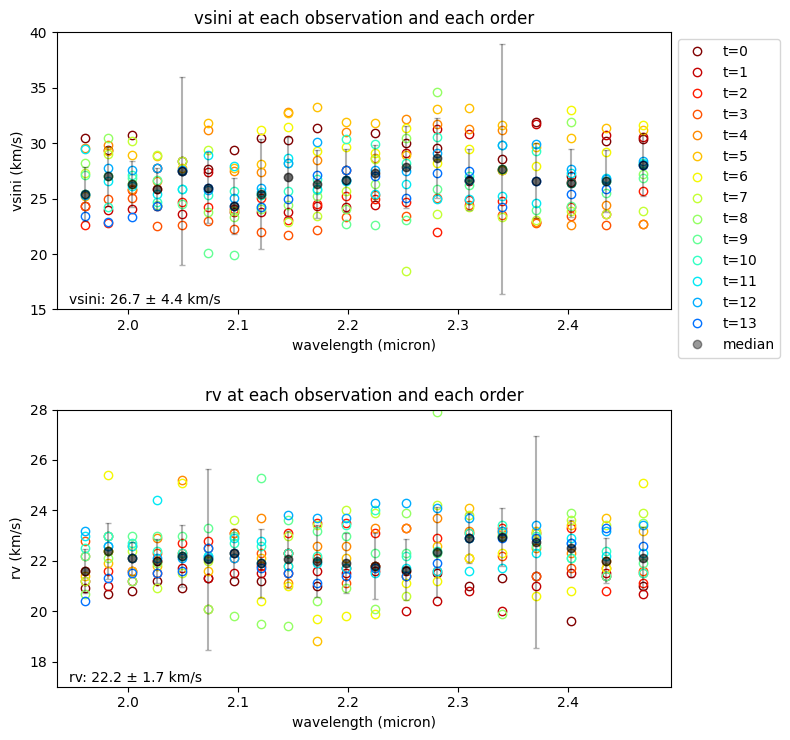}
    \caption{Fitted $v\sin i$ and rv for each observation and each spectral order for WISE 1049B, IGRINS K band.}
    \label{fig:fitparamK_B}
\end{figure}

\begin{figure}
    \centering
    \includegraphics[width=1\columnwidth]{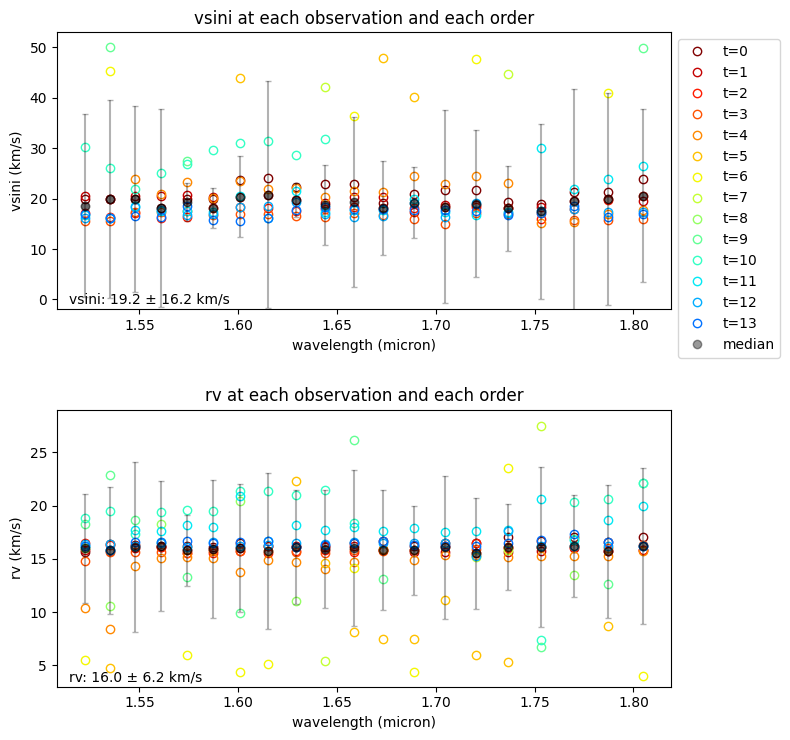}
    \caption{Same as Fig \ref{fig:fitparamH_B} but for WISE 1049A, H band.}
    \label{fig:fitparamH_A}
\end{figure}

\begin{figure}
    \centering
    \includegraphics[width=1\columnwidth]{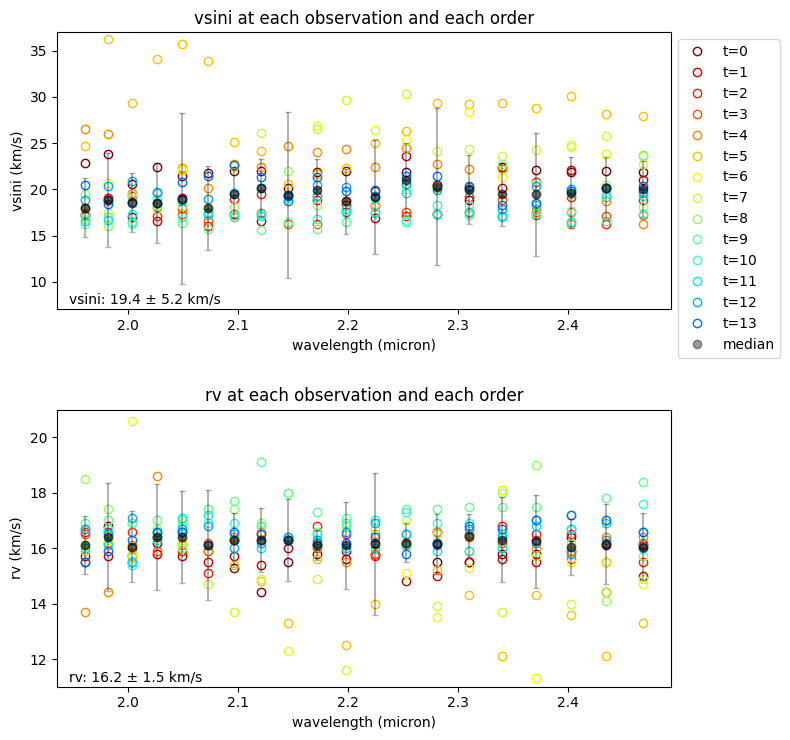}
    \caption{Same as Fig \ref{fig:fitparamK_B} but for WISE 1049A, K band.}
    \label{fig:fitparamK_A}
\end{figure}

\begin{figure*}
    \centering
    \begin{minipage}[b]{1\textwidth}
        \centering 
        \includegraphics[width=0.92\textwidth]{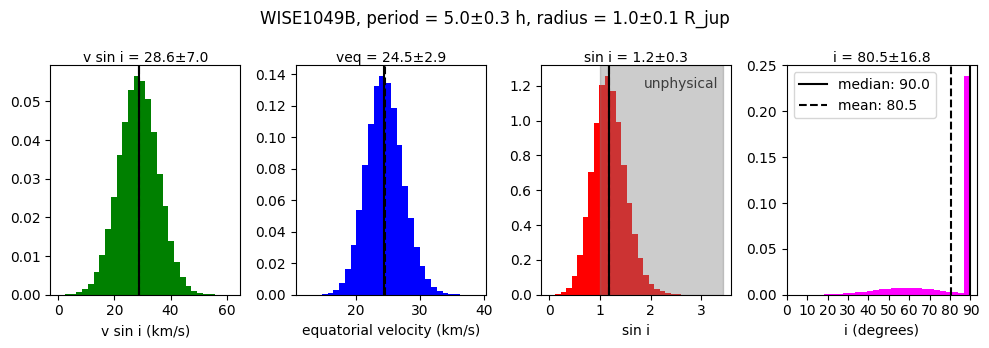}
    \end{minipage}
    \\[10pt]
    \begin{minipage}[b]{1\textwidth}
        \centering
        \includegraphics[width=0.92\textwidth]{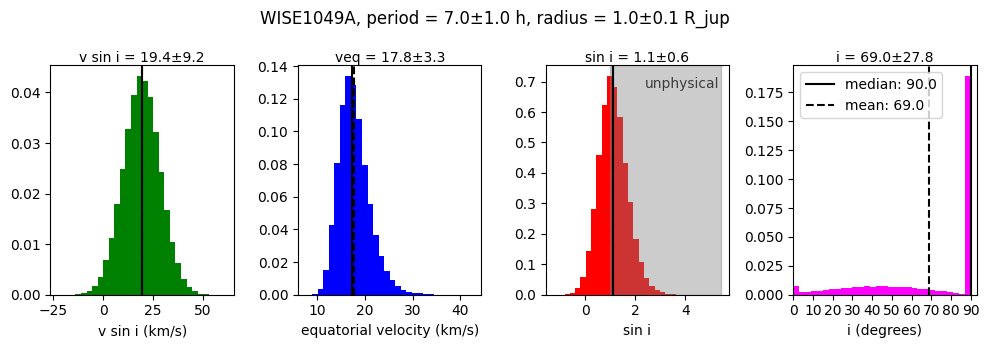}
    \end{minipage}
    \caption{Monte-Carlo simulation of the distribution of inclination for WISE 1049B (upper) and WISE 1049A (lower), given prior of $v\sin i$, period, and radius.}
    \label{fig:incMC}
\end{figure*}

%%%%%%%%%%%%%%%%%%%%%%%%%%%%%%%%%%%%%%%%%%%%%%%%%%
\section{Tests of maximum entropy Doppler imaging hyper-parameters}

\begin{figure*}
\centering
\includegraphics[width=0.9\textwidth]{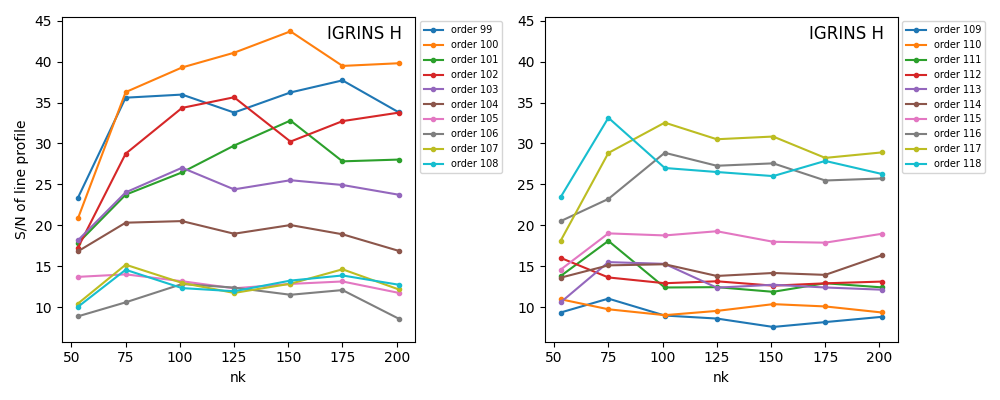}
\includegraphics[width=0.9\textwidth]{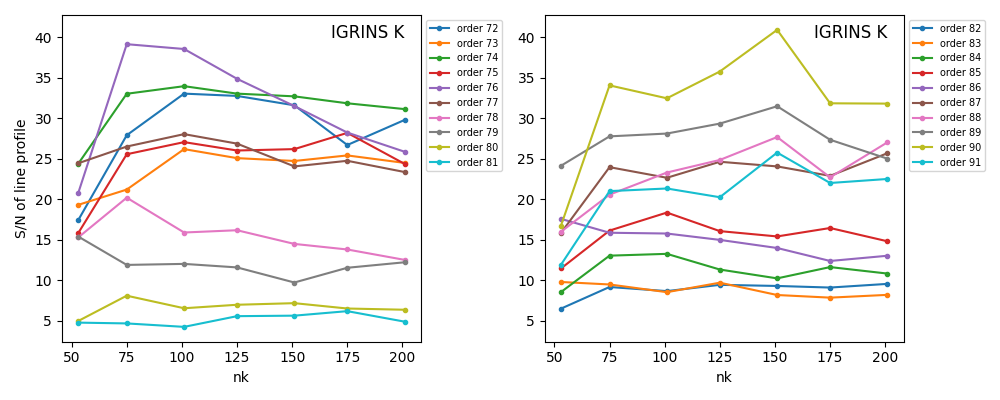}
\caption{SNR of LSD profiles vs $n_k$ (number of pixels used in the LSD kernel) for each spectral order in IGRINS H and K bands. The dotted line marks the chosen value for $n_k$ to maximize the SNR of line profiles. Only orders with SNR > 20 are selected for later Doppler imaging reconstruction.}
\label{fig:nktest}
\end{figure*}

\begin{figure*}
\centering
\includegraphics[width=0.7\textwidth]{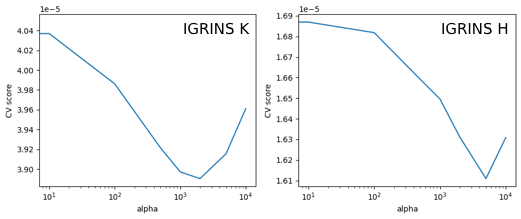}
\caption{Cross-validation score for different choice of $\alpha$ in maximum entropy Doppler image reconstruction. A lower CV score indicates a better balance between image smoothness and fit accuracy. For both IGRINS H and K bands, the best CV score is achieved around 2000-5000.}
\label{fig:cv}
\end{figure*}

%%%%%%%%%%%%%%%%%%%%%%%%%%%%%%%%%%%%%%%%%%%%%%%%%%
\section{Goodness-of-fit of Doppler map solutions}
\label{app:residual}

\begin{figure*}
\centering
\includegraphics[width=0.95\textwidth]{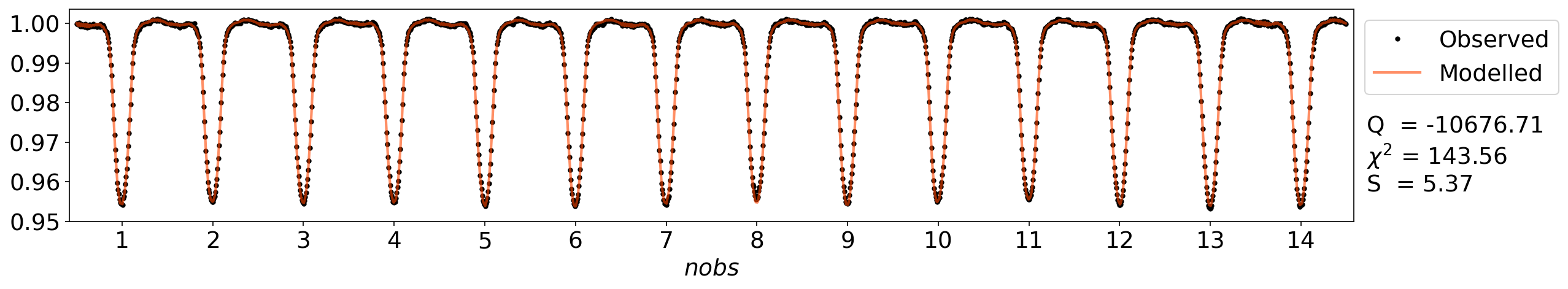}
\begin{minipage}{\textwidth}\centering
\vspace{10pt}
\includegraphics[width=0.93\textwidth]{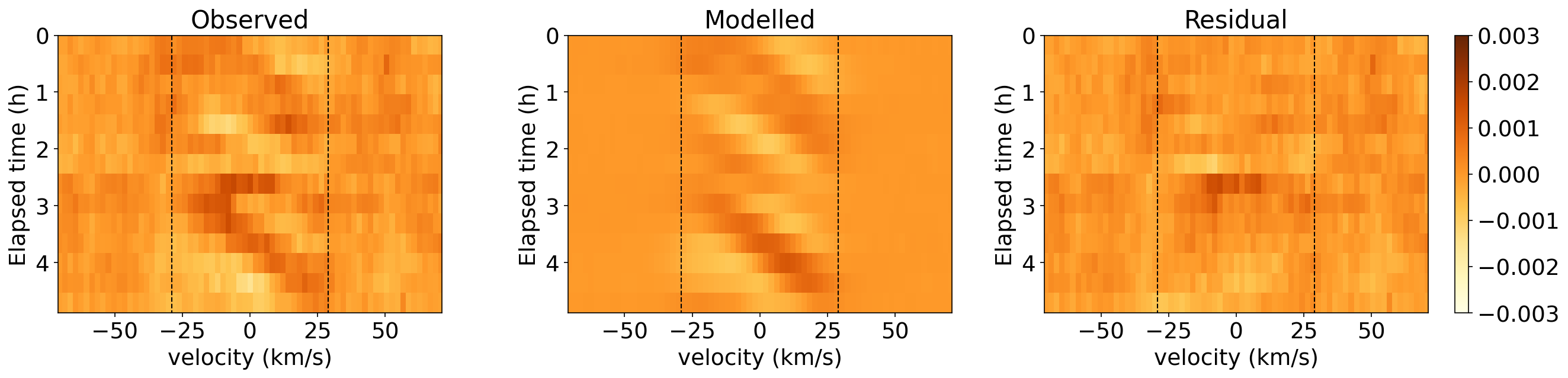}
\end{minipage}
\caption{Same as Fig. \ref{fig:residual_fit}, but for WISE 1049B H band on the night of Feb 9.}
\label{fig:res_B1H}
\end{figure*}

\begin{figure*}
\centering
\includegraphics[width=0.95\textwidth]{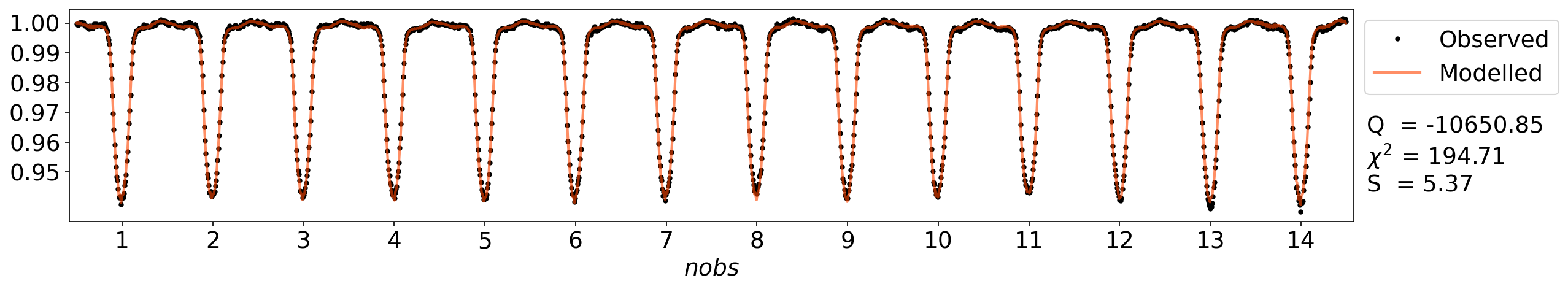}
\begin{minipage}{\textwidth}\centering
\vspace{10pt}
\includegraphics[width=0.93\textwidth]{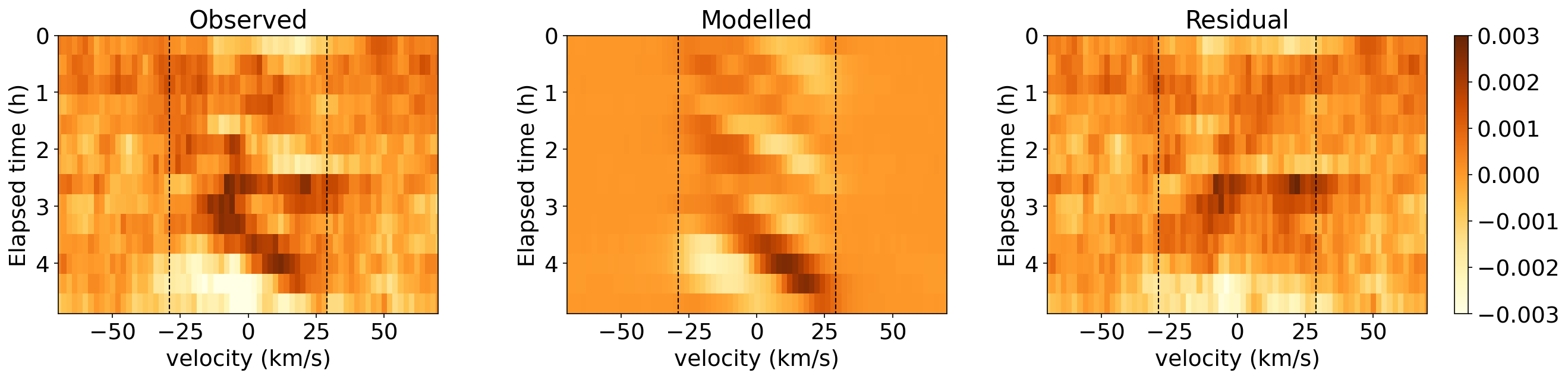}
\end{minipage}
\caption{Same as Fig. \ref{fig:residual_fit}, but for WISE 1049B K band on the night of Feb 9.}
\label{fig:res_B1K}
\end{figure*}

\begin{figure*}
\centering
\includegraphics[width=0.95\textwidth]{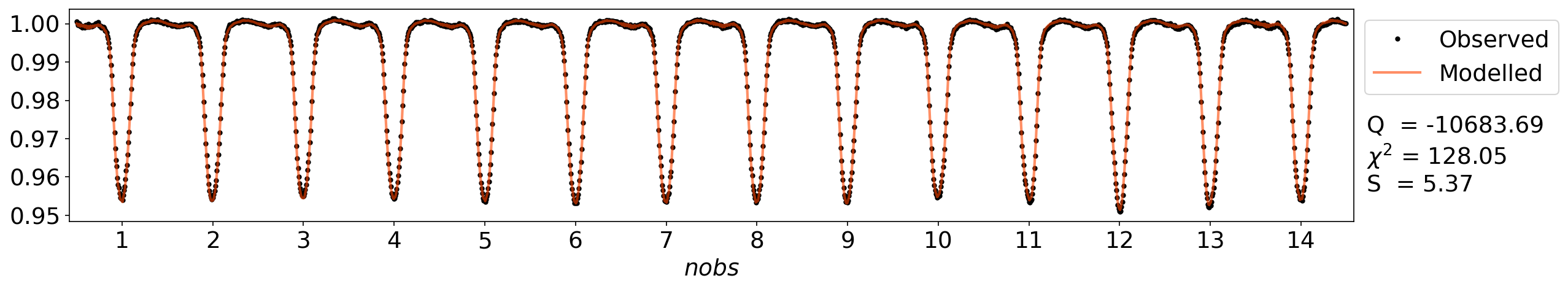}
\begin{minipage}{\textwidth}\centering
\vspace{10pt}
\includegraphics[width=0.93\textwidth]{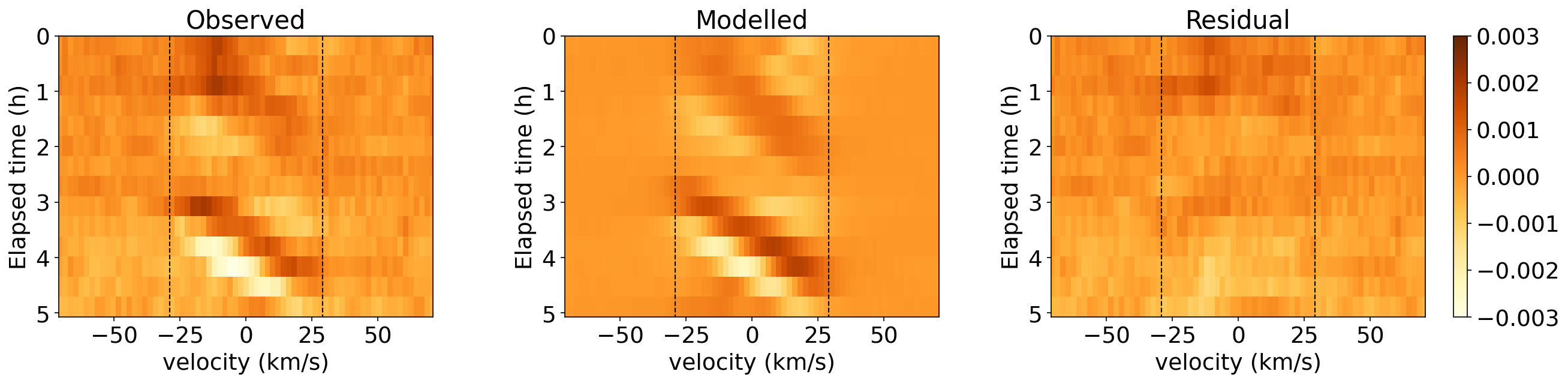}
\end{minipage}
\caption{Same as Fig. \ref{fig:residual_fit}, but for WISE 1049B H band on the night of Feb 11.}
\end{figure*}

\begin{figure*}
\centering
\includegraphics[width=0.95\textwidth]{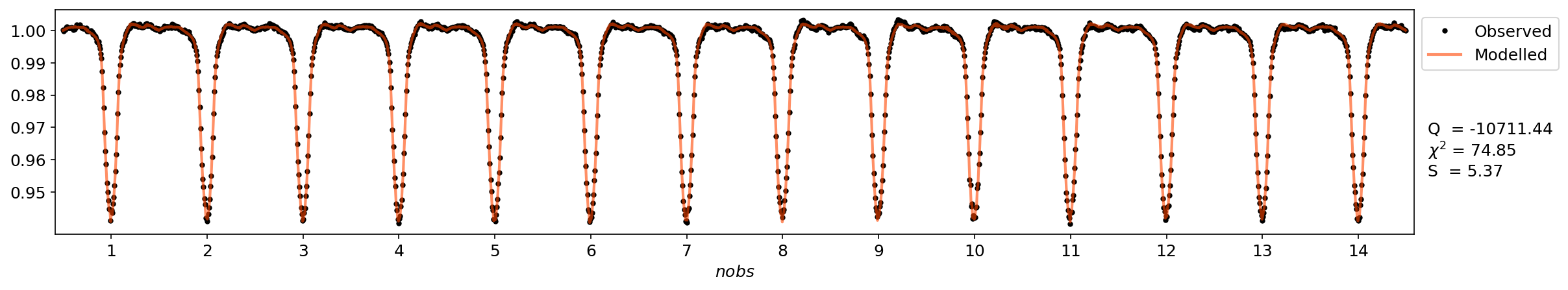}
\begin{minipage}{\textwidth}\centering
\vspace{10pt}
\includegraphics[width=0.93\textwidth]{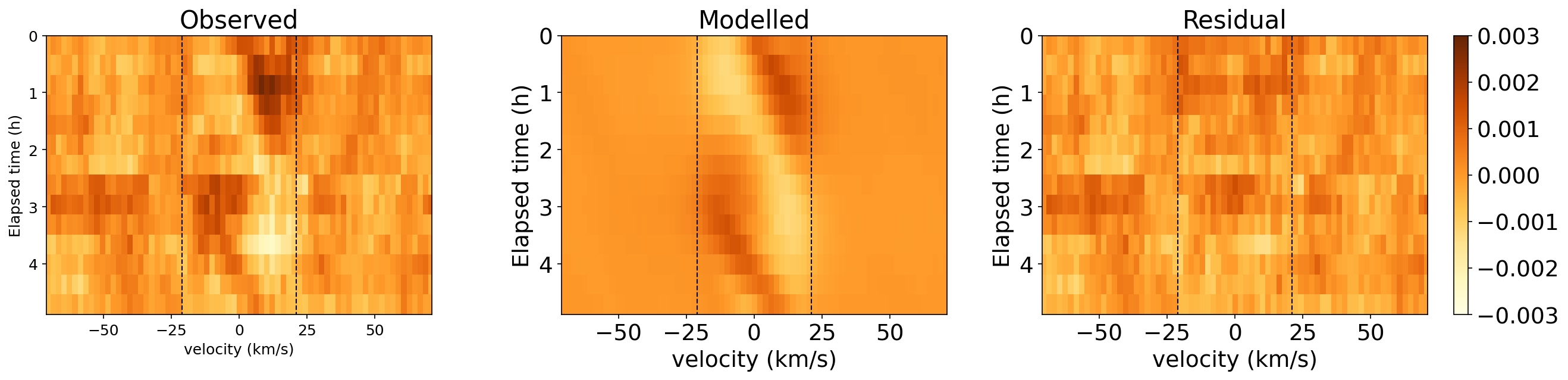}
\end{minipage}
\caption{Same as Fig. \ref{fig:residual_fit}, but for WISE 1049A H band on the night of Feb 9.}
\end{figure*}

\begin{figure*}
\centering
\includegraphics[width=0.95\textwidth]{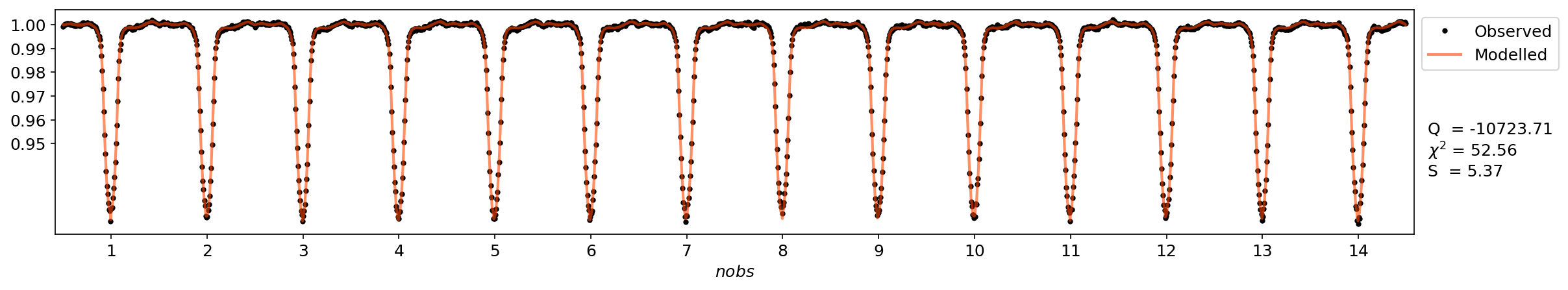}
\begin{minipage}{\textwidth}\centering
\vspace{10pt}
\includegraphics[width=0.93\textwidth]{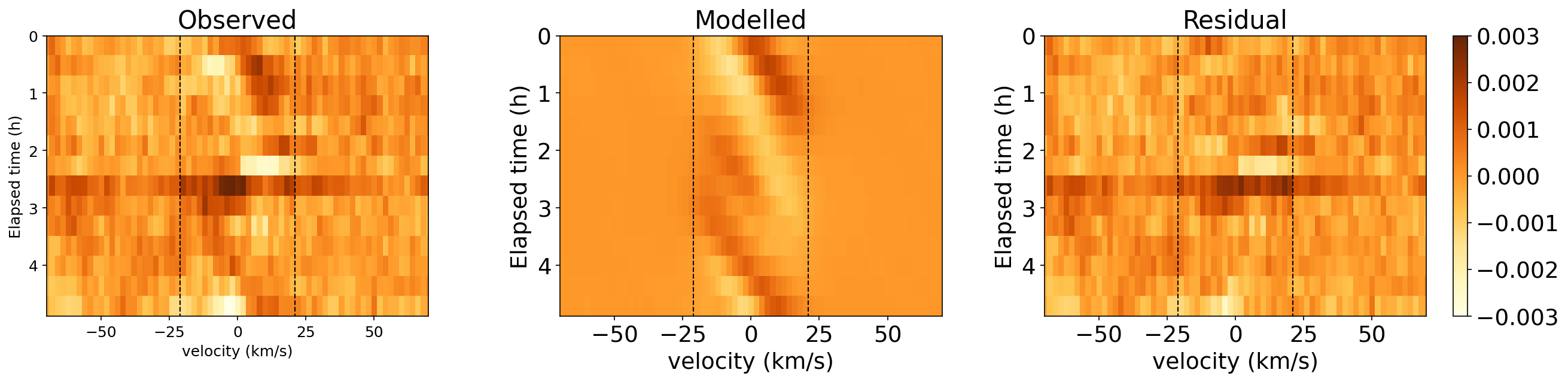}
\end{minipage}
\caption{Same as Fig. \ref{fig:residual_fit}, but for WISE 1049A K band on the night of Feb 9.}
\end{figure*}

\begin{figure*}
\centering
\includegraphics[width=0.95\textwidth]{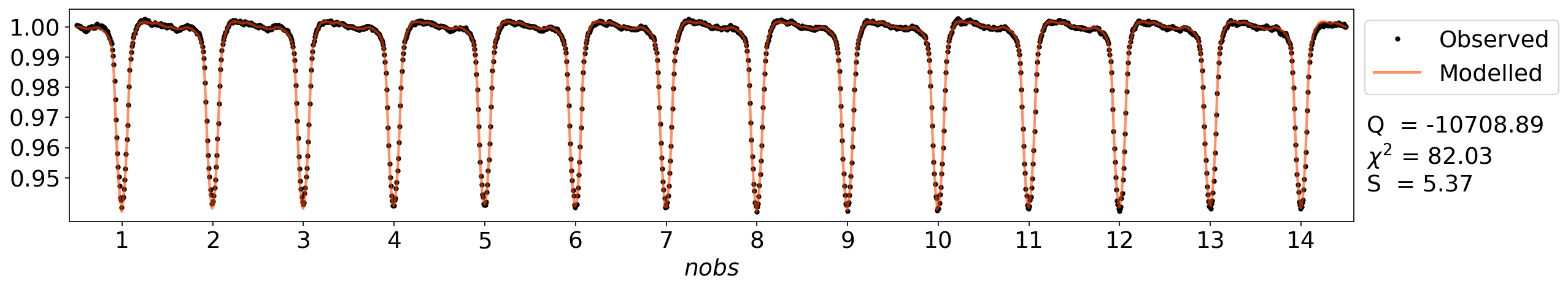}
\begin{minipage}{\textwidth}\centering
\vspace{10pt}
\includegraphics[width=0.93\textwidth]{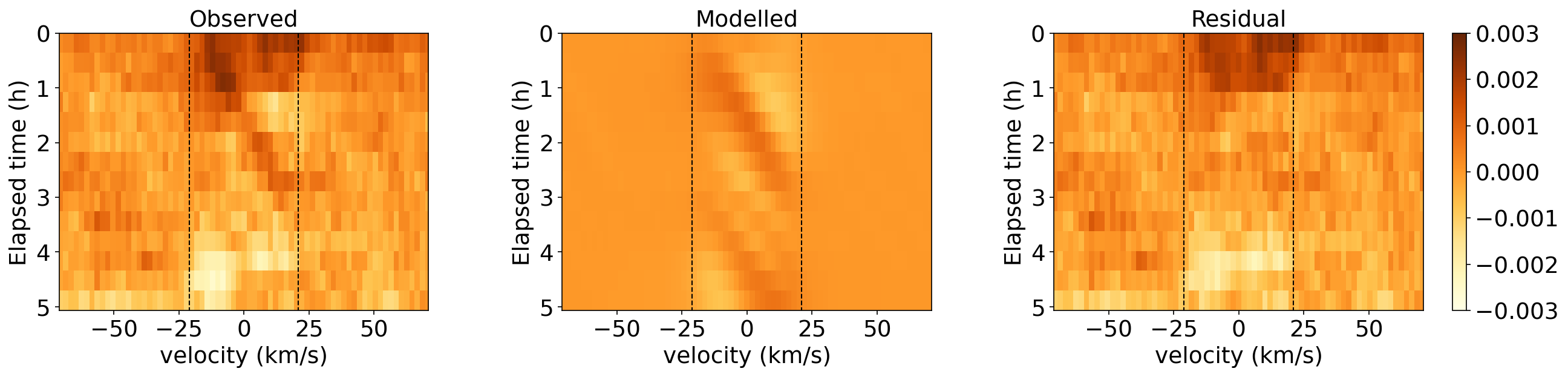}
\end{minipage}
\caption{Same as Fig. \ref{fig:residual_fit}, but for WISE 1049A H band on the night of Feb 11.}
\end{figure*}

\begin{figure*}
\centering
\includegraphics[width=0.95\textwidth]{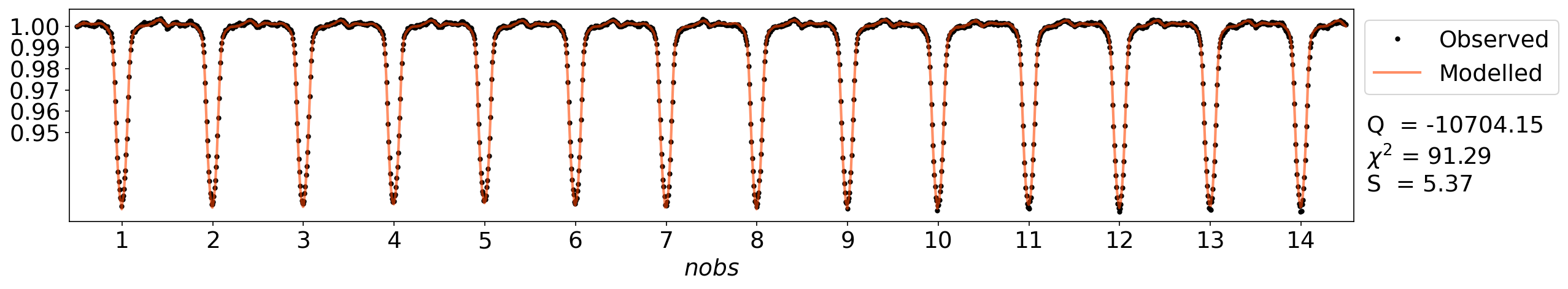}
\begin{minipage}{\textwidth}\centering
\vspace{10pt}
\includegraphics[width=0.93\textwidth]{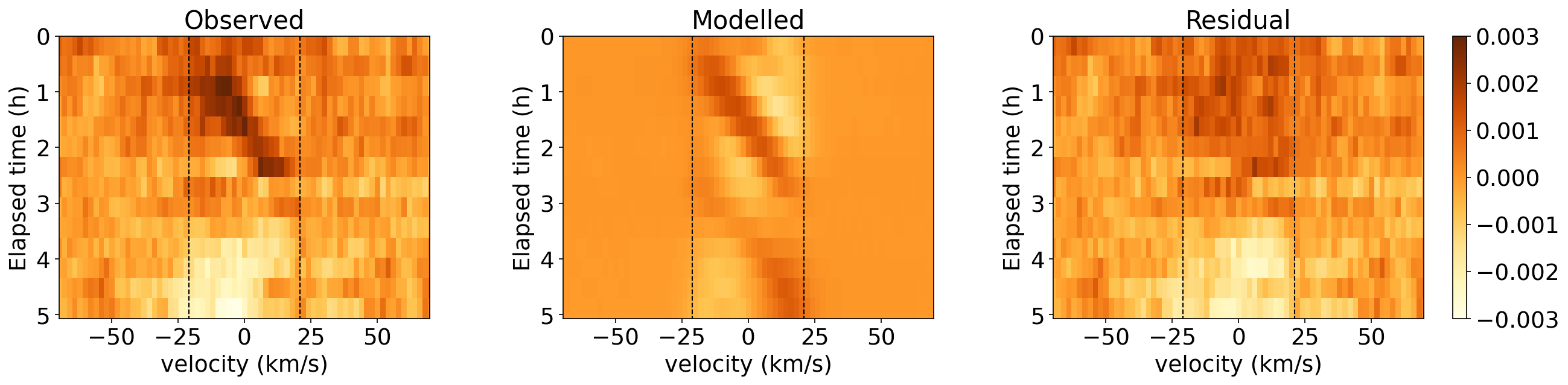}
\end{minipage}
\caption{Same as Fig. \ref{fig:residual_fit}, but for WISE 1049A K band on the night of Feb 11.}
\end{figure*}

\begin{figure*}
\centering
\includegraphics[width=0.95\textwidth]{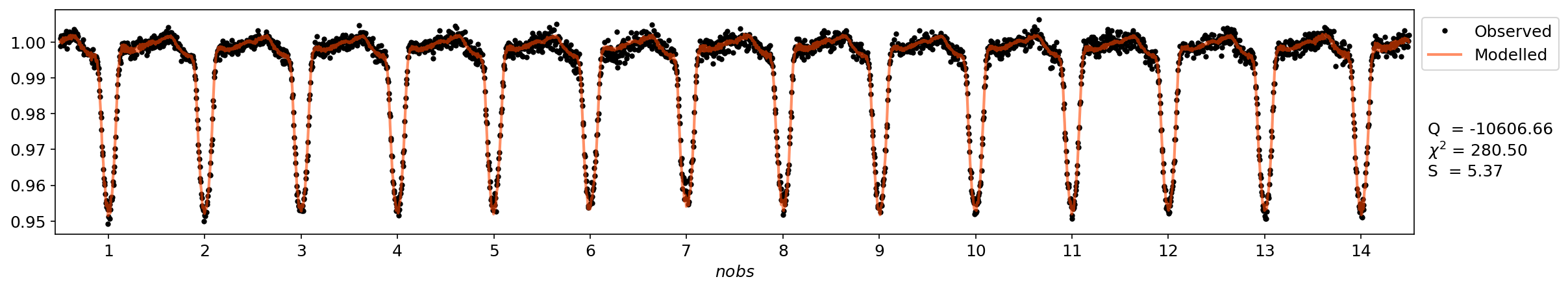}
\begin{minipage}{\textwidth}\centering
\vspace{10pt}
\includegraphics[width=0.93\textwidth]{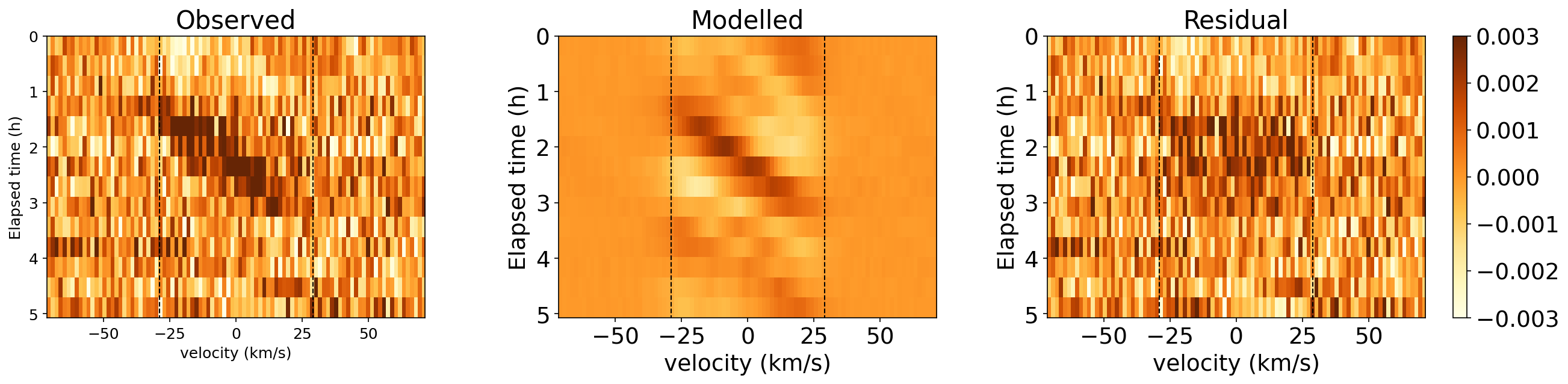}
\end{minipage}
\caption{Same as Fig. \ref{fig:residual_fit}, but for WISE 1049B CRIRES K band data from \citet{Crossfield2014}.}
\end{figure*}

%%%%%%%%%%%%%%%%%%%%%%%%%%%%%%%%%%%%%%%%%%%%%%%%%%
\section{\textsc{starry} Doppler maps}

\begin{figure*}
\centering
\includegraphics[width=\textwidth]{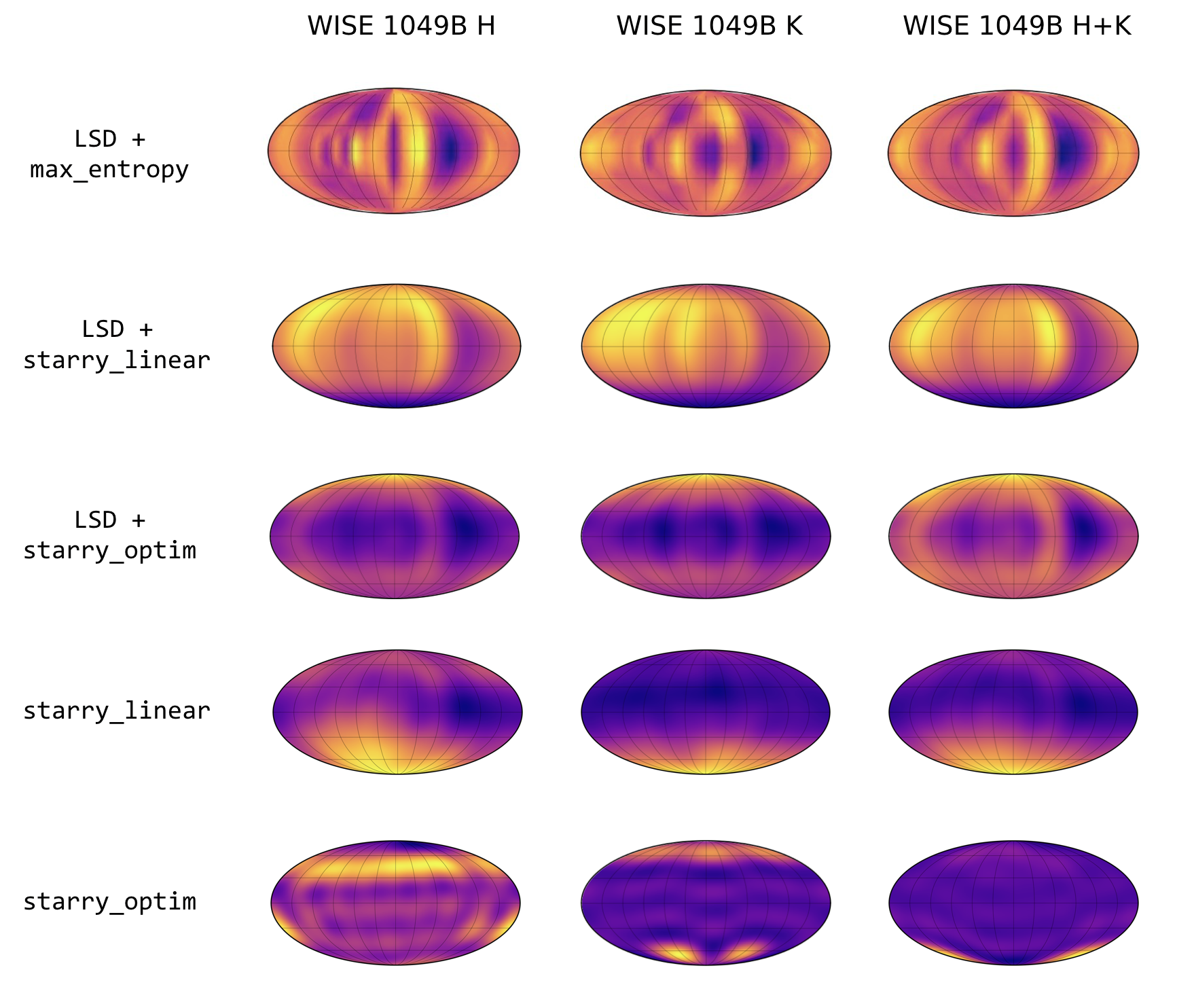}
\caption{WISE 1049B maps retrieved by the 4 different \texttt{starry} solving strategies: 1) LSD + linear solver, 2) LSD + optimization, 3) direct linear solver, and 4) direct optimization. The maximum entropy maps presented in Section \ref{sec:dime} are also shown on the top row for comparison. Overall, all 5 solvers successfully capture a prominent dark spot in the maps, except for the last one, i.e., directly \textsc{starry} optimization without LSD pre-processing, which didn't converge to a good solution. 
The longitude and size of the recovered dark feature are generally consistent among the methods, although the overall brightness contrast varies.}
\label{fig:solvers}
\end{figure*}

%%%%%%%%%%%%%%%%%%%%%%%%%%%%%%%%%%%%%%%%%%%%%%%%%%

% Don't change these lines
\bsp	% typesetting comment
\label{lastpage}
\end{document}